\documentclass[a4paper,11pt]{article}
\pdfoutput=1 
\usepackage{graphicx}

\usepackage{aas_macros} 
\usepackage{jcappub} 
\usepackage{empheq}
\usepackage{subcaption}
\usepackage{amsmath}
\usepackage{upgreek}
\usepackage{booktabs}
\usepackage{siunitx}

\usepackage{makecell}
\usepackage{tikz} 

\usepackage[T1]{fontenc} 

\usepackage{comment}
\usepackage{todonotes} 

\definecolor{deepgreen}{rgb}{0.2,0.8,0.2}

\definecolor{deepblue}{rgb}{0.2,0.2,0.8}

\definecolor{deepred}{rgb}{0.8,0.2,0.2}

\newcommand{\Sec}[1]{section~\ref{#1}}
\newcommand{\Secs}[1]{sections~\ref{#1}}
\newcommand{\Eq}[1]{eq.~\ref{#1}}
\newcommand{\Eqs}[1]{eqs.~\ref{#1}}
\newcommand{\Fig}[1]{figure~\ref{#1}}

\usepackage{subfiles}

\newcommand{\di}{\text{d}}
\newcommand{\pare}[1]{\left(#1\right)}
\newcommand{\parea}[1]{\left[#1\right]}
\newcommand{\avg}[1]{\left<#1\right>}

\newcommand{\bth}{\boldsymbol{\theta}}

\newcommand{\phiab}{\tilde{\phi}^\text{inst}}
\newcommand{\bea}{\begin{align}}
\newcommand{\eea}{\end{align}}

\newcommand{\bx}{\bold{x}}

\newcommand{\lb}{\ell_\text{b}}
\newcommand{\fc}{f_\text{c}}

\newcommand{\estim}[2]{\left(\frac{#1}{#2}\right)}

\newcommand{\vect}[1]{\boldsymbol{\mathbf{#1}}}
\newcommand{\dd}{\mathrm{d}}

\title{Extended-Path Intensity Correlation: Microarcsecond Astrometry with an Arcsecond Field of View}

\author[a]{Marios Galanis,}
\emailAdd{mgalanis@perimeterinstitute.ca}
\affiliation[a]{Perimeter Institute for Theoretical Physics, Waterloo, Ontario N2L 2Y5, Canada}

\author[b,c]{Ken Van Tilburg,}
\emailAdd{kenvt@nyu.edu}
\affiliation[b]{Center for Cosmology and Particle Physics, Department of Physics, New York University,
New York, NY 10003, USA}
\affiliation[c]{Center for Computational Astrophysics, Flatiron Institute, New York, NY 10010, USA}

\author[d]{Masha Baryakhtar,}
\emailAdd{mbaryakh@uw.edu}
\affiliation[d]{Department of Physics, University of Washington, Seattle WA 98195, USA}

\author[b]{and Neal Weiner}
\emailAdd{neal.weiner@nyu.edu }

\abstract{
We develop in detail a recently proposed optical-path modification of astronomical intensity interferometers. Extended-Path Intensity Correlation (EPIC)~\cite{shortpaper} introduces a tunable path extension, enabling differential astrometry of multiple compact sources such as stars and quasars at separations of up to a few arcseconds. Combined with other recent technological advances in spectroscopy and fast single-photon detection, a ground-based intensity interferometer array can achieve microarcsecond resolution and even better light-centroiding accuracy on bright sources of magnitude $m \lesssim 15$. We lay out the theory and technical requirements of EPIC, and discuss the scientific potential. Promising applications include astrometric lensing of stars and quasar images, binary-orbit characterization, exoplanet detection, Galactic acceleration measurements and calibration of the cosmic distance ladder. The introduction of the path extension thus significantly increases the scope of intensity interferometry while reaching unprecedented levels of relative astrometric precision. 
}

\begin{document}
\maketitle

\flushbottom

\section{Introduction}
\label{sec:intro}

Astrometry is the oldest science: the motions of celestial objects across the sky have been used for millennia to understand the cosmos and our place in it. In the present day, precision measurements of stellar velocities in our Galaxy and Local Group have provided invaluable information on galactic dynamics and structure formation. Relative velocities of stars in the central region of the Milky Way unveiled the supermassive black hole Sagittarius A*, while stellar parallax established the first rungs of the distance ladder used to discover the accelerated expansion of the Universe. In recent years, advances in astrometry have made possible the identification of isolated black holes~\cite{OGLE:2022gdj,2022ApJ...933L..23L,2022ApJ...937L..24M} and exoplanets~\cite{2022arXiv220605595G}, discoveries of new stellar substructures~\cite{Bonaca:2018fek,Necib:2019zka,2020ARA&A..58..205H}, a clearer picture of the formation and evolution of the Milky Way~\cite{Helmi:2020otr}, stringent tests of gravity~\cite{anderson1978tests,fomalont2009progress,crosta2006microarcsecond,2018A&A...618L..10G}, and searches for time-domain lensing by small dark matter halos and compact objects~\cite{van2018halometry}, to name but a few applications.
Future measurements may reveal the gravitational potential of the Milky Way~\cite{buschmann2021galactic}, constraints on dark matter substructure~\cite{Mondino:2020rkn}, and low-frequency gravitational waves~\cite{pyne1995gravitational,book2011astrometric,klioner2018gaia,moore2017astrometric}. 
These applications are enabled by the precision astrometry of space-based telescopes such as \textit{Hipparcos}~\cite{2000A&A...355L..27H}, \textit{Gaia}~\cite{prusti2016gaia}, \textit{Hubble Space Telescope (HST)}~\cite{1991PASP..103..317B}, and \textit{James Webb Space Telescope (JWST)}~\cite{2006SSRv..123..485G}, and of ground-based amplitude interferometers in the optical and radio bands such as GRAVITY at VLT~\cite{abuter2017first}, the Center for High Angular Resolution Astronomy (CHARA)~\cite{2005ApJ...628..453T}, and the Event Horizon Telescope (EHT)~\cite{EventHorizonTelescope:2019dse,EventHorizonTelescope:2019uob}, which have reached light-centroiding precisions on the order of tens of microarcseconds.

These observatories are all contending with a practical obstacle: diffraction-limited resolution, approximately equal to the wavelength of light divided by the observatory's effective size. For imaging telescopes, this size is the aperture diameter, whereas for interferometers it is the baseline distance: the separation between telescopes or receivers. 
Imaging telescopes can achieve a precision on the light centroid of a point source that far surpasses the angular resolution limit, as long as the exposure is sufficiently long, the point spread function (PSF) is stable, and other systematics are kept under control. 
In very long baseline interferometry (VLBI) in the radio band (up to millimeter wavelengths), receivers record the amplitude and phase of the electromagnetic field at different corners of the globe~\cite{EventHorizonTelescope:2019uob}. The data is then correlated offline, yielding information at an angular scale of tens of microarcseconds. The size of ground-based VLBI is already maximized, while shorter wavelengths present their own difficulties. Astronomical interferometers in the infrared and optical bands have to physically recombine the light from separate telescopes (since receiver electronics cannot keep track of the rapidly oscillating phase) and keep their separations stable to a fraction of wavelength, limiting their baselines to hundreds of meters at most~\cite{2003SPIE.4838...79C,2005ApJ...628..453T}.

Intensity interferometry combines the best of both worlds---short wavelengths yet long baselines (with milder tolerance) and off-line correlation. The technique relies on second-order coherence: interference of the intensities, rather than the amplitudes, of incoming light, so only the photon arrival times have to be recorded and correlated. It was conceived theoretically by Hanbury Brown and Twiss in 1954~\cite{1954PMag...45..663B}, who only two years later demonstrated the effect experimentally in the laboratory, using light from a mercury arc lamp~\cite{1956Natur.177...27B}. Later in 1956, with a rudimentary setup and poor viewing conditions, Hanbury Brown used intensity interferometry to provide the first measurement of the $0.0063\,\mathrm{arcsec}$ angular diameter of Sirius~\cite{1956Natur.178.1046H}, a spectacular initiation of a new method in observational astronomy. The program was then extended~\cite{hbtI,hbtII,hbtIII,hbtIV}, and culminated with the Narrabri Stellar Intensity Interferometer (NSII)~\cite{1967MNRAS.137..375H,1967MNRAS.137..393H,1974iiaa.book.....B}. In the mid-1970s, the NSII reported the angular diameters of 32 bright stars down to $30\,\mathrm{\mu as}$ precision~\cite{1974MNRAS.167..121H,1976ApJ...203..417C}, as well as orbital parameters, masses, and line-of-sight distance of a close stellar binary system~\cite{1971MNRAS.151..161H}. The advent of ever larger arrays of Imaging Atmospheric Cherenkov Telescopes (IACTs)~\cite{weekes2002veritas,hinton2004status,baixeras2004commissioning,anderhub2013design,cta2011design} to detect high-energy gamma rays has spurred a recent revival of intensity interferometry~\cite{nunez2012imaging,2019arXiv190803164K,2014JKAS...47..235T,nunez2015capabilities,dravins2016intensity,matthews2019astrophysical,abeysekara2020demonstration,kieda2021veritas,zampieri2021stellar,2012NewAR..56..143D}; in addition new fast photodetector technology has enabled the development of smaller telescope arrays with competitive sensitivity~\cite{horch2022observations}.

The advantages of intensity interferometry are unfortunately paired with two significant drawbacks. The first is the signal-to-noise ratio (SNR), as intensity correlations only receive contributions from ``nearly coincident'' photons in at least two receivers.  Since the intensity of starlight typically fluctuates on femtosecond time scales, the much longer timing resolution of photon counters smears out the intensity correlations and reduces the fringe contrast; the NSII measurements were limited to very bright stars of magnitude below 3. The second limitation of traditional intensity interferometry is the extremely narrow field of view (FOV), which is naturally of the same parametric size as the angular resolution itself. The combination of the small SNR and FOV have limited astronomical intensity interferometry almost exclusively to studies of the emission morphology of isolated bright stars, and a few very closely separated multiple-star systems~\cite{1971MNRAS.151..161H,2018MNRAS.480..245G}. 

\begin{figure}
\centering
\includegraphics[width=0.8\textwidth]{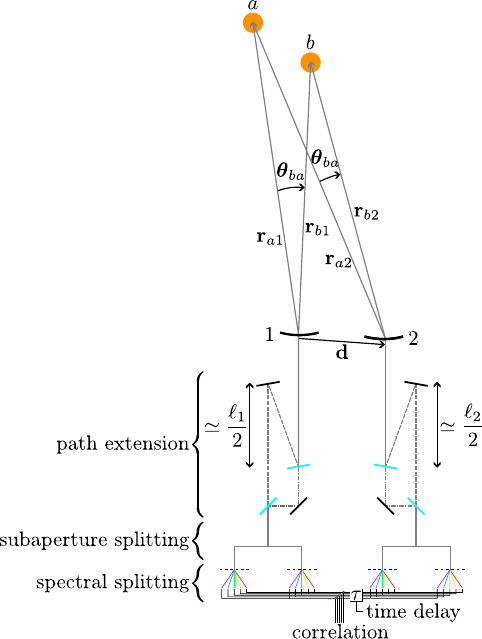}
\caption{Diagram of an extended-path intensity correlator (EPIC). Two sources $i = a,b$ separated by an angle $\vect{\theta}_{ba} = \hat{\vect{\theta}}_b - \hat{\vect{\theta}}_a$ are simultaneously observed by two telescopes $p = 1,2$; the respective distances are labeled as $\vect{r}_{ip}$. In each telescope, the light from both sources is collected and collimated into the ``path extension'' stage, where a beamsplitter directs the light along two paths, one longer than the other by a variable amount $\ell_p$, and a second beamsplitter recombines the beam. The light is then  split into multiple subapertures, and a dispersive element spectrally splits the light. Ultrafast single-photon detectors record the arrival times of photons in each spectral channel, which are ultimately correlated offline after applying a time offset $\tau$. The intensity correlations provide information on the separation $\vect{\theta}_{ba}$ of the two sources as well as their angular sizes.}\label{fig:basics}
\end{figure}

In this work, we detail the \emph{Extended-Path Intensity Correlation} (EPIC) observatory design which addresses both the low SNR and small FOV drawbacks of intensity interferometry~\cite{shortpaper}; a schematic of the design is shown in \Fig{fig:basics}. We outline how recent technological advances in multichannel spectroscopic readout and fast single photon detection open up intensity interferometry to much fainter sources. The advantages of a multichannel spectroscopic readout were already known by Hanbury Brown~\cite{1974iiaa.book.....B}, and its pairing with modern photodetectors was laid out clearly in ref.~\cite{2014JKAS...47..235T}, which argued that one could achieve limiting R-band magnitudes of about 14. Spectroscopic splitting along with the introduction of beamsplitters to improve the sensitivity and information gathered from the telescopes was also proposed in \cite{Stankus:2020hbc,Chen:2022ccn}. Here, we summarize further recent improvements in photodetection technology and provide a detailed quantitative description of a multichannel spectroscopic readout in the context of intensity interferometry.

To overcome the very narrow field of view, EPIC uses a tunable hardware modification to the constituent telescopes of an astronomical intensity interferometer array. This setup requires a pair of adjustable beamsplitters and mirrors that can add a differential path extension to the incoming light, thus superposing intensity fluctuations from sources at different locations on the celestial sphere~\cite{shortpaper}.\footnote{A related technique called ``time-delay interferometry'' has been proposed previously in the context of gravitational radiation sensing with (amplitude) interferometric detectors of unequal arm lengths, where a path extension (i.e.~geometric time delay) is introduced to eliminate laser phase noise~\cite{tinto2014time}. Likewise, ref.~\cite{franson1989bell} proposed to test quantum mechanics in photon coincidence experiments with a logically similar optical setup.} The resulting variable path extension (and wavefront correction) dramatically widens the FOV of ground-based intensity interferometry, up to a few arcseconds, the isoplanatic angle of the atmosphere. This alteration would drastically increase the scope of intensity interferometry, enabling differential astrometry at an unprecedented precision and accuracy for a wide range of compact celestial light sources.

In \Sec{sec:theory}, we review the basic theory of intensity interferometry (\Sec{sec:hbt_review}), describe the addition of the optical path extension (\Sec{sec:mirror}), discuss typical celestial sources of finite angular extent (\Sec{sec:finite}), calculate the idealized signal-to-noise ratio and sensitivity (\Sec{sec:snr}), and calculate field of view limitations from atmospheric aberrations (\Sec{sec:atm}). We outline the technical requirements in \Sec{sec:tech}, including telescope design and array layout (\Sec{sec:design}), observational procedure (\Sec{sec:observation}), photodetection (\Sec{sec:photodetectors}), dispersive elements for spectroscopy (\Sec{sec:grating}), and optical tolerances (\Sec{sec:opt_tolerances}). We propose a sample of scientific applications with EPIC in \Sec{sec:applications}: binary orbits~(\Sec{sec:binaries}), exoplanets~(\Sec{sec:exoplanets}), Galactic accelerations~(\Sec{sec:acceleration}), stellar microlensing~(\Sec{sec:stellar_microlensing}), and strongly lensed quasars~(\Sec{sec:quasar_microlensing}). We conclude in \Sec{sec:conclusions}. The appendices contain a list of the symbols used throughout the text (appendix~\ref{app:notation}), details of the derivations summarized in the text (appendix~\ref{app:derivations}), and a wave-optics treatment of our setup's optical elements (appendix~\ref{app:waveoptics}).

\section{Theory}
\label{sec:theory}

The basic intensity interferometry setup consists of at least two telescopes collecting photons---and precisely recording their arrival times---from a small region of the sky. This is to be contrasted with \emph{amplitude} interferometers, which record the electric field amplitudes as a function of time at a pair of receivers separated by a baseline $\vect{d}$. The two-point correlation function of  electric field amplitudes is proportional to the ``complex visibility'': the Fourier transform of the time-averaged intensity distribution of the image at angular wavenumber $\overline{k} \vect{d}_\perp$ conventionally decomposed into $(u,v)$ coordinates, where $\overline{k}$ is the mean wavenumber of the electromagnetic wave, and $\vect{d}_\perp$ is the baseline vector projection perpendicular to the line of sight. Through photon collection, an \emph{intensity} interferometer samples intensities (the square of the field amplitudes) at separate locations and compares their fluctuations. The excess correlation of the intensities $C(\vect{d},\tau)$ at points separated by distance $\vect{d}$ and time $\tau$ is proportional to a specific four-point function of the electromagnetic field: the square of the modulus of the complex visibility at angular wavenumber $\overline{k} \vect{d}_\perp$. The overall phase information is lost, which drastically degrades \emph{global} astrometric precision, but information about relative phases is retained, thus providing excellent \emph{differential} astrometry.

\begin{figure}
	\centering
        \includegraphics[width=0.48\textwidth]{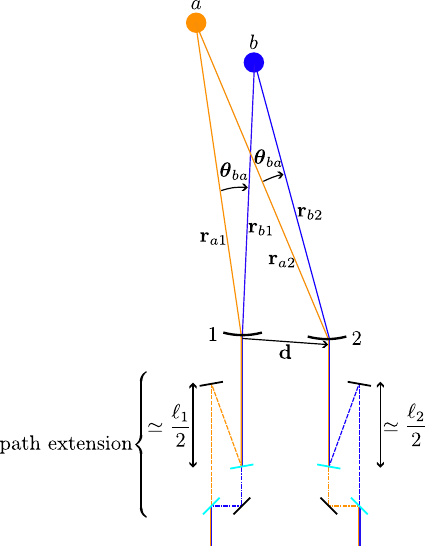}
	\includegraphics[width=0.48\textwidth]{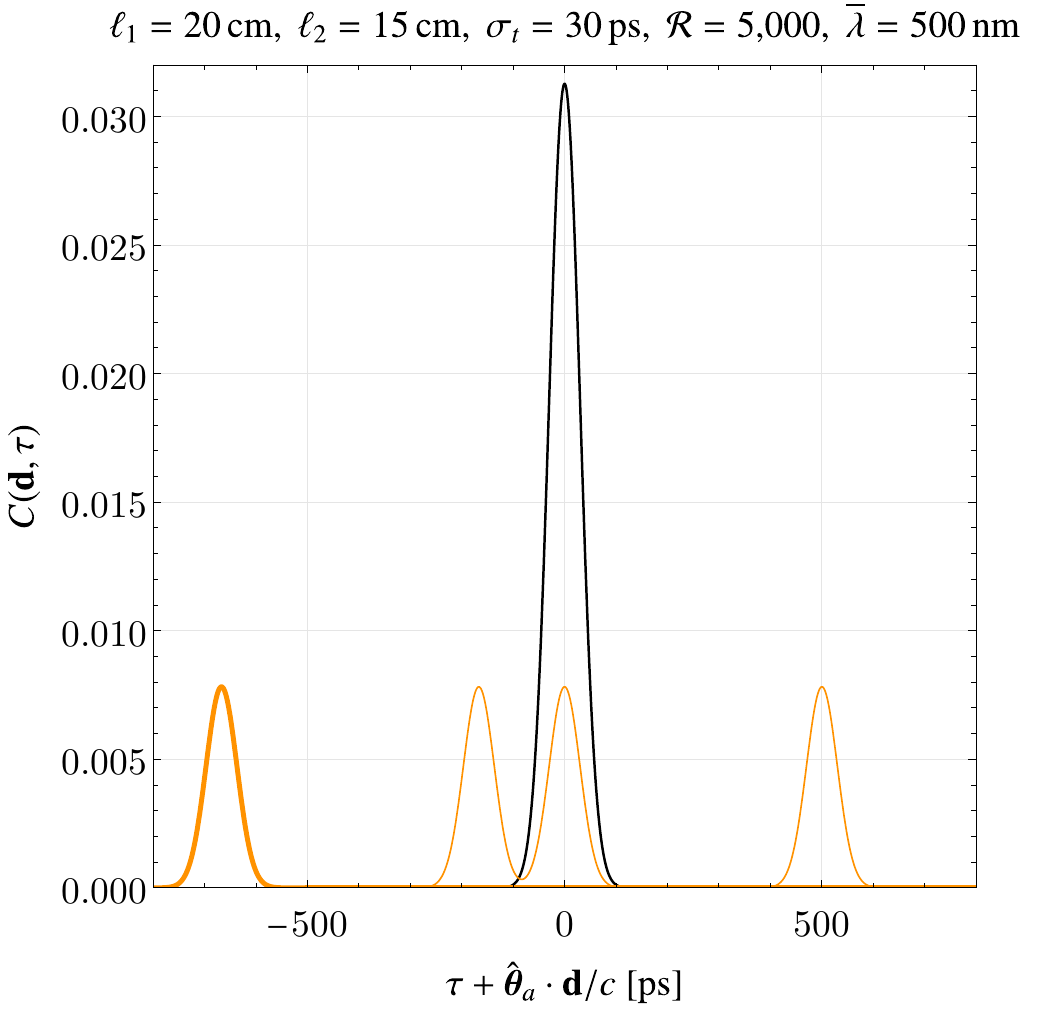}
	\caption{{\it Left:}~Same as figure~\ref{fig:basics}, but with the two-photon amplitude corresponding to the fringe in \Eqs{eq:ellopt1} and~\ref{eq:ellopt2} color coded in the path extension stage: orange corresponds to the amplitude from source $a$, blue from source $b$. The path extensions are tuned such that $\ell_1 + \ell_2 = (r_{a2} + r_{b1}) - (r_{a1} + r_{b2} ) = \vect{\theta}_{ba} \cdot \vect{d}$.
	~{\it Right:}~Excess fractional intensity correlation $C(\vect{d},\tau)$ as a function timing offset $\tau$ for a single point-like source at celestial position $\hat{\vect{\theta}}_a$, for timing resolution $\sigma_t = 30 \, \text{ps}$, and central wavenumber $\overline{k} = 2\pi / 500 \, \text{nm}$ and spectral resolution $\mathcal{R} \equiv \overline{k}/\sigma_k = 5{,}000$. The black curve shows the intensity fringe of \Eq{eq:C1pointsource} for an intensity interferometer without path extension, while the orange curve shows the four fringes of \Eq{eq:C2pointsource} for an \emph{extended-path} intensity correlator. The path extension retains the original main fringe at 1/4 the contrast, but also produces three equally bright ghost fringes, as if there were 3 other images located at $\hat{\vect{\theta}}_a + \frac{\ell_1}{d} \hat{\vect{d}} $, $\hat{\vect{\theta}}_a - \frac{\ell_2}{d} \hat{\vect{d}} $, $\hat{\vect{\theta}}_a - \frac{\ell_2-\ell_1}{d} \hat{\vect{d}} $, respectively. The fringe corresponding to the paths from source $a$ shown in the {\it left} panel is the leftmost fringe highlighted in bold.}\label{fig:C1}
\end{figure}

Our main new idea is to enlarge the scope (literally and figuratively) of intensity interferometry by introducing a variable, probabilistic path extension inside each constituent telescope of the intensity interferometer array while keeping the same angular resolution and light-centroiding precision. This is done by introducing additional adjustable optical elements as shown schematically in figure~\ref{fig:C1}~{\it(left)}. In a conventional intensity interferometer, the excess intensity correlation  between a pair of telescopes for a point-like source at sky location $\hat{\vect{\theta}}_a$ occurs near a time offset of $\tau = - \vect{d} \cdot \hat{\vect{\theta}}_a / c$ with $c$ the speed of light, depicted in figure~\ref{fig:C1}~{\it(right)} by the black line. In an Extended-Path Intensity Correlator (EPIC), the path extension stage at telescope $p$ splits the beam into two, one geometrically longer than the other by an amount $\ell_p$. The intensity correlation fringe is thus split into $2^2$ fringes, each of $1/4$ the fringe contrast, depicted as orange lines in figure~\ref{fig:C1}~{\it(right)}: the ``original'' fringe at $c \tau = - \vect{d} \cdot \hat{\vect{\theta}}_a $, one only delayed in telescope 1 at $c \tau = - \vect{d} \cdot \hat{\vect{\theta}}_a - \ell_1$, one only delayed in telescope 2 at $c \tau = - \vect{d} \cdot \hat{\vect{\theta}}_a + \ell_2$, and one delayed in both at $c \tau = - \vect{d} \cdot \hat{\vect{\theta}}_a + \ell_2 - \ell_1$. This procedure can be used to ``superpose'' the intensity correlation fringes of two sources $a$ and $b$, at locations $\hat{\vect{\theta}}_a$ and $\hat{\vect{\theta}}_b$, to measure their relative angular separation $\vect{\theta}_{ba} \equiv \hat{\vect{\theta}}_b- \hat{\vect{\theta}}_a$ with extreme precision. For example, the intensity correlation fringe of source $a$ delayed by $\ell_1$ in telescope 1, and that of source $b$ delayed by $\ell_2$ in telescope 2, will coincide (and yield mutual intensity interference) when:
\begin{alignat}{2}
\boxed{\ell_1 + \ell_2 = \vect{\theta}_{ba} \cdot \vect{d}.} \label{eq:path_extension}
\end{alignat}
Other permutations of fringes are also possible. In this way, the maximum angular separation for intensity interferometry can be enlarged up to the $\mathcal{O}(\mathrm{arcsec})$ isoplanatic angle determined by atmospheric aberrations while retaining exquisite angular resolution. Hence, the dynamic range (i.e.~angular separation divided by resolution) of EPIC is many orders of magnitude larger than that of traditional intensity interferometry. 

In \Sec{sec:hbt_review}, we present a classical description of intensity interferometry in idealized scenarios, based on but significantly extending the review of ref.~\cite{BaymNotes}. More sophisticated treatments, including a quantum description and the first- and second-order coherence properties of light, can be found e.g.~in ref.~\cite{mandel1995optical}. For a detailed review on optical stellar interferometry, see e.g.~ref.~\cite{labeyrie2006introduction}; for recent developments see e.g.~refs.~\cite{karl2022comparing,bojer2022quantitative}. In \Sec{sec:mirror}, we describe the path extension theory of EPIC. We review basic properties of the most relevant astrophysical sources in \Sec{sec:finite} in order to estimate the signal-to-noise ratio and light-centroiding precision in \Sec{sec:snr}. We calculate the atmospheric limitations on EPIC in \Sec{sec:atm}.

\subsection{Intensity interferometry review}
\label{sec:hbt_review}
\subsubsection{General}
Suppose we have a single-photon detector recording a time series of photon counts impinging on a telescope. 
We will assume the light in the telescope is a superposition of a very large number $N_\theta$ of discrete emitters $i$, each emitting a large number $N_k$ of modes  $\alpha$ with random phases $\phi^\text{em}_{i\alpha}$, giving an electric field at detector $p$,
\begin{align}
E_p(t_p) \equiv \sum_{i=1}^{N_\theta} \sum_{\alpha = 1}^{N_k} E_{i\alpha} \exp\left\lbrace i \left[k_\alpha \left(c t_p - r_{ip}\right) + \phi^\text{em}_{i\alpha} +\widetilde{\phi}^{(p)}_{i\alpha}(t_p) \right] \right\rbrace \equiv \sum_{i,\alpha} \widetilde{E}_{i\alpha}^{(p)}
\label{eq:efield}
\end{align}
at a time $t_p$ kept by the clock associated with each photodetector $p$. In the above, $k_\alpha$ is the wavenumber of each mode $\alpha$;  $r_{ip} = |\vect{r}_{ip}|$ is the length of the vector $\vect{r}_{ip} \equiv  \vect{r}_i - \vect{r}_p$ pointing from the photodetector $p$ to the emitter $i$. Additional phases that may accumulate along the line of sight due to atmospheric fluctuations and telescope imperfections are denoted by $\widetilde{\phi}^{(p)}_{i\alpha}(t_p)$ but will be suppressed until \Sec{sec:atm} and \Sec{sec:opt_tolerances}. As will be discussed in \Secs{sec:atm} and~\ref{sec:tech}, these path-dependent phases will ultimately limit the field of view of the intensity interferometry setup and place constraints on the telescope quality. Without loss of generality, the field amplitude $E_{i\alpha}$ associated with each mode can be taken to be real; we defined the complex amplitudes $\widetilde{E}_{i\alpha}^{(p)}$ in the last part of \Eq{eq:efield} to simplify subsequent notation. Finally, we treat the field amplitudes as scalars for clarity of exposition; for unpolarized sources, intensity correlations signals are reduced by a factor of $1/2$, which will be accounted for in \Sec{sec:snr}.

Throughout this work, we will assume that the detector separations are much smaller than the (transversely projected) separations of the sources, which are in turn taken to be much smaller than the line-of-sight distances (small angle approximation). This implies that the angular separation of two sources as observed from vantage point $p$,
\begin{align}
\vect{\theta}_{ij} \equiv \hat{\vect{r}}_{ip} -  \hat{\vect{r}}_{jp} \simeq \hat{\vect{\theta}}_i - \hat{\vect{\theta}}_j,
\end{align}
is independent of $p$, and that $|\vect{\theta}_{ij}| \ll 1$. Note that this also implies that electric field magnitude $E_{i\alpha}$ is independent of $p$ for each mode. See figure~\ref{fig:basics} for the basic geometry in the case of just two sources $i = a,b$.

\begin{figure}
	\centering
	\includegraphics[width=1\textwidth, trim = 40 5 58 10, clip]{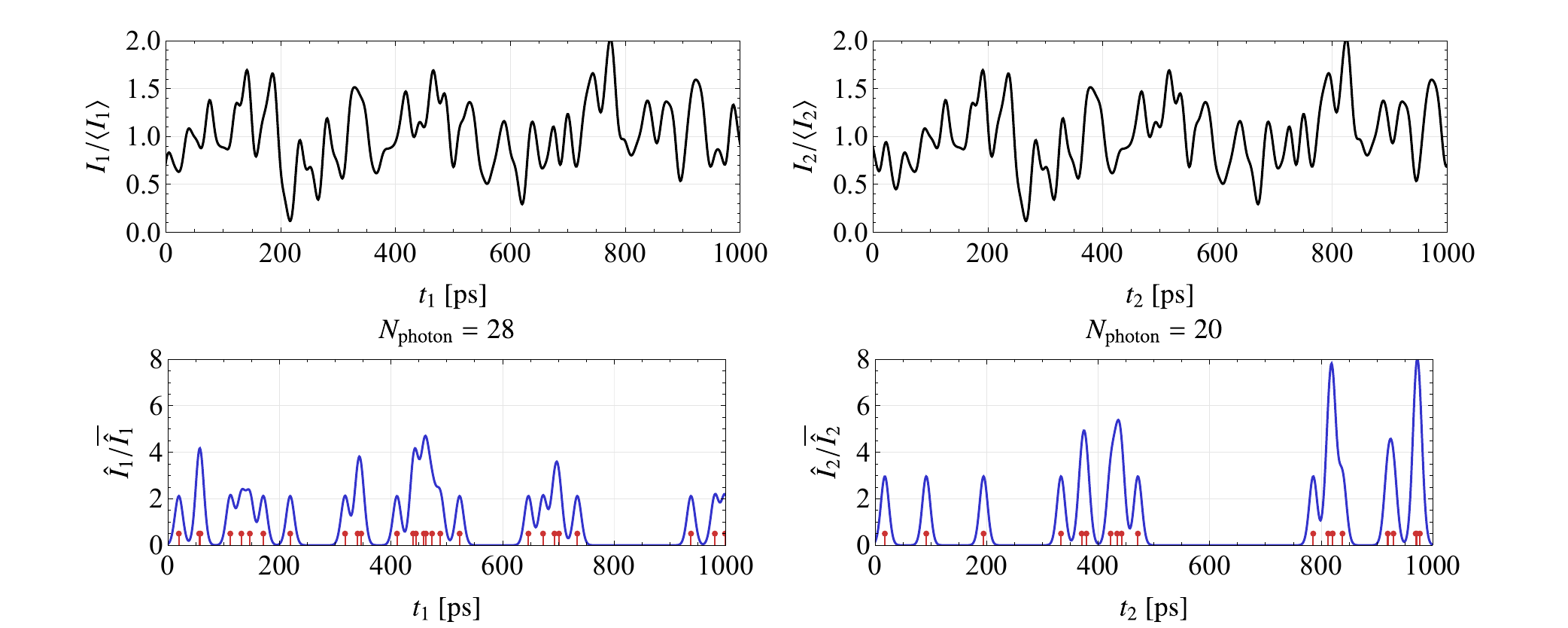}
	\includegraphics[width=1.1\textwidth, trim = 80 0 0 0, clip]{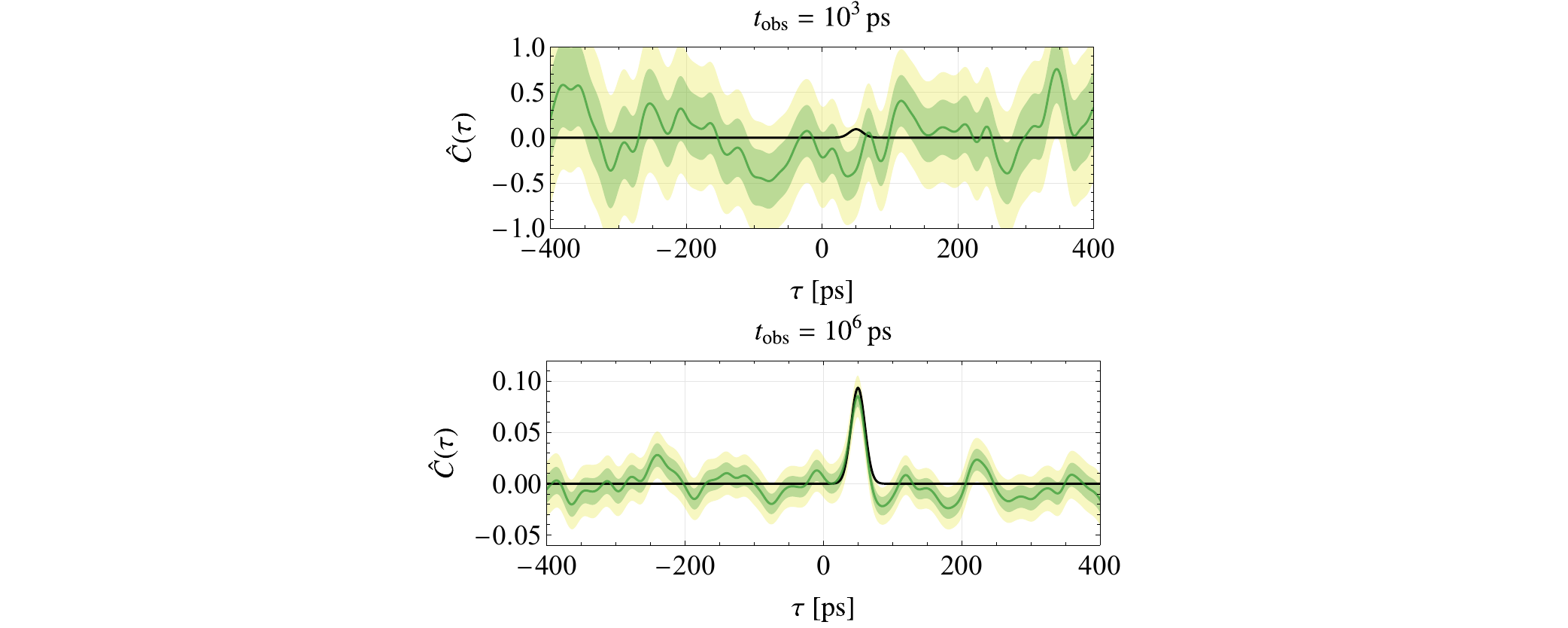}
	\caption{
		\textit{First row:} Instantaneous intensity $I_p$ normalized to the true mean intensity $\langle I_p \rangle$ from a single source as observed by two telescopes labeled $p=1,2$ (\textit{left}, \textit{right}), for light with a Gaussian spectrum with mean wavenumber $\overline{k} = 2\pi / 500\, \mathrm{nm}$ and standard deviation $\sigma_k = \overline{k}/5000$. The true instantaneous intensity has order-unity fractional fluctuations, a coherence time of order $1/c\sigma_k$, and is \emph{identical} in both telescopes up to a time shift of $t_2 - t_1 = 50 \, \mathrm{ps}$ due to the source's wavefronts arriving earlier in telescope 1.
		\textit{Second row:} Photon arrival times (red sticks) in telescope 1 \textit{(left)} and 2 \textit{(right)} as a function of time. The photon detection probability is directly proportional to the instantaneous intensity from the top panel(s). We plot in blue the estimated intensities $\hat{I}_p$ (eq.~\ref{eq:Ihat1}) divided by their estimated time averages $\overline{\hat{I}_p}$ (cfr.~eq.~\ref{eq:averaging}), assuming each telescope has a timing resolution of $\sigma_t / \sqrt{2}$ with $\sigma_t = 10\, \mathrm{ps}$.
		\textit{Third Row:}  Reconstructed excess fractional intensity correlation $\hat{C}$ (green line, eq.~\ref{eq:C2estimate}) as a function of timing offset $\tau = t_2 - t_1$ based on the estimated intensities from the middle panels with an observation time of $t_\mathrm{obs} = 10^3\,\mathrm{ps}$; green and yellow bands indicate $68\%$ and $95\%$ CL intervals. The true excess correlation signal $C = \langle I_1 I_2 \rangle / \langle I_1 \rangle \langle I_2 \rangle  -1$ (black) is too small to be detected. \textit{Fourth row:} Same as in the third row, but for a much longer total observation time $t_\mathrm{obs} = 10^6\,\mathrm{ps}$, whereafter the excess correlation at $\tau = 50\,\mathrm{ps}$ is detected at high significance.
	}\label{fig:time_series}
\end{figure}

The photodetectors record intensities $I_p(t_p) = \frac{1}{2} |E_p(t_p)|^2$, with expected values:
\begin{align}
\avg{I} \equiv \avg{I_p(t_p)}_{\phi^\text{em}} = \frac{1}{2} \sum_{i,\alpha} \left(E_{i\alpha}\right)^2\rightarrow \int \di k \int \di \Omega \, \frac{\di I}{\di k \, \di \Omega}, \label{eq:I1}
\end{align}
where the average is taken over the emitter phases $\phi^\text{em}_{i\alpha}$, leaving only the ``diagonal'' terms. On the right, we have taken the continuum limit, i.e.~$N_\theta, N_k \to \infty$, giving an integral over wavenumbers $k$ and solid angle $\Omega$ over the field of view of the telescope. The expected differential intensity $\frac{\di I}{\di k \, \di \Omega}$ is to be understood as a convolution of the telescope-photodetector response with the source's intrinsic spectrum and image intensity. With our approximations and assumptions of identical photodetectors and telescopes, the intensities are independent of $p$. An illustration of the intensity impinging on a detector with finite time resolution is given in figure~\ref{fig:time_series}.

The intensity correlation between two detectors $p = 1,2$ is:
\begin{align}
\avg{I_1(t_1) I_2(t_2)}_{\phi^\text{em}} &= \frac{1}{4} \sum_{\substack{i_1,j_1\\i_2,j_2}}\sum_{\substack{\alpha_1,\beta_1\\\alpha_2,\beta_2}} \widetilde{E}^{(1)}_{i_1\alpha_1} \widetilde{E}^{(1)*}_{j_1\beta_1}\widetilde{E}^{(2)}_{i_2\alpha_2} \widetilde{E}_{j_2\beta_2}^{(2)*}  \nonumber\\
&= \left(\frac{1}{2} \sum_{i,\alpha} \left(E_{i\alpha}\right)^2 \right)^2 + \left|\frac{1}{2}  \sum_{i,\alpha} \widetilde{E}^{(1)}_{i\alpha} \widetilde{E}^{(2)*}_{i\alpha} \right|^2  - \frac{1}{4} \sum_{i,\alpha} \left(E_{i\alpha}\right)^4 \label{eq:II1} \\
&\to \avg{I_1}\avg{I_2} + \left| \int \di k \int \di \Omega \, \frac{\di I}{\di k \, \di \Omega} e^{i\left[ck(t_1 - t_2) - k\hat{\vect{\theta}} \cdot \vect{d} \right]} \right|^2 .\nonumber
\end{align}
For the first term in the second line, we have kept only ``diagonal'' terms from the first line with $i_1 = j_1$, $i_2 = j_2$, $\alpha_1 = \beta_1$, and $\alpha_2 = \beta_2$. For the second term, we kept ``off-diagonal terms'' with $i_1 = j_2$, $\alpha_1 = \beta_2$, $i_2 = j_1$, and $\alpha_2 = \beta_1$. The third term of the second line subtracts off terms that we overcounted, but this is a negligible contribution in the continuum limit taken in the third line of \Eq{eq:II1}. The second term in the third line can be recognized as the modulus-squared of the complex visibility function. We also introduced the baseline vector $\vect{d}$, leading to a difference in propagation distance $r_{i2}- r_{i1}$ to the two photodetectors:
\begin{align}
\vect{d} \equiv  \vect{r}_2 - \vect{r}_1 = -\vect{r}_{i2} + \vect{r}_{i1} ; \qquad r_{i2}- r_{i1} = - \hat{\vect{\theta}}_i \cdot \vect{d},\label{eq:d}
\end{align}
with the latter relation holding for all sources $i$.

We can write the ``excess fractional intensity correlation'' (relative to random chance) as:
\begin{align}
C(\vect{d},\tau) &\equiv \frac{\avg{I_1(t) {I_2(t+\tau)}}}{\avg{I_1}\avg{I_2}} - 1 \label{eq:C1}  \\
&= \int \frac{\di^2 b_1}{A_1} \int \frac{\di^2 b_2}{A_2} \int\frac{ \di (\Delta t) }{\sqrt{2\pi}\sigma_t} e^{\frac{-(\Delta t)^2}{2\sigma_t^2}}
\times \left| \int \di k \int \di \Omega \, \frac{\di \widetilde{I}}{\di k \, \di \Omega} e^{i\left[k\left(c(\Delta t - \tau) - \hat{\vect{\theta}} \cdot (\vect{d}+\vect{\Delta b})\right)  \right]} \right|^2, \nonumber
\end{align}
with $\di \widetilde{I}  \equiv \di I / \avg{I}$, shown for the example of a single source in figure~\ref{fig:time_series}. In the second line of \Eq{eq:C1}, we have introduced two ``averaging kernels'' over $\vect{b}_1$ and $\vect{b}_2$ that take into account the finite aperture sizes $A_1$ and $A_2$ of the telescopes associated with photodetectors 1 and 2, effectively smearing out the baseline $\vect{d}$ between the centers of the telescopes by an amount $\vect{\Delta b} \equiv \vect{b}_2 - \vect{b}_1$ for each pair of points $\vect{b}_1$ and $\vect{b}_2$ on the respective telescopes' apertures. A third averaging kernel over $\Delta t$ takes into account the relative timing error of the clocks associated with photodetectors $1$ and $2$, around their mean offset of $\tau \equiv t_2 - t_1$. Its timing spread $\sigma_t$ is
\begin{align}
\sigma_t^2 = \sigma_{t_{\text{res},1}}^2 + \sigma_{t_{\text{res},2}}^2 + \sigma_{t_\text{sync}}^2, \label{eq:sigmat}
\end{align}
the (quadrature) sum of the respective timing resolutions and the relative synchronization error $\sigma_{t_\text{sync}}$.

For a sufficiently narrow frequency spectrum around $k = \overline{k}$, and with negligible smearing and phase noise, the excess correlation of \Eq{eq:C1} is proportional to the \emph{square modulus of the Fourier transform of the image}, at an angular wavevector of $k \vect{d}_\perp$, thus providing a fiducial angular resolution
\begin{align}
\sigma_{\theta_\text{res}} \equiv \frac{1}{\overline{k} d} = \frac{\overline{\lambda}}{2\pi d} \approx \underbrace{7.96 \times 10^{-12} \, \text{rad}}_{1.64\,\mu\text{as}}  \left( \frac{\overline{\lambda}}{500\,\text{nm}}\right)  \left( \frac{10\,\text{km}}{d}\right). \label{eq:sigmatheta}
\end{align}
in the direction parallel to $\vect{d}_\perp \equiv \vect{d} (1-\hat{\vect{d}} \cdot \hat{\overline{\vect{\theta}}})$ (with $\hat{\overline{\vect{\theta}}}$ the location of the light centroid of the image).
This resolution is much better than that of very-long-baseline amplitude interferometry in the radio band (e.g.~EHT~\cite{EventHorizonTelescope:2019uob}) at much longer wavelengths (typically $\lambda \sim 1\,\text{mm}$), and than that of optical/infrared Michelson interferometry (e.g.~Keck~\cite{2003SPIE.4838...79C}, CHARA~\cite{2005ApJ...628..453T}, GRAVITY~\cite{2017A&A...602A..94G}) which is restricted to much shorter baselines ($d\lesssim 300\,\mathrm{m}$) because the light from both apertures needs to be physically recombined. With intensity interferometry, the correlation can be computed offline, and the detectors can in principle be placed at opposite sides of the Earth. However, as we shall see in \Secs{sec:finite} and~\ref{sec:snr}, for $d \gg 10\,\text{km}$ and naively even better angular resolution, the excess correlation will be suppressed because all sufficiently bright sources for intensity interferometry have an angular size larger than $10^{-12}\,\text{rad}$, so the fiducial estimate of \Eq{eq:sigmatheta} is (practically) a best-case scenario. 

\subsubsection{Examples}\label{sec:hbt_review_ex}
Before we move on to extended-path intensity correlation (EPIC) in section~\ref{sec:mirror}, it is instructive to consider a handful of examples to gain intuition about intensity interferometry and to understand its limitation without the path extension.

The simplest example is that of a single point-like source $a$ at a location $\hat{\vect{\theta}}_a$ on the celestial sphere with a narrow gaussian spectrum  of width $\sigma_k$ around a mean wavenumber $\overline{k}$:
\begin{align}
\frac{\di I}{\di k \, \di \Omega} = \frac{I(\overline{k})}{\sqrt{2\pi}\sigma_k}e^{\frac{-(k-\overline{k})^2}{2\sigma_k^2}} \delta^2\left(\hat{\vect{\theta}} - \hat{\vect{\theta}}_a \right). \label{eq:Ipointsource}
\end{align}
Assuming that the planes of the apertures are oriented (sufficiently) perpendicular to the location of the source, i.e.~$|\vect{b}_p \cdot \hat{\vect{\theta}}_a| \ll c \sigma_t$, the aperture smearing can be ignored, and we find after evaluating \Eq{eq:C1}:
\begin{align}
C(\vect{d},\tau)  &= \int \di (\Delta t) \frac{1}{\sqrt{2\pi}\sigma_t} e^{\frac{-(\Delta t)^2}{2\sigma_t^2}} \left|\exp\left\lbrace i \overline{k} \left(c(\Delta t - \tau) - \hat{\vect{\theta}}_a \cdot \vect{d} \right) - \frac{\sigma_k^2}{2} \left(c(\Delta t - \tau) - \hat{\vect{\theta}}_a \cdot \vect{d}\right)^2\right\rbrace \right|^2 \nonumber \\
&= \frac{1}{\sqrt{1+2 c^2 \sigma_k^2 \sigma_t^2}} \exp \left\lbrace - \frac{\sigma_k^2}{1+2c^2\sigma_k^2 \sigma_t^2} \left(c\tau + \hat{\vect{\theta}}_a \cdot \vect{d} \right)^2\right\rbrace. \label{eq:C1pointsource}
\end{align}
Thus, we find that the intensity fluctuations at the two photodetectors are maximally correlated as long as the timing offset is chosen to be $\tau \simeq - \hat{\vect{\theta}}_a \cdot \vect{d} / c$, commensurate to the differential wavefront arrival time at both photodetectors. Practically, the coherence time of the light is significantly shorter than the relative timing precision, so $c \sigma_k \sigma_t \gg 1$, as we will discuss in section~\ref{sec:snr}. Hence, the offset $\tau$ needs to be chosen around its optimal value to a precision of $\sigma_t$ or better, which then produces a maximal intensity correlation of $C(\vect{d},\tau = - \hat{\vect{\theta}}_a \cdot \vect{d}/c) \simeq 1/(\sqrt{2} c \sigma_k \sigma_t)$.

For our next example, let us take two point-like sources $a$ and $b$ at locations $\hat{\vect{\theta}}_a$ and $\hat{\vect{\theta}}_b$ with total flux $\avg{I} = I(\overline{k})$ and fractional fluxes $\widetilde{I}_a$ and $\widetilde{I}_b$ (by construction, $\widetilde{I}_a + \widetilde{I}_b = 1$):
\begin{align}
\frac{\di \widetilde{I}}{\di k \, \di \Omega} = \frac{1}{\sqrt{2\pi}\sigma_k}e^{\frac{-(k-\overline{k})^2}{2\sigma_k^2}} \left[ \sum_{i=a,b} \widetilde{I}_i \, \delta^2\left(\hat{\vect{\theta}} - \hat{\vect{\theta}}_i \right)\right]. \label{eq:Ipointsource2}
\end{align}
We will again neglect the finite aperture spread, despite the fact that $\hat{\vect{\theta}}_i \cdot \vect{b}_p = 0$ cannot be maintained for both sources simultaneously, as more severe suppression effects will typically kick in for $\vect{\theta}_{ba} \equiv \hat{\vect{\theta}}_b - \hat{\vect{\theta}}_a \neq 0$. We evaluate \Eq{eq:C1}:
\begin{align}
C(\vect{d},\tau)  &= \int \di (\Delta t) \frac{1}{\sqrt{2\pi}\sigma_t}  e^{\frac{-(\Delta t)^2}{2\sigma_t^2}} \left|\sum_{i=a,b} \widetilde{I}_i \, e^{ i \overline{k} \left(c (\Delta t - \tau) - \hat{\vect{\theta}}_i \cdot \vect{d} \right) - \frac{\sigma_k^2}{2} \left(c (\Delta t - \tau) - \hat{\vect{\theta}}_i \cdot \vect{d}\right)^2}\right|^2 \nonumber \\
&= \frac{1}{\sqrt{1+2c^2 \sigma_k^2 \sigma_t^2}}\Bigg\lbrace \widetilde{I}_a^2   e^{ \frac{-\sigma_k^2}{1+2c^2 \sigma_k^2 \sigma_t^2} \left(c \tau + \hat{\vect{\theta}}_a \cdot \vect{d} \right)^2} + \widetilde{I}_b^2  e^{\frac{-\sigma_k^2}{1+2c^2 \sigma_k^2 \sigma_t^2} \left(c \tau + \hat{\vect{\theta}}_b \cdot \vect{d} \right)^2}   \label{eq:C1pointsource2} \\
&\phantom{=} 
 +2\widetilde{I}_a \widetilde{I}_b \cos\left[\overline{k} \vect{d} \cdot \vect{\theta}_{ba} \right] e^{-\sigma_k^2 (\vect{\theta}_{ba} \cdot \vect{d})^2 / 4} e^{ \frac{-\sigma_k^2}{1+2c^2 \sigma_k^2 \sigma_t^2} \left[ c \tau + (\hat{\vect{\theta}}_a + \hat{\vect{\theta}}_b) \cdot \vect{d}/2 \right]^2} \Bigg \rbrace \nonumber
\end{align}
The generalization of \Eq{eq:C1pointsource2} to an arbitrary number of point-like sources is straightforward and has a similar form.

\begin{figure}
\centering
	\begin{subfigure}[b]{1\textwidth}
	\centering
	\includegraphics[trim = 0 0 0 0, clip, width=\textwidth]{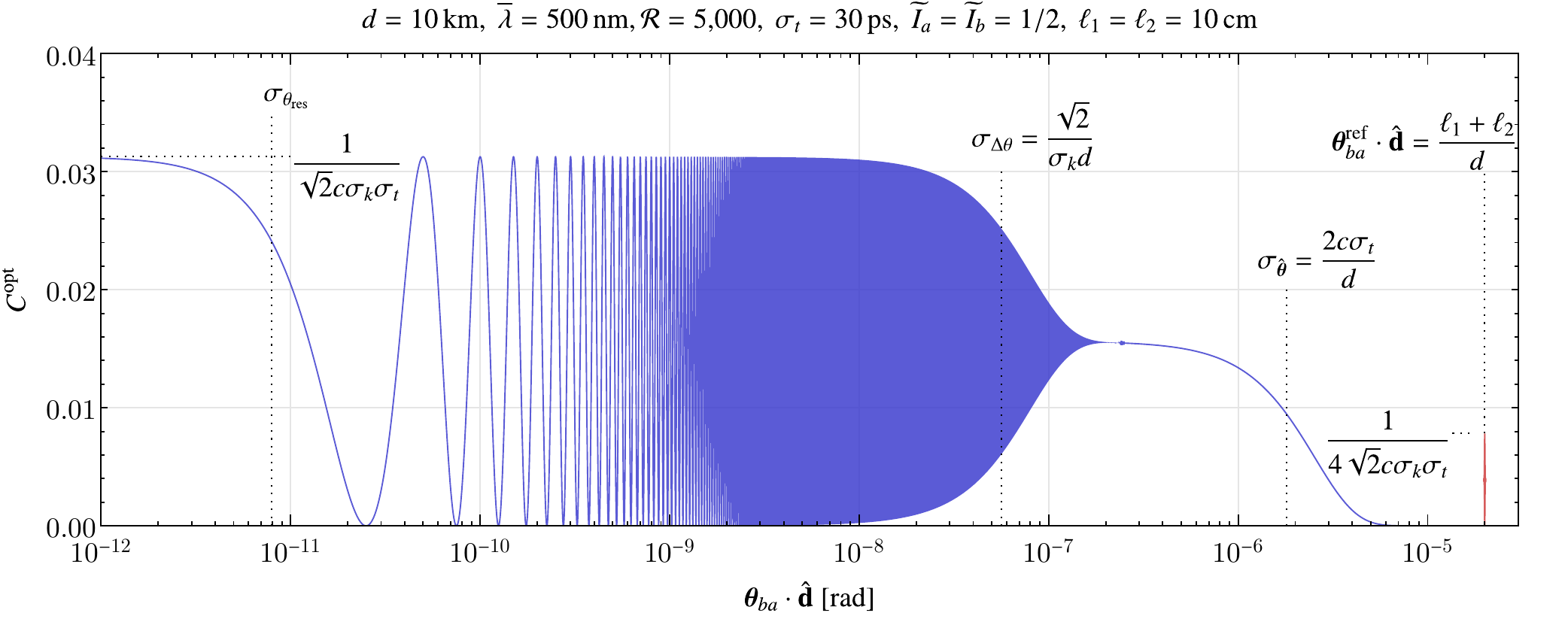}
	\caption{}
	\label{fig:C2a}
	\end{subfigure}
\hfill
	\begin{subfigure}[b]{1\textwidth}
	\centering
	\includegraphics[trim = 0 0 0 30, clip, width=\textwidth]{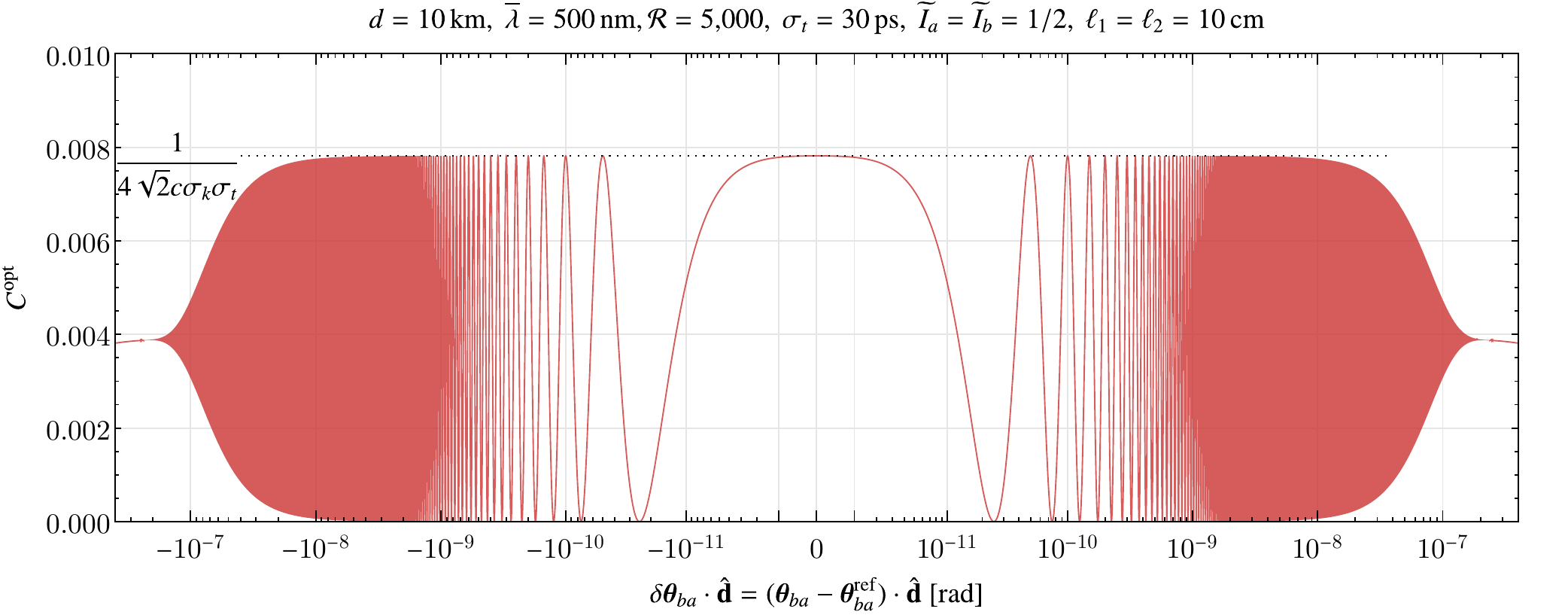}
	\caption{}
	\label{fig:C2b}
	\end{subfigure}
\caption{Excess correlation $C^\text{opt}(\vect{d})$ as a function of $\vect{\theta}_{ba} \cdot \hat{\vect{d}}$ for fiducial parameters of the baseline separation $d = 10\,\text{km}$, mean wavenumber $\overline{k} = 2\pi / 500\,\text{nm}$ of the light, and relative timing precision $\sigma_t = 30\,\text{ps}$, for two equally bright sources: $\widetilde{I}_a = \widetilde{I}_b = 1/2$. The blue curve in figure~\ref{fig:C2a} is \Eq{eq:C1pointsource2opt} for intensity interferometry \emph{without} any path extension. The red curve (spike) in figure~\ref{fig:C2a} is the excess correlation of \Eq{eq:C2pointsource2opt} for EPIC, with the same parameters but including geometric delays of $\ell_1 = \ell_2 = 10 \, \text{cm}$, producing a ghost intensity correlation fringe of  $1/4$ the contrast at a reference angle $\vect{\theta}_{ba}^\text{ref} \cdot \hat{\vect{d}} = 2 \times 10^{-5}\,\text{rad}$, but otherwise identical to the main fringe. A zoom-in of this ghost fringe around $\vect{\theta}_{ba}\approx \vect{\theta}_{ba}^\text{ref}$ is shown in figure~\ref{fig:C2b}. The characteristic scales for angular resolution $\sigma_{\theta_\text{res}}$ (\Eq{eq:sigmatheta}), dynamic field of view $\sigma_{\Delta \theta}$ (\Eq{eq:sigmaDeltatheta}), and global astrometric precision $\sigma_{\hat{\vect{\theta}}}$ (\Eq{eq:sigmahattheta}) are also indicated.} \label{fig:C2}
\end{figure}

We find that the excess correlation of \Eq{eq:C1pointsource2} has ``one-source'' contributions, proportional to $\widetilde{I}_i^2$, of exactly the same functional form as in \Eq{eq:C1pointsource}. The last line of \Eq{eq:C1pointsource2} contains the ``two-source'' contribution, proportional to $\widetilde{I}_a \widetilde{I}_b$. This is the term most sensitive to the separation of the sources, as it contains the most rapidly varying phase, namely $\overline{k} \vect{d} \cdot \vect{\theta}_{ba}$ in the cosine argument. The two-source contribution is maximized when the timing offset is taken to be $\tau = \tau^\text{opt} = - (\hat{\vect{\theta}}_a + \hat{\vect{\theta}}_b) \cdot \vect{d} / 2 c$, and we find in the limit $c \sigma_k \sigma_t \gg 1$:
\begin{align}
C^\text{opt}(\vect{d}) = \frac{1}{\sqrt{2}c \sigma_k \sigma_t}\Bigg\lbrace
 (\widetilde{I}_a^2+\widetilde{I}_b^2)  e^{\frac{-(\vect{\theta}_{ba} \cdot \vect{d})^2}{8\sigma_t^2}}  + 2\widetilde{I}_a \widetilde{I}_b \cos\left[\overline{k} \vect{d} \cdot \vect{\theta}_{ba} \right] e^{\frac{-\sigma_k^2 (\vect{\theta}_{ba} \cdot \vect{d})^2}{4}}
 \Bigg\rbrace. \label{eq:C1pointsource2opt}
\end{align}
In the above \Eq{eq:C1pointsource2opt}, we recognize three angular scales. The first and smallest is the angular resolution scale $\sigma_{\theta_\text{res}} = 1/\overline{k} d$ from \Eq{eq:sigmatheta}, which sets the fringe separation and thus determines the angular resolution on $\vect{\theta}_{ba} \cdot \hat{\vect{d}}$. The second angular scale is
\begin{align}
\sigma_{\Delta \theta} = \frac{\sqrt{2}}{\sigma_k d} \approx \underbrace{5.63 \times 10^{-8} \, \text{rad}}_{11.6 \,\text{mas}} \left(\frac{\overline{k}/\sigma_k }{1{,}000}\right)  \left( \frac{\overline{\lambda}}{500\,\text{nm}}\right)  \left( \frac{10\,\text{km}}{d}\right) , \label{eq:sigmaDeltatheta}
\end{align}
which sets the dynamical range for $\vect{\theta}_{ba} \cdot \hat{\vect{d}}$ over which the fringes are not exponentially suppressed, i.e.~the effective field of view. The third and typically largest angular scale is:
\begin{align}
\sigma_{\hat{\vect{\theta}}} = \frac{2 c \sigma_t}{d} \approx \underbrace{6.00\times 10^{-7} \, \text{rad}}_{124\,\text{mas}} \left(\frac{\sigma_t}{10\,\mathrm{ps}}\right)  \left( \frac{10\,\text{km}}{d}\right) , \label{eq:sigmahattheta}
\end{align} 
which determines the fiducial, \emph{global} astrometric precision on $\hat{\vect{\theta}}_i \cdot \hat{\vect{d}}$ from the ``one-source'' contributions to the excess intensity correlation, independent of any pairwise angular separations $\vect{\theta}_{ij}$. In figure~\ref{fig:C2}, we plot the excess correlation of \Eq{eq:C1pointsource2opt} for two equally bright sources, and indicate the three angular scales from \Eqs{eq:sigmatheta},~\ref{eq:sigmaDeltatheta}, and~\ref{eq:sigmahattheta} for some representative parameters.

\subsection{Extended-path intensity correlation}
\label{sec:epic}
\label{sec:mirror}

\subsubsection{General}

The limited (angular) dynamic range of intensity interferometry severely limits its use for differential astrometric measurements. If one wants to harness the full resolving power at a resolution $\sigma_{\theta_\text{res}} \sim \mathcal{O}(1\,\mu\text{as})$ (see \Eq{eq:sigmatheta}), the effective field of view is also tiny, of order $\sigma_{\Delta\theta} \lesssim \mathcal{O}(10\,\text{mas})$ if one has no more than $\overline{k}/\sigma_k \sim 10^4$ fringes (see \Eq{eq:sigmaDeltatheta}), which is already restrictive on the spectral width. Very few bright sources, other than tight binaries, are expected have such small relative separations. Furthermore, it is practically difficult to discriminate between the $n^\text{th}$ fringe and the $(n\pm1)^\text{th}$ fringe for large $n \sim k \vect{d} \cdot \vect{\theta}_{ba} / 2\pi \gg 1$.

\begin{figure}
\centering
	\begin{subfigure}[b]{0.45\textwidth}
	\centering
	\includegraphics[width=\textwidth]{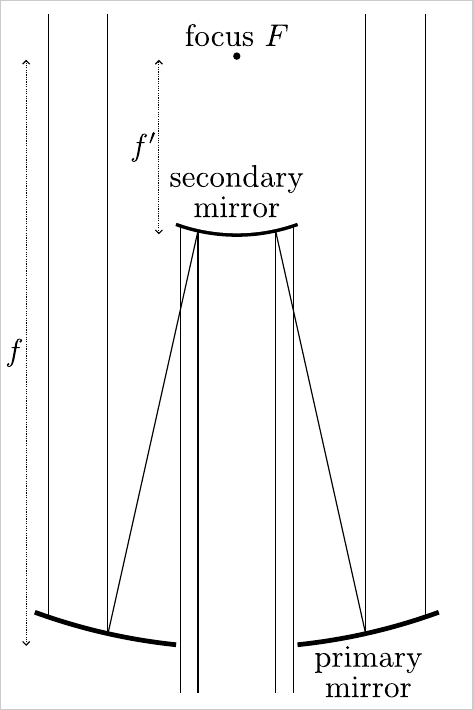}
	\caption{Light gathering and collimation}
	\label{fig:telescope_1}
	\end{subfigure}
\hfill
	\begin{subfigure}[b]{0.45\textwidth}
	\centering
	\includegraphics[width=\textwidth]{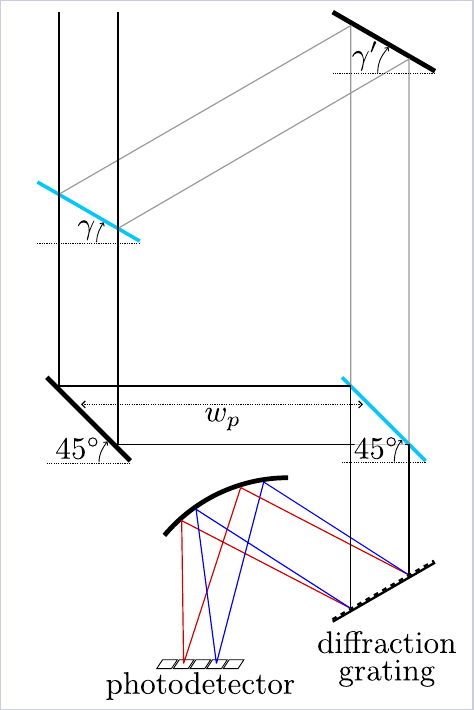}
	\caption{Path extension and spectral splitting}
	\label{fig:telescope_2}
	\end{subfigure}
\caption{Schematic optical path for the path extension in a single telescope of an extended-path intensity correlator. Panel~\ref{fig:telescope_1}: Light from a compact source is collected by a large primary mirror with focal length $f$, and collimated down to a narrow beam by a secondary mirror with focal length $f'$ sharing the same focal point $F$ as the primary. (Actual setups may have several foci, and a slit near a real pupil to select light only from the source of interest.) Panel~\ref{fig:telescope_2}: A semi-transparent flat plate whose normal is at an angle $\gamma$ relative to the collimated beam, splits the light into two optical paths (gray and black) of different path lengths (for $\gamma \simeq \gamma' \neq 45^\circ$) but undistorted wavefronts (for $\gamma = \gamma'$). The paths are recombined by another semi-transparent plate, after which the collimated beam is spectrally split by a reflective diffraction grating, and focused onto a photodetector whose pixels are read out as different spectral channels. 
}
\label{fig:telescope}
\end{figure}

EPIC solves this small-separation problem, and allows for effective fields of view on the order of a \emph{few arcseconds}. It is accomplished by modifying the optical path in at least one of the telescopes by introducing a differential path extension illustrated in figure~\ref{fig:telescope}. Light from the source(s) impinges on the telescope's large primary mirror (with collecting area $A_p$), and is then re-collimated to a narrow beam (figure~\ref{fig:telescope_1}). This collimated light then enters the path-extension stage (top half of figure~\ref{fig:telescope_2}), where the beam is split and recombined by a pair of partially reflective plates, with electric-field reflectivity and transmission of $\epsilon$ and $\sqrt{1-\epsilon^2}$, respectively. For $\gamma = \gamma'$, the two paths have a differential geometric path length of:
\begin{align}
\ell_p = \frac{w_p}{\sin 2 \gamma} \left(1 + \cos 2 \gamma - \sin 2 \gamma\right)\quad \text{for } 0 < \gamma < \pi/2, \label{eq:ell}
\end{align}
which can be positive ($\gamma < \pi/4$) or negative ($\gamma > \pi/4$).
It will be advantageous to have $\gamma'$ deviate slightly from $\gamma$, as discussed around \Eqs{eq:ellopt1} and~\ref{eq:ellopt2}. After the path-extension stage, the light is spectrally split (bottom half of right panel), and focused onto a photodetector array, whose individual pixels each read out one spectral channel, centered at some wavenumber $\overline{k}$ with narrow bandwidth $\sigma_k$. We discuss photodetector capabilities in \Sec{sec:photodetectors} and spectroscopic elements in \Sec{sec:grating}.

In such a setup, with optical losses and plate thicknesses ignored, the electric field as seen at the location of (a single pixel of) photodetector $p$ is:
\begin{align}
E_p(t_p) &= \epsilon \sqrt{1-\epsilon^2} \sum_{i=1}^{N_\theta} \sum_{\alpha = 1}^{N_k} \sum_{n=0}^1  e^{-i n k_\alpha  \ell_p}  E_{i\alpha} \exp\left\lbrace i \left[k_\alpha (t_p - r_{ip}) + \phi^\text{em}_{i\alpha}  \right] \right\rbrace, \label{eq:Efield2}
\end{align}
where $n=0$ corresponds to the ``left path'' in figure~\ref{fig:telescope_2}, and $n=1$ to the ``right path'', which have a differential geometric propagation phase due to their difference in length by $\ell_p$ from \Eq{eq:ell}. The expected intensity calculation proceeds exactly as in \Eq{eq:I1}, and becomes in the continuum limit:
\begin{align}
\avg{I_p(t_p)} = 2 \epsilon^2 \left(1-\epsilon^2\right) \int \di k \int \di \Omega \, \frac{\di I}{\di k \, \di \Omega} \left(1 + \cos k \ell_p \right). \label{eq:I2}
\end{align}
The latter factor $(1 + \cos k \ell_p)$ is due to interference of the left and right paths: it is maximal when the $\ell_p$ is an integer number of wavelengths, and vanishes when $\ell_p$ is a half-integer number of wavelengths (corresponding to when the photodetector is at the ``dark port'' of the internal interferometer).
The calculation of the intensity correlation is analogous to that of \Eq{eq:II1}, with the addition of a $\epsilon \sqrt{1-\epsilon^2} (1 + e^{-i k \ell_p})$ factor for every electric field $E_p$:
\begin{align}
&\avg{I_1(t_1) I_2(t_2)} = \avg{I_1}\avg{I_2} + \epsilon^4 \left(1-\epsilon^2\right)^2 \label{eq:II2} \\
& \times \left| \int \di k \int \di \Omega \, \frac{\di I}{\di k \, \di \Omega}  \big(1+e^{- i k \ell_1}\big)\big(1+e^{i k \ell_2}\big)   e^{i k\left((t_1 - t_2) - \hat{\vect{\theta}} \cdot \vect{d} \right) } \right|^2. \nonumber
\end{align}
Putting together \Eqs{eq:C1},~\ref{eq:I2}, and~\ref{eq:II2}, and including aperture smearing and finite relative timing resolution effects, thus yields the excess fractional intensity correlation:
\begin{empheq}[box=\fbox]{align}
&C(\vect{d},\tau) = \frac{\epsilon^4 \left(1-\epsilon^2\right)^2 }{\avg{I_1}\avg{I_2}} \int \frac{\di^2 b_1}{A_1} \int \frac{\di^2 b_2}{A_2} \int \di (\Delta t) \frac{1}{\sqrt{2\pi}\sigma_t} e^{\frac{-(\Delta t)^2}{2\sigma_t^2}}\label{eq:C2}\\
& \times \left| \int \di k \int \di \Omega \, \frac{\di I}{\di k \, \di \Omega}\big(1+e^{- i k \ell_1}\big)\big(1+e^{ i k \ell_2}\big) e^{i k\left(c (\Delta t - \tau) - \hat{\vect{\theta}} \cdot (\vect{d}+\vect{\Delta b})\right) } \right|^2 \nonumber.
\end{empheq}
Equation~\ref{eq:C2} is the central formula for extended-path intensity correlation. 

\subsubsection{Examples}\label{sec:mirror_ex}
To understand the consequences of the path extension in \Eq{eq:C2}, let us again consider a single point source with a differential intensity distribution from \Eq{eq:Ipointsource}. The effective transmitted intensity at each telescope is, from \Eq{eq:I2}:
\begin{align}
\avg{I_p} = 2 \epsilon^2 (1 - \epsilon^2) I(\overline{k}) \Big[1 +e^{-\sigma_k^2 \ell_p^2 / 2} \cos \overline{k} \ell_p  \Big] \simeq 2 \epsilon^2 (1 - \epsilon^2) I(\overline{k}). \label{eq:I2pointsource}
\end{align}
As we will see shortly, to effectively increase the field of view EPIC beyond that of conventional intensity interferometry, it is necessary to take $\sigma_k \ell \gg 1$, so that the last approximation in \Eq{eq:I2pointsource} will hold to high precision.  The excess intensity correlation for this source can be calculated from \Eq{eq:II2}, with the same approximation and with apertures perpendicular to the line of sight ($\hat{\vect{\theta}}_a \cdot \vect{b}_p$):
\begin{align}
C(\vect{d},\tau) 
&\simeq \frac{e^{\frac{-\sigma_k^2 \left( c \tau + \hat{\vect{\theta}}_a \cdot \vect{d} \right)^2}{1+2c^2 \sigma_k^2 \sigma_t^2}} + e^{\frac{-\sigma_k^2 \left( c \tau + \hat{\vect{\theta}}_a \cdot \vect{d} + \ell_1  \right)^2}{1+2c^2 \sigma_k^2 \sigma_t^2}} + e^{\frac{-\sigma_k^2 \left( c \tau + \hat{\vect{\theta}}_a \cdot \vect{d} - \ell_2  \right)^2}{1+2c^2 \sigma_k^2 \sigma_t^2}}+ e^{\frac{-\sigma_k^2\left( c \tau + \hat{\vect{\theta}}_a \cdot \vect{d} - \ell_2 + \ell_1  \right)^2}{1+2c^2 \sigma_k^2 \sigma_t^2} }}{4\sqrt{1+2c^2 \sigma_k^2 \sigma_t^2}}  . \label{eq:C2pointsource}
\end{align}
Comparing \Eqs{eq:C1pointsource} and~\ref{eq:C2pointsource}, or the black and orange curves in figure~\ref{fig:C1}, we can see that the effect of the path extension is to reduce the contrast of the ``main'' intensity fringe, at $c \tau \simeq - \hat{\vect{\theta}}\cdot \vect{d}$, by a factor of 4. More importantly, however, the initial fringe is now also split into four, including three other equally bright ``ghost fringes'', as if there were three duplicate images of the source at locations $\hat{\vect{\theta}}_{a}^\text{ghost} = \hat{\vect{\theta}}_a +  \frac{\lbrace  \ell_1, - \ell_2, -\ell_2 + \ell_1 \rbrace }{d} \hat{\vect{d}} $, respectively. When $\sigma_k \ell_p \gg 1$, these ghost fringes are far outside the effective field of view (relative to the main fringe) of regular intensity interferometry (cfr.~\Eq{eq:sigmaDeltatheta}).

With the behavior of the extended-path intensity correlator understood for a single point source, let us see what the excess intensity correlator looks like for two point sources under influence of the path extension; evaluation of the fractional intensity correlation formula of \Eq{eq:II2} for the two sources of \Eq{eq:Ipointsource2} gives:
\begin{align}
&C(\vect{d},\tau) = \frac{1}{4\sqrt{1+2c^2 \sigma_k^2 \sigma_t^2}} \int \frac{\di^2 b_1}{A_1}\int \frac{\di^2 b_2}{A_2} \Bigg \lbrace \sum_{i=a,b} \widetilde{I}_i^2 \sum_\ell  e^{ \frac{-\sigma_k^2}{1+2c^2 \sigma_k^2 \sigma_t^2} \left[c \tau + \hat{\vect{\theta}}_i \cdot (\vect{d}+\Delta\vect{b}) + \ell  \right]^2}    \label{eq:C2pointsource2} \\
& +2 \widetilde{I}_a \widetilde{I}_b  \sum_{\ell,\ell'} \cos\left[\overline{k}  \left((\vect{d} + \Delta\vect{b}) \cdot \vect{\theta}_{ba}  -\ell + \ell' \right)\right] 
e^{\frac{-\sigma_k^2}{4}\left[\vect{\theta}_{ba} \cdot (\vect{d}+\Delta\vect{b}) - \ell + \ell' \right]^2}   
e^{\frac{-\sigma_k^2 \left[c \tau + \frac{\hat{\vect{\theta}}_a + \hat{\vect{\theta}}_b}{2} \cdot (\vect{d} + \vect{\Delta b}  ) +\frac{ \ell + \ell'}{2} \right]^2}{1+2c^2 \sigma_k^2 \sigma_t^2}  } \Bigg \rbrace. \nonumber
\end{align}
Each sum over $\ell$ and $\ell'$ runs over the list $\lbrace 0, \ell_1, -\ell_2, -\ell_2 + \ell_1 \rbrace$ with $\ell_1$ and $\ell_2$ as in \Eq{eq:ell}. The possibilities for $\ell$ correspond to whether the light from source $a$ is extended nowhere, in telescope 1 only, in telescope 2 only, or in both telescopes, respectively; $\ell'$ runs over the same possibilities for source $b$. For $\vect{\theta}_{ba}\cdot \hat{\vect{d}} \gg \sigma_{\Delta \theta}  \equiv \sqrt{2}/{\sigma_k d}$, the ``main'' differential astrometry fringe between sources $a$ and $b$ (corresponding to $\ell=\ell'=0$), or really any fringe with $\sigma_k \left| \vect{\theta}_{ba} \cdot \vect{d} - \ell + \ell' \right| \gg 1$, is exponentially suppressed  (see \Eq{eq:sigmaDeltatheta} and figure~\ref{fig:C1}).

However, even when the separation between the main images of the source is large, the apparent separation between their respective ghost images need not be. By tuning $\ell - \ell' \simeq \vect{\theta}_{ba}\cdot \hat{\vect{d}}$, we can overcome the field-of-view limitation of intensity interferometry. Furthermore, by making $\ell_p$ depend on $\vect{b}_p$, for example by a slight tilting of one of the mirrors ($\gamma \neq \gamma'$), aperture smearing effects can be negated completely. The optimal geometric path extensions for differential astrometry with EPIC are achieved at e.g.~$\ell = + \ell_1^\text{opt}$ and $\ell' = - \ell_2^\text{opt}$ with:
\begin{align}
\ell_1^\text{opt}(\vect{b}_1) &= \vect{\theta}^\text{ref}_{ba} \cdot \big[\rho \vect{d} - \vect{b}_1 \big], \label{eq:ellopt1}\\ 
\ell_2^\text{opt}(\vect{b}_2) &= \vect{\theta}_{ba}^\text{ref} \cdot \big[(1-\rho) \vect{d} + \vect{b}_2 \big]; \label{eq:ellopt2}
\end{align}
where $\rho$ is arbitrary, and other fringe combinations of $\ell$ and $\ell'$ are also possible. This particular fringe corresponds to the amplitude depicted in figure~\ref{fig:C1}.
The reference separation vector $\vect{\theta}_{ba}^\text{ref}$ should be chosen close to the actual separation, i.e.~$\delta \vect{\theta}_{ba} \equiv \vect{\theta}_{ba} - \vect{\theta}_{ba}^\text{ref}$ has length much less than $\sigma_{\Delta \theta}$. For maximum sensitivity, one should choose the optimal timing offset $\tau = \tau^\text{opt}$ that maximizes the differential astrometry fringe:
\begin{align}
\tau^\text{opt} = -  \frac{\hat{\vect{\theta}}_a + \hat{\vect{\theta}}_b - (1-2\rho) \vect{\theta}^\text{ref}_{ba}}{2} \cdot \frac{\vect{d}}{c}.\label{eq:tauopt}
\end{align}
Finally, we obtain the excess fractional intensity correlation for two point sources in this optimal configuration:
\begin{align}
C^\text{opt}(\vect{d}) \simeq  \frac{1}{4\sqrt{2} c \sigma_k \sigma_t} & \Bigg \lbrace \widetilde{I}_a^2 +  \widetilde{I}_b^2 + 2 \widetilde{I}_a \widetilde{I}_b \cos\left[ \overline{k} \vect{d} \cdot \vect{\delta\theta}_{ba} \right] e^{\frac{-\sigma_k^2}{4} \left(\vect{d} \cdot \vect{\delta \theta}_{ba} \right)^2}  \Bigg \rbrace. \label{eq:C2pointsource2opt}
\end{align}
The utility of EPIC is apparent upon comparison of \Eqs{eq:C1pointsource2opt} and~\ref{eq:C2pointsource2opt}. Without the path extension, the two-source fringes are only visible for an ultra-narrow field: $|\vect{\theta}_{ba} \cdot \hat{\vect{d}}| \lesssim \sigma_{\Delta \theta}$. With the path extension, the two-source fringes are visible whenever $|\vect{\delta \theta}_{ba} \cdot \hat{\vect{d}}| \lesssim \sigma_{\Delta \theta}$. In other words, the dynamic range has not increased, but instead of the usable field of view being centered around the locus where $\vect{\theta}_{ba} \cdot \hat{\vect{d}} \approx 0$, it is centered around $\vect{\theta}_{ba} \cdot \hat{\vect{d}} \approx \vect{\theta}^\text{ref}_{ba} \cdot \hat{\vect{d}} = \frac{\ell_1 + \ell_2}{d}$. From the point of view of the excess intensity correlation, the path extensions resulting from the beamsplitter in each telescopes according to \Eqs{eq:ellopt1} and~\ref{eq:ellopt2} effectively produce \emph{co-located ghost images} of sources $a$ and $b$, despite their true images located at different points on the celestial sphere.  Furthermore, the $\vect{b}_p$-dependence of $\ell^\text{opt}_b$ can correct for the aperture smearing at each telescope.

\subsection{Finite source size} \label{sec:finite}
In \Secs{sec:hbt_review_ex} and~\ref{sec:mirror_ex}, we have given examples of intensity correlation signals for \emph{point-like} sources. However, once the intensity interferometer resolves either or both of the sources ($\sigma_{\theta_\text{res}} \lesssim \theta_j$ with $\theta_j$ the characteristic angular size of the source), the visibility is reduced. This is an important effect, because the differential light-centroiding precision between two sources scales favorably with baseline distance $d$ until the sources are resolved, so the optimal operating point will be $\sigma_{\theta_\text{res}} \sim \theta_j$ (see section~\ref{sec:snr}). We review the details of this finite-source-size effect on the intensity fringes here for the two main sources of interest---stars and quasars---but refer the reader to the classic references~\cite{hbtIII,hbtIV} for more background, as determining stellar radii has been the primary use case of intensity interferometry so far.

We write the differential surface brightness per unit wavenumber $k$ in a single spectral channel of bandwidth $\sigma_k$ centered at $k = \overline{k}$ as:
\begin{align}
\frac{ \di I}{\di k \, \di \Omega} = e^{\frac{-(k-\overline{k})^2}{2\sigma_k^2}} \sum_j^\text{sources}\left. \frac{ \di I_j}{\di k \, \di \Omega}\right|_\text{thermal},
\end{align}
where the pre-factor approximates the effect of a (Gaussian) spectral filter, and we assume that the sources $j$ are optically thick thermal emitters, whose intrinsic differential surface brightness
\begin{align}
\left.\frac{ \di I_j}{\di k \, \di \Omega}\right|_\text{thermal} = \frac{\hbar c^2}{4\pi^3} \frac{k^3}{e^{\hbar c k/ k_\mathrm{B} T_j(\hat{\vect{\theta}}-\hat{\vect{\theta}}_j)}-1}
\end{align}
depends only on the temperature function $T_j(\hat{\vect{\theta}}-\hat{\vect{\theta}}_j)$ with $\hat{\vect{\theta}}_j$ taken to be the light centroid of the source and $k_\mathrm{B}$ the Boltzmann constant. The excess fractional intensity correlation signals of \Eqs{eq:C1} and~\ref{eq:C2} contain integrals of the form:
\begin{align}
\int \di k \int \di \Omega \, \frac{\di I}{\di k \, \di \Omega} \, e^{-i k \hat{\vect{\theta}}\cdot \vect{d}} \simeq I(\overline{k}) \int \di k \, \frac{e^{\frac{-(k-\overline{k})^2}{2\sigma_k^2}}}{\sqrt{2\pi}\sigma_k} \,\sum_j \, e^{-i k \hat{\vect{\theta}}_j \cdot \vect{d}} \, \widetilde{I}_j(\overline{k}) \, \mathcal{F}_j(\overline{k}, \vect{d}) \label{eq:finite_source_supp}
\end{align}
where we have defined the total flux $I(\overline{k})$ in the channel centered at $\overline{k}$
\begin{align}
I (\overline{k})  \equiv \sum_j I_j(\overline{k}) = \sum_j \int \di \Omega \, \frac{\hbar c^2}{2^{3/2} \pi^{5/2} } \frac{\sigma_k \overline{k}^3}{e^{\hbar c \overline{k}/k_\mathrm{B} T_j(\hat{\vect{\theta}})} - 1}, \label{eq:flux}
\end{align}
and the fractional fluxes $\widetilde{I}_j(\overline{k}) \equiv I_j(\overline{k}) / I(\overline{k})$. The ``form factor''  is
\begin{align}
\mathcal{F}_j(\overline{k}, \vect{d}) \equiv \frac{ \int \di \Omega \, e^{-i \overline{k} \vect{d} \cdot \hat{\vect{\theta}}} \left[ e^{\hbar c \overline{k}/k_\mathrm{B} T_j(\hat{\vect{\theta}})} - 1\right]^{-1}}{ \int \di \Omega  \left[ e^{\hbar c \overline{k}/k_\mathrm{B} T_j(\hat{\vect{\theta}})} - 1\right]^{-1}} \label{eq:form}
\end{align}
and is equal to unity for point-like sources, i.e.~$\mathcal{F}_j(\overline{k}, \vect{d}) \to 1$ as $\overline{k} d \theta_j \to 1$, and is the implicit limit we have worked in for the examples in \Secs{sec:hbt_review_ex} and  \ref{sec:mirror_ex}. 

Incorporating finite-size effects is straightforward upon inspection of \Eq{eq:finite_source_supp}: for every (fractional) flux factor $\widetilde{I}_j(\overline{k})$ of a source $j$ in a spectral channel centered at $\overline{k}$, one is to insert also the form factor $\mathcal{F}_j(\overline{k}, \vect{d})$ from \Eq{eq:form}. For example, the excess fractional correlation for two sources $a$ and $b$ from \Eq{eq:C2pointsource2opt} is modified to:
\begin{align}
C^\text{opt}(\vect{d}) &\simeq  \frac{1}{4\sqrt{2} c \sigma_k \sigma_t}  \Bigg \lbrace \widetilde{I}_a(\overline{k})^2 \left|\mathcal{F}_a(\overline{k}, \vect{d})\right|^2 +  \widetilde{I}_b(\overline{k})^2 \left|\mathcal{F}_b(\overline{k}, \vect{d})\right|^2  \label{eq:C2pointsource2opt_finite} \\
&\phantom{\simeq}+ 2 \, \widetilde{I}_a(\overline{k})\,  \widetilde{I}_b(\overline{k}) \, \text{Re} \left\lbrace\mathcal{F}_a(\overline{k}, \vect{d}) \mathcal{F}^*_b(\overline{k}, \vect{d}) e^{i\left[ \overline{k} \vect{d} \cdot \vect{\delta\theta}_{ba} \right]} \right\rbrace
e^{\frac{-\sigma_k^2}{4} \left(\vect{d} \cdot \vect{\delta \theta}_{ba} \right)^2}  \Bigg \rbrace. \nonumber
\end{align}
The suppression of the one-source terms $\left|\mathcal{F}_j(\overline{k}, \vect{d})\right|^2 \to 0$ as $k d \theta_j \to \infty$ in the first line of \Eq{eq:C2pointsource2opt_finite} is how stellar diameters can be determined. Not surprisingly, the finite source sizes also suppress the two-source term in the second line. Note that $\mathcal{F}_a \mathcal{F}_b^*$ is in general a complex number that can give an additive contribution to the phase of interest, namely $\overline{k} \vect{d} \cdot \vect{\delta\theta}_{ba}$. For sources that are reflection-symmetric along the axis defined by $\vect{d}$ projected onto the celestial sphere, which is guaranteed for \emph{any} baseline orientation as long as the source is spherically symmetric, the form factor is real and does not contribute to the phase. Even if $\mathcal{F}_j$ is not real, its phase can be modeled and estimated from the one- and two-source intensity fringes. We assume that this is done at sufficient fidelity and will ignore this phase hereafter.

As we will see in section~\ref{sec:snr}, differential astrometry with intensity interferometry is best performed on compact sources with high surface brightness, so stars and quasars are the most suitable targets. Let us compute the flux and form factor of \Eqs{eq:flux} and~\ref{eq:form} for these two types of objects. For simplicity, we model stars as single-temperature disks
\begin{align}
T_\text{s}(\hat{\vect{\theta}}) = T_\text{s} \, \Theta_\text{H}( \theta_\text{s} - \theta)
\end{align}
where the temperature equals $T_\text{s}$ on the disk of angular radius $\theta_\text{s} = R_\text{s} / D_\text{s}$, with $R_\text{s}$ and $D_\text{s}$ the physical stellar radius and line-of-sight distance. This gives a flux per spectral channel of
\begin{align}
I_\text{s}(\overline{k}) = \frac{\hbar c^2}{(2\pi)^{3/2}} \frac{\sigma_k \overline{k}^3 \theta_\text{s}^2}{e^{\hbar c \overline{k}/k_\mathrm{B} T_\text{s} - 1}},
\end{align}
and a simple form factor
\begin{align}
\mathcal{F}_\text{s}(\overline{k}, \vect{d}) = \frac{2 J_1 ( \overline{k} d \theta_\text{s})}{\overline{k} d \theta_\text{s}} \label{eq:formS}
\end{align}
which scales as $-\sqrt{8/\pi} \cos(\overline{k} d \theta_\text{s} + \pi/4) /( \overline{k} d \theta_\text{s})^{3/2}$ at $\overline{k} d \theta_\text{s} \gg 1$.

For quasars, we assume the emission is also optically thick with a local temperature function of:
\begin{align}
T_\mathrm{q}(\hat{\vect{\theta}}) = \frac{T_{500}}{1+z_\mathrm{q}} \left(\frac{R_{500}}{R}\right)^{3/4} = T_{500} \left(\frac{\theta_{500}}{\theta}\right)^{3/4} \label{eq:Tq}
\end{align}
where $T_{500} = 28{,}776 \, \mathrm{K}$ is the temperature corresponding to a wavelength of $\lambda = 500\,\mathrm{nm}$, and $R_{500}$ is the physical radius (in the quasar's local frame) at which this temperature is achieved. The apparent angular radius $\theta_{500} = R_{500}/[D_\mathrm{q}(1+z_\mathrm{q})^{4/3}]$ is where this temperature is achieved in the \emph{observer's} frame, after accounting for the redshift of the quasar $z_\mathrm{q}$ and its angular diameter distance $D_\mathrm{q}$. This model is a simplified version of the emission from a face-on accretion disk in the model of ref.~\cite{1973A&A....24..337S}, with temperature function in the quasar's frame:
\begin{align}
	T_\text{q}(R) = T_{\mathrm{q},0} \left(\frac{R_\text{isco}}{R} \right)^{3/4}\left[1-\left(\frac{R_\text{isco}}{R}\right)^{1/2} \right]^{1/4},
\end{align}
with $T_{\mathrm{q},0}$ is a reference temperature given by 
\begin{align}
T_{	\text{q},0} &= \left( \frac{3 G^2 M_\text{BH}^2 m_p f_\text{Edd}}{2 c \sigma_\text{B} \sigma_\text{T} \eta_\text{q} R_\text{isco}^3} \right)^{1/4} \approx 3.33 \times 10^5 \, \text{K}  \left(\frac{f_\text{Edd}}{\eta_\text{q}} \right)^{1/4} \left(\frac{10^9 \, M_\odot}{M_\text{BH}}\right)^{1/4} \left(\frac{G M_\mathrm{BH}/c^2}{R_\text{isco}} \right)^{1/4} \label{eq:Tq2}
\end{align}
with $M_\text{BH}$ the black hole mass, $m_p$ the proton mass, $f_\text{Edd}$ the bolometric luminosity relative to the Eddington luminosity, $\eta_\text{q}$ the bolometric luminosity relative to the accretion rate, $\sigma_\text{B} = (\pi^2/60) k_\text{B}^4/\hbar^3 c^2$ the Stefan–Boltzmann constant, $\sigma_\text{T}  = (8 \pi/3) (\hbar^2/c) \alpha^2 / m_e^2$ the Thomson scattering cross-section, and $R_\text{isco}$ the innermost stable circular orbit within the accretion disk (equaling $6 G M_\text{BH}/c^2$ for a non-spinning black hole, and $G M_\text{BH}/c^2$ for a maximally spinning black hole). The apparent angular radius of the accretion disk is $\theta_\text{isco} = R_\text{isco} / D_\text{q}$.

In the simplified model of \Eq{eq:Tq}, the observed flux from the quasar per spectral channel is then:
\begin{align}
I_\mathrm{q}(\overline{k}) = \frac{\xi_\mathrm{q} \hbar c^2}{(2\pi)^{3/2}} \,\sigma_k \overline{k}^{1/3} \, \left(\frac{k_\mathrm{B}T_{500}}{\hbar c} \right)^{8/3} \, \theta_\mathrm{500}^2, \label{eq:Iq}
\end{align}
with $\xi_\mathrm{q} = \Gamma(11/3)\zeta(8/3) \approx 5.15$, and the form factor is:
\begin{align}
\mathcal{F}_\mathrm{q}(\overline{k}, \vect{d}) = \xi^{-1}_\mathrm{q} \int_0^\infty \dd y\, \frac{2 y}{\exp y^{3/4} - 1} J_0\left[ \left(\frac{k_\mathrm{B}T_{500}}{\hbar c} \right)^{4/3} \,  \overline{k}^{-1/3} \, d \, \theta_{500} \, y  \right].\label{eq:formQ}
\end{align}
This form factor depends on $\overline{k}$ and $d$ through the combination $a = (k_\mathrm{B} T_{500}/\hbar c)^{4/3}\, \overline{k}^{-1/3}\, d \, \theta_{500}$; when $a \ll 1$, the form factor goes to unity, while for $a \gg 1$, it is approximately $\mathcal{F}_\mathrm{q} \approx 0.3 a^{-5/4}$. This is a slightly less severe suppression ($\propto d^{-5/4}$) than the form factor scaling for a star ($\propto d^{-3/2}$).

\subsection{Noise and light-centroiding precision}
\label{sec:snr}
In this section, we estimate the statistical noise on our primary observable---the excess intensity correlation of \Eq{eq:C2}---allowing us to calculate the signal-to-noise ratio (SNR) and light-centroiding precision for different sources. We postpone a discussion of phase errors and other systematic errors to section~\ref{sec:tech}.

The data collected by each photodetector $p$ are arrival times $\lbrace t_i \rbrace$ for $N_p$ photons, from which one can build a simple estimator $\hat{I}_p$ (the hat indicates that it is a data-driven \emph{estimate}) of the instantaneous intensity at time $t_p$:
\begin{align}
\hat{I}_p(t_p) = \frac{\hbar c \overline{k}}{\eta_p A_p} \sum_{i=1}^{N_p} \frac{e^{\frac{-(t_p-t_i)^2}{\sigma_t^2}}}{\sqrt{\pi} \sigma_t}. \label{eq:Ihat1}
\end{align}
Above, $\overline{k}$ is the mean wavevector of the light (assumed to be observed in a fractionally narrow band), $\eta_p$ is the photodetection efficiency (excluding transmissivity and reflectivity factors from the splitter and combiner in section~\ref{sec:mirror}), and $A_p$ is the telescope's aperture area. The effective timing precision \emph{per photodetector} is taken to be $\sigma_t/\sqrt{2}$ with $\sigma_t$ from \Eq{eq:sigmat}, so that relative photon arrival times, i.e.~$t_2 - t_1$, have relative timing precision $\sigma_t$. An unbiased, optimal estimator of the excess fractional intensity correlation between two detectors, time averaged over a time interval $t_\text{obs}$, is then:
\begin{align}
\hat{C}(\vect{d},\tau) = \frac{\overline{ \hat{I}_1(t) \hat{I}_2(t+\tau)}}{\overline{\hat{I}_1(t) } ~ \overline{\hat{I}_2(t+\tau) }} - 1 . \label{eq:C2estimate}
\end{align}
The overline signifies time-averaging according to:
\begin{alignat}{1}
	\overline{ f(t) } \equiv \int_{t-t_\text{obs}/2}^{t+t_\text{obs}/2} \frac{\di t'}{t_\text{obs}} \, f(t'), \label{eq:averaging}
\end{alignat}
for any time-dependent observable $f(t)$. The expectation value of the estimator in \Eq{eq:C2estimate} is $\langle \hat{C} \rangle= C$ with $C$ from \Eq{eq:C2}; further details can be found in appendix~\ref{app:snr}. (We assume that the time dependence of $C$ during $t_\text{obs}$ is negligible.)

The estimator for the intensity correlation in \Eq{eq:C2estimate} only receives contributions when photodetectors 1 and 2 each record photons with relative times of arrival (given a fixed offset $\tau$) separated by $\mathcal{O}(\sigma_t)$ or less. Most photons detected at photodetector 1 will not have such a ``simultaneous'' photon at photodetector 2, and the excess fractional correlation is small numerically (of order $C \sim 1/c \sigma_k \sigma_t \ll 1$), so obtaining a high SNR requires sufficiently bright sources and long observation times, so that $N_p \gg 1$. In what follows, we can therefore safely neglect Poisson fluctuations in the \emph{total number} $N_p$ of photons, and only consider the arrival times $\lbrace t_i \rbrace$ as random variables. We further assume
\begin{align}
\frac{1}{c\overline{k}} \ll \frac{1}{c\sigma_k} \ll \sigma_t \ll  t_\text{reset} \ll\frac{t_\mathrm{obs}}{N_p}, \label{eq:snrassumptions}
\end{align}
though modifications of the formulae below for weak or reversed hierarchies are not difficult.
Equation~\ref{eq:snrassumptions} is equivalent to assuming a strong hierarchy of time scales, in ascending order: the period of light ($1/c\overline{k}$), the coherence time of the light ($1/c\sigma_k$), the relative timing precision ($\sigma_t$), the reset time of the photodetector ($t_\text{reset}$), and the typical time between photon counts ($t_\text{obs}/N_p$).
With those assumptions, it is straightforward to calculate the expected estimated intensity:
\begin{align}
 \avg{ \overline{\hat{I}_p} }  = \frac{\hbar c \overline{k}}{\eta_p A_p}\frac{N_p}{t_\text{obs}} \equiv \langle I_p \rangle, \label{eq:Ihat2}
\end{align}
and, more tediously, the statistics-limited standard deviation on $\hat{C}$:
\begin{align}
\sigma_{\hat{C}} &\simeq \sqrt{\frac{\text{Var}\left\lbrace\overline{\hat{I}_1 \hat{I}_2}\right\rbrace}{\langle I_1 \rangle \langle I_2 \rangle}} \simeq \sqrt{\frac{t_\text{obs}}{\sqrt{4\pi} \sigma_t}} \frac{1}{\sqrt{N_1 N_2}} = \sqrt{\frac{1}{\sqrt{4\pi} \sigma_t t_\text{obs}}} \frac{\hbar c \overline{k}}{\sqrt{\eta_1 \eta_2 A_1 A_2}}\frac{1}{\sqrt{\langle I_1 \rangle  \langle I_2 \rangle}}. \label{eq:sigmaC}
\end{align}
We provide derivations of \Eqs{eq:Ihat2} and~\ref{eq:sigmaC} in appendix~\ref{app:snr}.

\begin{figure}
\centering
\includegraphics[width=0.7\textwidth]{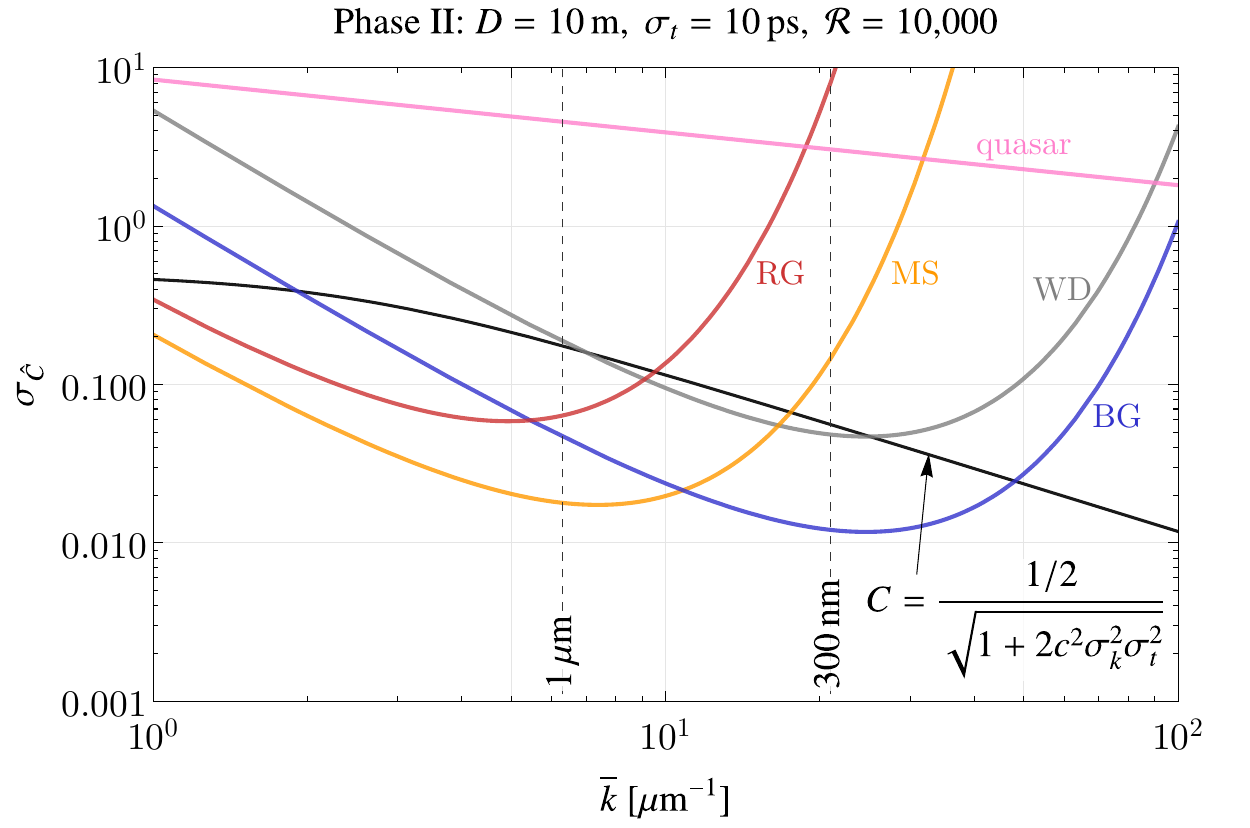}
\caption{Single-channel precision $\sigma_{\hat{C}}$ from \Eq{eq:sigmaC} on the excess intensity correlation, as a function of wavenumber $\overline{k}$ (for reference, the wavelength range of $\overline{\lambda} = [300\,\text{nm},1\,\mathrm{\mu m}]$ is indicated by black, dashed, vertical lines). The light gray, yellow, red, and blue curves correspond to a white dwarf (``WD'', $T = 20{,}000\,\text{K}$, $D = 10\,\text{pc}$, $R = 10^{-2}\,R_\odot$), Sun-like main sequence star (``MS'', $T = 6{,}000\,\text{K}$, $D = 100\,\text{pc}$, $R = 1\,R_\odot$), red giant (``RG'', $T = 4{,}000\,\text{K}$, $D = 5\,\text{kpc}$, $R = 10^{2}\,R_\odot$), and blue giant (``BG'', $T = 20{,}000\,\text{K}$, $D = 5\,\text{kpc}$, $R = 10 \,R_\odot$), respectively. The pink curve shows $\sigma_{\hat{C}}$ for a quasar at redshift $z_\mathrm{q} = 1$ with optical radius $R_{500} = 10^{16}\,\mathrm{cm}$ (\Eq{eq:Tq}). 
We assume telescope parameters from Phase II in table~\ref{tab:phases}, and an integration time $t_\text{obs} = 10^4\,\text{s}$. The solid black line indicates the single-source excess fractional correlation (\Eq{eq:C1pointsource}). The integrated SNR of \Eq{eq:SNR} between $1\,\mathrm{\mu m}$ and $300\,\text{nm}$ is $\lbrace 93, 432, 371, 91, 2.2 \rbrace$ for these sources $\lbrace \text{WD}, \text{MS}, \text{RG}, \text{BG}, \text{quasar} \rbrace$.
}\label{fig:sigmaC}
\end{figure}

We can estimate $\sigma_{\hat{C}}$ numerically by using the flux-per-channel formula from \Eq{eq:flux}.
If one has $n_p^\text{arr}$ telescopes at site $p=1,2$, the noise is reduced by a factor of $1/\sqrt{n_1^\text{arr} n_2^\text{arr}}$ by averaging at each of the two sites. Finally, we correct for the suppression due to the two uncorrelated polarizations of light from unpolarized sources by multiplying by $1/2$ the total signal-to-noise ratio (SNR) on the excess fractional intensity correlation:
\begin{align}
\text{SNR} = \frac{1}{2}\sqrt{ {n_1^\text{arr} n_2^\text{arr}}\sum_{\overline{k}}\left(\frac{C(\vect{d},\tau)}{\sigma_{\hat{C}}} \right)^2_{\overline{k}}},\label{eq:SNR}
\end{align}
where the sum is over channels with different $\overline{k}$ separated from each other by more than $\sigma_k$, e.g.~with a logarithmic spectral spacing between channels of the form $e^{2n\sigma_k/\overline{k}}$ for integer $n$.
In figure~\ref{fig:sigmaC}, we plot $\sigma_{\hat{C}}$  from \Eq{eq:sigmaC} for a single pair of telescopes ($n_p^\text{arr} = 1$) as a function of $\overline{k}$ for thermal spectra of four different reference stars, convoluted with a narrow spectral filter of width $\sigma_k$. For the parameters plotted in figure~\ref{fig:sigmaC}, the integrated signal-to-noise ratio over $6{,}020$ logarithmically spaced channels between $\overline{\lambda} = 1\,\mu\text{m}$ and $\overline{\lambda} = 300\,\text{nm}$ from \Eq{eq:SNR} is $\text{SNR} = \lbrace 93, 432, 371, 91 \rbrace$ for the white dwarf, main sequence, red giant, and blue giant stars, respectively. Likewise, for the fiducial quasar shown in pink in figure~\ref{fig:sigmaC}, the integrated $\text{SNR}$ equals $2.2$, despite $\sigma_{\hat{C}} > C \simeq 1/2\sqrt{2}c \sigma_k \sigma_t$ in any one spectral channel.

For differential astrometry using EPIC, the relevant term in $C$ is reduced by a factor of $1/4$ (cfr.~\Eq{eq:C2pointsource2opt}), while the observed fluxes $\propto \langle I_p \rangle$ are reduced by a factor of $1/2$ per source if $\epsilon = 1/\sqrt{2}$ (cfr.~\Eq{eq:I2pointsource}), reducing the total SNR by a factor of $1/8$ (compared to not having any path extension at all). We note though that compared to figure~\ref{fig:sigmaC}, this flux loss is compensated by observing \emph{two} sources simultaneously. In addition, with the use of a polarizing beamsplitter, one could theoretically recover a factor of $\sqrt{2}$ in SNR and light-centroiding precision~\cite{1974iiaa.book.....B,2022SPIE12183E..0GM}. Likewise, one could recover an extra factor of 2 from the second output port of the beam recombiner (by also recording light going to the right from the last semi-transparent plate in figure~\ref{fig:telescope_2}).

The main application of EPIC is to measure the separation---and especially their change over time---between the light centroids of two point-like sources. This can be achieved at a precision better than $\sigma_{\theta_\text{res}}$ from \Eq{eq:sigmatheta} if the differential astrometry fringe of \Eq{eq:C2pointsource2opt} is detected at high $\text{SNR}$. Denote by $\sigma_{\delta \theta}$ the precision to which the light-centroid separation $\vect{\theta}_{ba} \cdot \hat{\vect{d}}$ along the baseline direction can be measured. A standard error propagation calculation on \Eq{eq:C2pointsource2opt_finite} then yields a light-centroiding precision of
\begin{align}
\sigma_{\delta \theta} &= 2 \left[ n_1^\text{arr} n_2^\text{arr}  \sum_{\overline{k}} \left( \frac{\sigma_{\theta_\text{res}} \sigma_{\hat{C}}}{\left|\sin(\overline{k} \vect{d} \cdot \vect{\delta \theta}_{ba} )\right|} \frac{2\sqrt{2}c \sigma_k \sigma_t}{\widetilde{I}_a \widetilde{I}_b  \left|\mathcal{F}_a \mathcal{F}^*_b \right|}\right)^{-2} \right]^{-1/2} \label{eq:sigmaCentroid1}
\end{align}
when combining all channels labeled by $\overline{k}$, between $k_\text{min} < \overline{k} < k_\text{max}$. In the limiting case of two identical detectors ($n^\mathrm{arr}_1 = n^\mathrm{arr}_2 = 1$, $\eta_1 = \eta_2 = \eta$, $A_1 = A_2 = A$, etc.), fixed  spectral resolution ($\overline{k}/\sigma_k $ independent of $\overline{k}$), and identical, unpolarized, sources $a$ and $b$, we find:
\begin{align}
\sigma_{\delta \theta} \simeq \frac{2^{13/2} \pi^{5/4} \hbar^3 c^3}{k_\mathrm{B}^3} \frac{1}{\eta A d_\parallel} \sqrt{\frac{\sigma_t}{t_\text{obs}}}\sqrt{\frac{\sigma_k}{\overline{k}}} \begin{cases} 
\frac{1}{T_\text{s}^3 \theta_\text{s}^2} \left[\int_{x_\mathrm{min,s}}^{x_\mathrm{max,s}} \di x \, \frac{x^5}{(e^x -1)^2} \mathcal{F}^4_\text{s}(x k_\mathrm{B} T_\text{s}/\hbar c ,\vect{d}) \right]^{-1/2}\\
\frac{1}{T_{500}^3 \theta_{500}^2} \left[\int_{x_\mathrm{min,500}}^{x_\mathrm{max,500}}\di x \, x^{-1/3} \mathcal{F}^4_\text{q}(x k_\mathrm{B} T_{500}/\hbar c, \vect{d}) \right]^{-1/2},
\end{cases} \label{eq:sigmaCentroid2}
\end{align}
where $x_\mathrm{min,s} \equiv \hbar c k_\text{min}/k_\mathrm{B}T_\text{s}$ and $x_\mathrm{min,500} \equiv \hbar c k_\text{min}/k_\mathrm{B}T_\text{500}$, and similarly for $x_\mathrm{max,s}$ and $x_\mathrm{max,500}$.
To get to \Eq{eq:sigmaCentroid2} from \Eq{eq:sigmaCentroid1}, we approximated $\sin^2(\overline{k} \vect{d} \cdot \vect{\delta \theta}_{ba})  \sim 1/2$, and turned the sum over $\overline{k}$ into an integral according to $\sum_{\overline{k}} \to \int \di \overline{k} / 2 \sigma_k$. The functions $\mathcal{F}_\text{s}$ and $\mathcal{F}_\text{q}$ are defined in \Eqs{eq:formS} and~\ref{eq:formQ}, respectively. The effective baseline $d_\parallel$ is the magnitude of the baseline vector component of $\vect{d}$ parallel to the source separation $\vect{\theta}_{ba}$.

\begin{figure}
\centering
\includegraphics[width=0.7\textwidth]{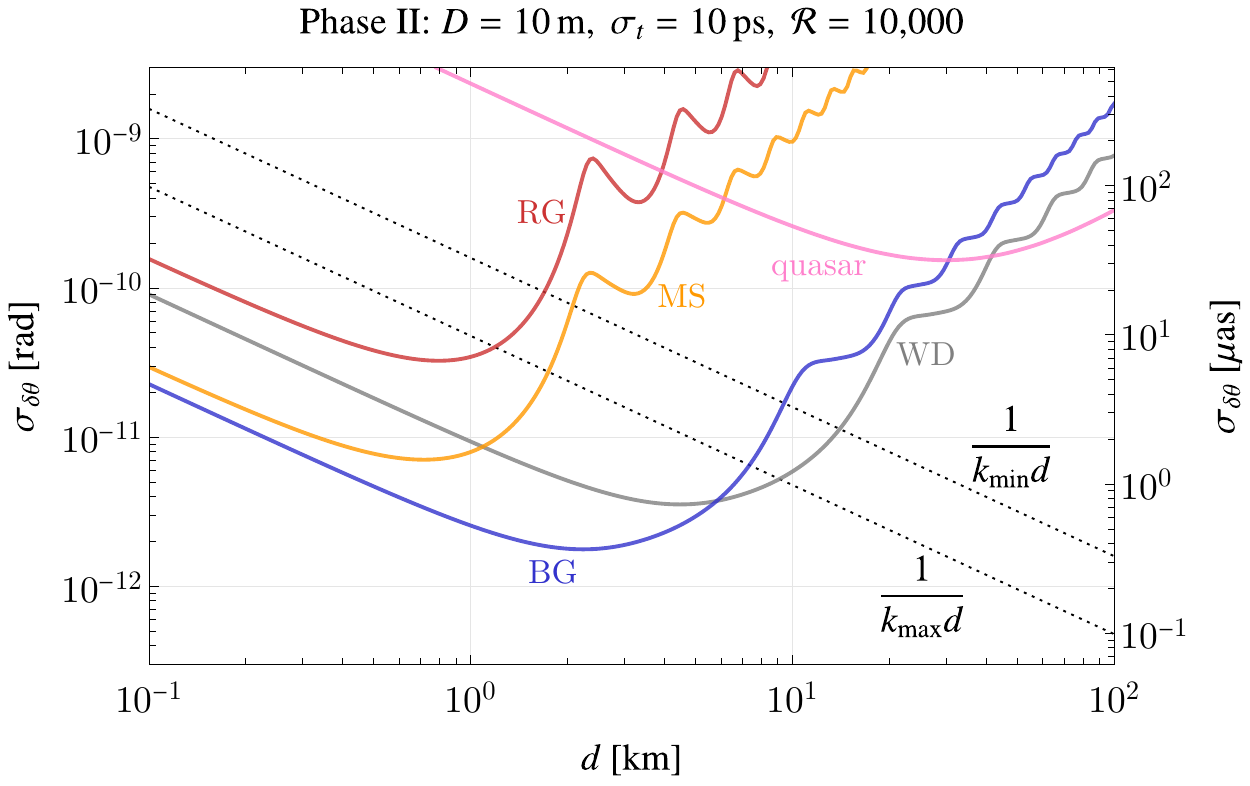}
\caption{Light-centroiding precision $\sigma_{\delta \theta}$ from \Eq{eq:sigmaCentroid2} as a function of baseline distance $d$ (assuming optimal alignment, i.e.~$\vect{d} \cdot \vect{\theta}_{ba} = d \theta_{ba}$) between \emph{identical} stars and quasars, indicated by color and contours (in units of $\mu \mathrm{as}$). The parameters for the telescopes (Phase II in table~\ref{tab:phases}), stars, and quasars are the same as in figure~\ref{fig:sigmaC}.} \label{fig:thetaCentroid1}
\end{figure}

The factors outside the curly brackets in \Eq{eq:sigmaCentroid2} are in the observer's control: higher efficiencies $\eta$, larger collecting areas $A$, better timing precision $\sigma_t$, longer observation times $t_\text{obs}$, and higher spectral resolution $\overline{k}/\sigma_k$ generally improve the light-centroiding precision. Inside the curly brackets, we have provided the expression relevant for astrometry between two identical stars (top line) of temperature $T_\text{s}$ and angular radius $\theta_\text{s}$, and two identical quasar images (bottom line) with apparent angular optical radius of $\theta_{500}$.cLight centroiding precision improves for higher-temperature sources (higher $T_\text{s}$), and for sources that subtend larger angles on the sky (larger $\theta_\text{s}$ in the case of stars, or $\theta_{500}$ for quasars). 

\begin{figure}
\centering
\includegraphics[height=0.7\textwidth, trim = 0 0 0 0]{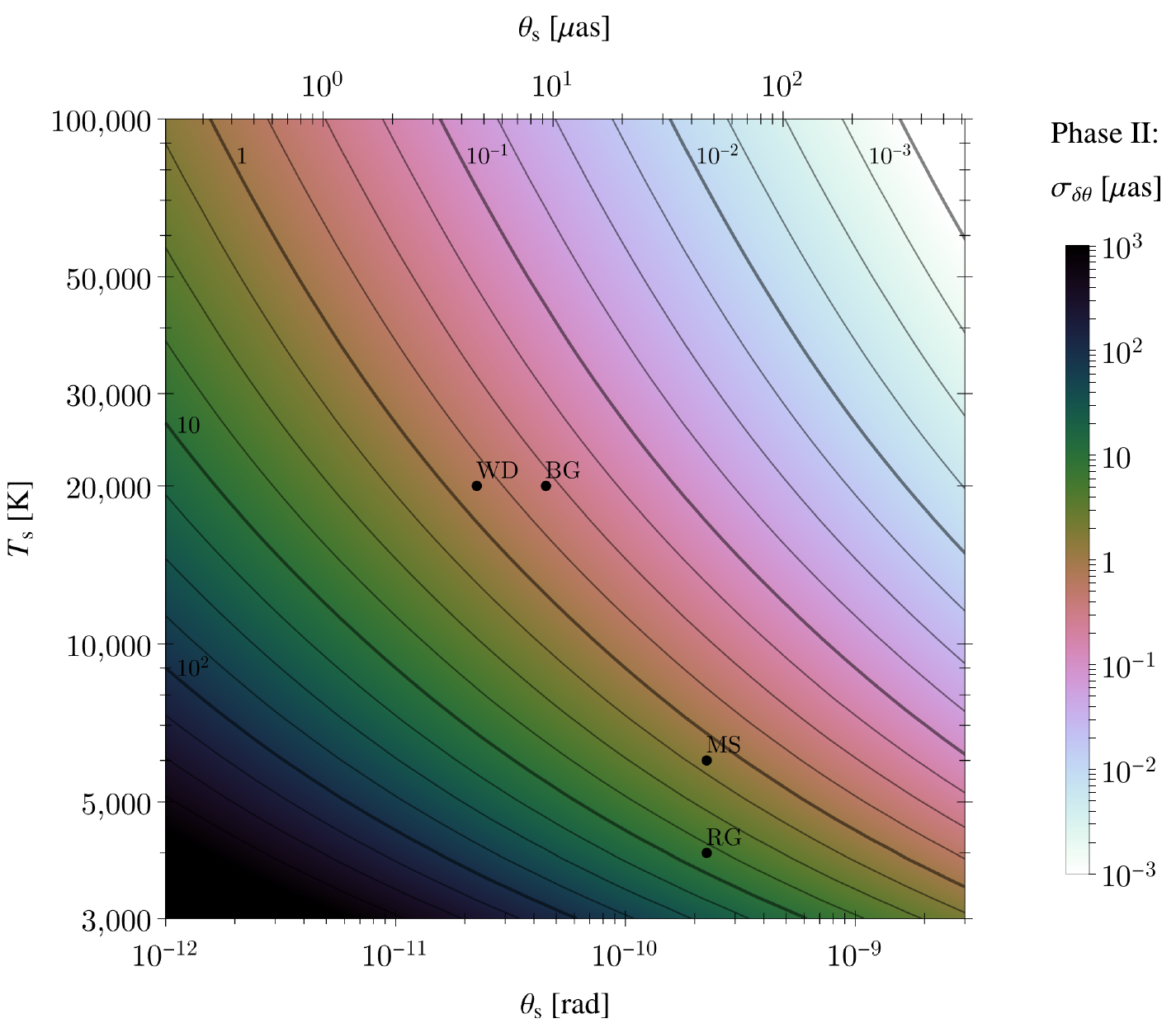}
\caption{Light-centroiding precision $\sigma_{\delta \theta}$ from \Eq{eq:sigmaCentroid2} between two identical stars of angular size $\theta_\text{s} = R_\text{s} / D_\text{s}$ and temperature $T_\text{s}$, at the optimal baseline distance $d^\text{opt}$ for a Phase II EPIC (see table~\ref{tab:phases}).}\label{fig:thetaCentroid3}
\end{figure}

In figure~\ref{fig:thetaCentroid1}, we plot the light-centroiding precision of \Eq{eq:sigmaCentroid2} for the same 4 stars and fiducial quasar, and instrumental parameters as in figure~\ref{fig:sigmaC}; e.g.~the gray curve is the differential astrometric light-centroiding precision between two identical white dwarfs of $T_\text{s} = 10^4 \, \text{K}$ with $R_\text{s} = 10^{-2} \, R_\odot$ at $D_\text{s}  = 10 \, \text{pc}$. We can see that the astrometric error scales as $\sigma_{\delta \theta} \propto 1/d_\parallel$ until the sources are resolved, so there is an optimum baseline distance $d_\text{opt}$. For quasars, the optimum baseline distance is just a simple linear function of $\theta_{500}$:
\begin{align}
d_\text{opt,q} \approx 2.31\,\mathrm{km} \, \left(\frac{10^{-11}\,\mathrm{rad}}{\theta_{500}} \right) \label{eq:doptq}
\end{align}
where we assumed $k_\mathrm{min} = 2\pi/(1000\,\mathrm{nm})$ and $k_\mathrm{max} = 2\pi/(300\,\mathrm{nm})$. 
In figure~\ref{fig:thetaCentroid3}, we display the light-centroiding precision at this optimal baseline distance $d_\text{opt}$, assuming $d = d_\parallel$ for stellar sources as a function of $\theta_\text{s}$ and $T_\text{s}$. We show the equivalent quantity in figure~\ref{fig:thetaCentroid500} for quasars as a function of $\theta_{500}$, assuming the optimum baseline of \Eq{eq:doptq}.

\subsection{Atmospheric phases}
\label{sec:atm}

In this section, we outline the effects of non-negligible propagation phases on EPIC at ``large'' angles. We compute the size of the atmospheric aberration and its resulting limitation on the field of view of EPIC. 

Let us start with the same electric field as in \Eqs{eq:efield} and~\ref{eq:Efield2}, but retain the phases $\widetilde{\phi}^{(p)}_{i\alpha}(t_p)$ \emph{not} intrinsic to the source but  accumulated along the path from the source $i$ to the detector $p$ for a mode of frequency $k_\alpha$:
\begin{alignat}{2}
	E_p(t_p) &= \epsilon \sqrt{1-\epsilon^2} \sum_{i=1}^{N_\theta} \sum_{\alpha = 1}^{N_k} \sum_{n=0}^1  e^{-i n k_\alpha  \ell_p}  E_{i\alpha} \exp\left\lbrace i \left[k_\alpha (c t_p - r_{ip}) + \phi^\text{em}_{i\alpha} + \widetilde{\phi}^{(p)}_{i\alpha}(t_p)  \right] \right\rbrace, \label{eq:Efieldph}
\end{alignat}
In the continuum limit, the relevant extended-path intensity correlator is modified to:
\begin{alignat}{2}
	&C(\vect{d},\tau) = \frac{\epsilon^4 \left(1-\epsilon^2\right)^2 }{\avg{I_1}\avg{I_2}} \int \frac{\di^2 b_1}{A_1} \int \frac{\di^2 b_2}{A_2} \int \di (\Delta t) \frac{1}{\sqrt{2\pi}\sigma_t} e^{\frac{-(\Delta t)^2}{2\sigma_t^2}}\label{eq:C2ph}\\
	& \times \left| \int \di k \int \di \Omega \, \frac{\di I}{\di k \, \di \Omega}\big(1+e^{- i k \ell_1}\big)\big(1+e^{ i k \ell_2}\big) e^{i\left[k\left(c (\Delta t - \tau) - \hat{\vect{\theta}} \cdot (\vect{d}+\vect{\Delta b})\right) + \widetilde{\phi}^{(1)}(t,\hat{\vect{\theta}},k)- \widetilde{\phi}^{(2)}(t+\tau,\hat{\vect{\theta}},k) \right]} \right|^2 \nonumber,
\end{alignat}
which should be regarded as the generalization of \Eq{eq:C2} to account for propagation phases. The limit towards conventional intensity interferometry is attained in the limit $\ell_p \to 0$. The expected intensities $\langle I_p \rangle$ are unchanged from \Eq{eq:I2}. 

For two point-like sources $a$ and $b$, we find
\begin{alignat}{2}
	C^\text{opt}(\vect{d}) \simeq  \frac{1}{4\sqrt{2} c \sigma_k \sigma_t} & \Bigg \lbrace \widetilde{I}_a^2 +  \widetilde{I}_b^2 + 2 \widetilde{I}_a \widetilde{I}_b \cos\left[ \overline{k} \vect{d} \cdot \vect{\delta\theta}_{ba}  +  \Delta\widetilde{\phi}^{(1,2)}(\overline{k}, \hat{\vect{\theta}}_a,\hat{\vect{\theta}}_b)\right] e^{\frac{-\sigma_k^2}{4} \left(\vect{d} \cdot \vect{\delta \theta}_{ba} \right)^2}  \Bigg \rbrace \label{eq:C2pointsource2optph}
\end{alignat}
as the generalization of \ref{eq:C2pointsource2opt} for example. In the above, we have defined the doubly-differential noise phase:
\begin{align}
\Delta\widetilde{\phi}^{(1,2)}(\overline{k}, \hat{\vect{\theta}}_a,\hat{\vect{\theta}}_b) \equiv \widetilde{\phi}^{(1)}(\overline{k}, \hat{\vect{\theta}}_a)  + \widetilde{\phi}^{(2)}(\overline{k}, \hat{\vect{\theta}}_b) - \widetilde{\phi}^{(1)}(\overline{k}, \hat{\vect{\theta}}_b) - \widetilde{\phi}^{(2)}(\overline{k}, \hat{\vect{\theta}}_a), \label{eq:doublediffphase}
\end{align}
whose time dependence we have implicitly suppressed (since the relevant quantity is its time average).

We are particularly interested in calculating the atmospheric contribution to the doubly-differential phase of \Eq{eq:doublediffphase} due to the fluctuating index of refraction along the light paths.
Denote by $n[\overline{k},\vect{x}_{ip}(s)]$ the deviation of the index of refraction from its mean at wavenumber $\overline{k}$  along the path from detector $p$ to source $i$: $\vect{x}_{ip}(s) \equiv \vect{r}_p + \hat{\vect{\theta}}_i s$ with $0 < s < r_{ip}$. The total refractive phase accumulation along the path is:
\begin{align}
\widetilde{\phi}^{(p)}(\overline{k},\hat{\vect{\theta}}_i) = \overline{k} \int_0^{r_{ip}} \di s \, n[\overline{k},\vect{x}_{ip}(s)]. \label{eq:phi_atm}
\end{align}
We have explicitly subtracted out the mean index of refraction, so $\langle n[\overline{k},\vect{x}(s)] \rangle_\mathrm{atm}$ vanishes by construction along any path $\vect{x}(s)$ through the atmosphere, and thus also the corresponding $\langle \widetilde{\phi} \rangle_\mathrm{atm}$. This subtraction of the mean can be done to a sufficient fidelity, down to an effective precision less than a millimeter, corresponding to $\lesssim 3\,\mathrm{ps}$ delay~\cite{mendes2004high,hulley2007ray,stuhl2021atmospheric}.

However, there are \emph{fluctuations} in $n$, as measured by the structure function~\cite{kolmogorov1961classic,tatarskii1971effects,1981PrOpt..19..281R,2000JMOp...47.1111R}:
\begin{align}
D_n(r) \equiv \left\langle (n[\vect{r}'+\vect{r}]-n[\vect{r}'])^2 \right\rangle _\mathrm{atm} = C_n^2(h) r^{2/3} \quad \text{for }l_0 \ll r \ll L_0, \label{eq:Dn}
\end{align}
where we have dropped the dependence on $\overline{k}$ for notational simplicity.
The structure function does not depend on the reference position $\vect{r}'$ except through the slow dependence on altitude $h$ in the atmosphere. The two-thirds scaling follows from Kolmogorov's ansatz for turbulence, and is an \emph{overestimate} both above the outer scale $L_0 \sim 10^{1\pm1}\,\mathrm{m}$ (where $D_n(r)$ asymptotes to a constant $\langle n^2 \rangle$) and below the inner scale $l_0 \sim 10^{-2.5\pm 0.5} \, \mathrm{m}$ (where $D_n(r) \approx r^2$). The constant $C_n$ has dimensions of $\mathrm{[length]}^{-1/3}$, depends on the weather, and typically decreases with altitude. 
Typical values are $C_n \sim 10^{-8}\,\mathrm{m}^{-1/3}$ for altitudes between $1$--$10\,\mathrm{km}$, as depicted in figure~\ref{fig:SLC} for the SLC night model~\cite{miller1979turbulence}.
Also useful is the correlation function
\begin{align}
B_n(r) \equiv \left\langle n[\vect{r}'+\vect{r}] n[\vect{r}'] \right\rangle _\mathrm{atm}. \label{eq:Bn}
\end{align}
The structure function can be expressed in terms of the correlation function as $D_n(r) = 2 [B_n(0) - B_n(r)]$.

\begin{figure}
\centering
\includegraphics[width = 0.7\textwidth]{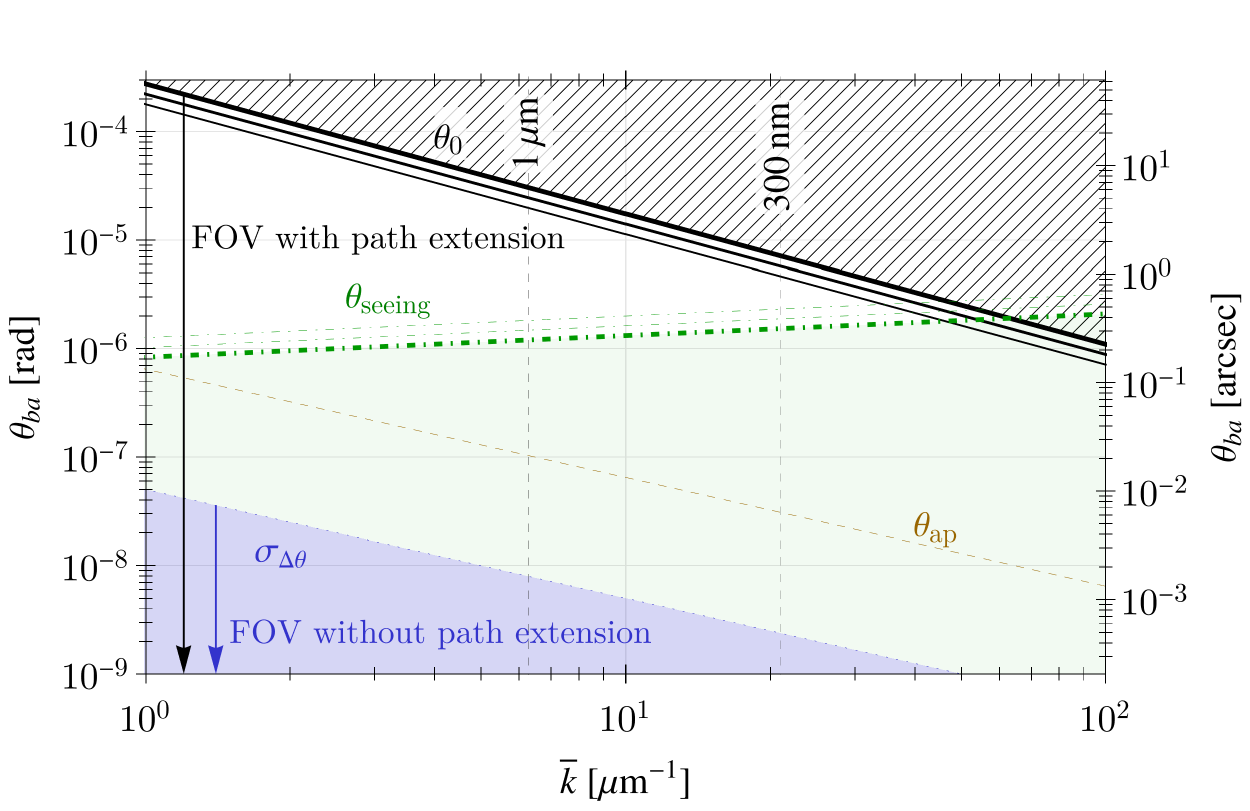}
	\caption{Isoplanatic patch angle $\theta_0$ (black solid lines) as a function of wavenumber $\overline{k}$ for three different zenith angles $\gamma = \lbrace 0^\circ,45^\circ,60^\circ\rbrace$. Atmospheric refraction suppresses the intensity fringe for differential astrometry at an angle $\theta_{ba}$ as $\exp\lbrace - (\theta_{ba}/\theta_0)^{5/3} \rbrace$ (cfr.~\Eq{eq:C2pointsource2optatm}). The blue dotted line indicates the dynamic range $\sigma_{\Delta \theta}$ of \Eq{eq:sigmaDeltatheta} for intensity interferometry without path extensions at $d = 100\, \text{km}$ and $\overline{k}/\sigma_k = 5{,}000$. The effective field of view (FOV) for traditional intensity interferometry is $\sigma_{\Delta \theta}$, while it is as large as $\theta_0$ for EPIC, an increase by more than three orders of magnitude at the chosen parameters. We plot the angular size $\theta_\text{seeing}$ of the seeing disk from \Eq{eq:thetaseeing} as green dash-dotted lines ($\gamma = \lbrace 60^\circ,45^\circ,0^\circ\rbrace$) above this line, two sources can be resolved from a ground-based observatory at an excellent site without adaptive optics. The brown dashed line indicates the critical angle $\theta_\mathrm{ap}$ above which wavefront errors need to be taken into account for a $4\,\text{m}$ aperture diameter (the $\vect{b}_p$ dependence in \Eq{eq:C2pointsource2optph}).}\label{fig:thetaIso}
\end{figure}

The doubly differential atmospheric refraction phase from \Eq{eq:doublediffphase} is a (rapidly) fluctuating quantity with expected standard deviation:
\begin{align}
&\sigma_\phi^\text{atm} \equiv \sqrt{\Big \langle \left[\Delta\widetilde{\phi}^{(1,2)}(\overline{k}, \hat{\vect{\theta}}_a,\hat{\vect{\theta}}_b)\right]^2 \Big \rangle_\mathrm{atm} } \simeq \sqrt{2} \left( \frac{\theta_{ba}}{\theta_0(\overline{k},\gamma)} \right)^{5/6}, \label{eq:sigma_phi_atm}
\end{align}
where $\gamma$ is the angle of the sources away from zenith (assumed to be the same for both detector locations, a reasonable assumption for baselines smaller than Earth's radius $d \ll R_\oplus$), and $\theta_0$ is the \emph{isoplanatic patch angle}~\cite{fried1982anisoplanatism,smith1993atmospheric}:
\begin{align}
\theta_0(\overline{k},\gamma) \equiv \left[ 2.9 \overline{k}^2 \int \dd s \,  C_n^2(s \cos \gamma) s^{5/3} \right]^{-3/5} \approx 2.7\,\mathrm{arcsec} \left( \frac{\overline{\lambda}}{500\,\text{nm}}\right)^{6/5} \cos^{5/8}(\gamma).\label{eq:isoplanatic}
\end{align}
In the latter numerical estimate, we have used the SLC night model. We derive \Eq{eq:sigma_phi_atm} in appendix~\ref{app:atm}. This fluctuating atmospheric phase suppresses the fringe of differential astrometry as $\int \di (\Delta \widetilde{\phi}) (\sqrt{2\pi}\sigma_\phi^\text{atm})^{-1} e^{-i \Delta \widetilde{\phi}^2/2 (\sigma_\phi^\text{atm})^2} e^{-i \Delta \widetilde{\phi} }= e^{- (\sigma_\phi^\text{atm})^2/2}$, which when applied to \Eq{eq:C2pointsource2opt} gives:
\begin{align}
C^\text{opt}(\vect{d}) \simeq  \frac{1}{4\sqrt{2} c \sigma_k \sigma_t} & \Bigg \lbrace \widetilde{I}_a^2 +  \widetilde{I}_b^2 + 2 \widetilde{I}_a \widetilde{I}_b \cos\left[ \overline{k} \vect{d} \cdot \vect{\delta\theta}_{ba}  \right] 
\exp\left\lbrace - \left[ \frac{\theta_{ba}}{\theta_0(\overline{k},\gamma)} \right]^{5/3} \right\rbrace
e^{\frac{-\sigma_k^2}{4} \left(\vect{d} \cdot \vect{\delta \theta}_{ba} \right)^2}  \Bigg \rbrace. \label{eq:C2pointsource2optatm}
\end{align}
with any other contributions to $\Delta\widetilde{\phi}^{(1,2)}(\overline{k}, \hat{\vect{\theta}}_a,\hat{\vect{\theta}}_b)$ other than from the atmosphere ignored. Thus, any source separation smaller than the isoplanatic patch angle, i.e.~$\theta_{ba} \lesssim \theta_0(\overline{k},\gamma)$, can be considered acceptable.

The atmospheric phase error $\sigma_\phi^\mathrm{atm}$ for intensity interferometry is dominated by atmospheric turbulence in the \emph{upper} atmosphere, as the quantity $h^{8/3} C_n^2(h)$ typically peaks near $h \sim 10\,\mathrm{km}$. Therefore, the altitude of the telescopes plays only a minor role, and the observations can be taken near sea level, even in comparatively poor atmospheric conditions. This is to be contrasted with the size of the seeing disk:
\begin{align}
\theta_\text{seeing}(\overline{k},\gamma) = 0.92 \, \overline{k}^{1/5} \left[\int \dd s \,  C_n^2(s \cos \gamma) \right]^{3/5} \approx \frac{0.29\,\text{arcsec}}{\cos^{3/5} \gamma} \left( \frac{500\,\text{nm}}{\overline{\lambda}}\right)^{1/5}, \label{eq:thetaseeing}
\end{align}
where the SLC night model was used again to obtain the (optimistic) numerical estimate.
At excellent observing sites, the seeing is smaller than the isoplanatic angle, so suitable sources can be resolved from the ground with some idea of their separation before EPIC is performed.

\section{Observatory Design}
\label{sec:tech}

In this section, we discuss the technical aspects of the extended-path intensity correlation technique. We begin by outlining the telescope design and array layout (\Sec{sec:design}), and sketch out the operational protocol of an EPIC observation (\Sec{sec:observation}). We focus on two technological advances on the original intensity interferometry setup that allow us to perform intensity interferometry (including EPIC) on fainter sources than previously possible: fast photodetectors and readout electronics (\Sec{sec:photodetectors}) and spectroscopic splitting (\Sec{sec:grating}). The required tolerances on the optically active surfaces and timing are summarized in \Sec{sec:opt_tolerances}; a detailed conceptual design report is left for future work.

\subsection{Overview}
\label{sec:design}
\begin{table}[tp]
    \centering
    \renewcommand{\arraystretch}{1.5} 
    \begin{tabular}{l | r r r r | S S S}
    \hline \hline
	& \multicolumn{1}{c}{$D$} & \multicolumn{1}{c}{$\sigma_t$} & \multicolumn{1}{c}{$\mathcal{R}$} & \multicolumn{1}{c|}{$n_\mathrm{arr}$} & \multicolumn{1}{c}{$\sigma_{\delta \theta}  \, [\mathrm{\mu as}]$} & \multicolumn{1}{c}{$\sigma_{\hat{\vect{\theta}}} \, [\mathrm{arcsec}]$} & \multicolumn{1}{c}{$\sigma_{\Delta \theta} \, [\mathrm{arcsec}]$} \\
	\hline
	Phase I  & {$4\,\mathrm{m}$} & {$30\,\mathrm{ps}$} & $5{,}000$ & 1 & 22.3 & 5.24 & 0.164 \\
	Phase II  & {$10\,\mathrm{m}$} & {$10\,\mathrm{ps}$} & $10{,}000$ & 1 & 1.46 & 1.75 & 0.327\\
	Phase III  & {$10\,\mathrm{m}$} & {$3\,\mathrm{ps}$} & $20{,}000$ & 10 & 0.0564 & 0.524 & 0.656\\
    \hline\hline
    \end{tabular}
    \caption{
	Fiducial EPIC array parameters in the near, medium, and long term (Phases I, II, and III): aperture diameter $D$, relative timing precision $\sigma_t$, spectral resolution $\mathcal{R}$, and number $n_\mathrm{arr}$ of telescope pairs sharing the same baseline. We assume a photodetection efficiency of $\eta = 0.5$ on top of the beamsplitters' reflectivity/transmission coefficients. We quantify their performance by the optimum light-centroiding precision $\sigma_{\delta \theta}$ after an integration time $t_\mathrm{obs} = 10^4 \,\mathrm{s}$ for a pair of Sun-like stars at a distance of $100\,\mathrm{pc}$ as well as the characteristic global astrometric precision $\sigma_{\hat{\vect{\theta}}}$ and angular dynamic range $\sigma_{\Delta \theta}$ (at the optimum centroiding baseline). For such source pair, the optimum (projected) baseline distance is $\hat{\vect{\theta}}_{ba} \cdot \vect{d} \approx 0.71\,\mathrm{km}$, corresponding to a fiducial angular resolution of $\sigma_{\theta_\mathrm{res}} \approx 23 \, \mathrm{\mu as}$ at $\overline{\lambda} = 500\,\mathrm{nm}$ for all three EPIC Phases.
    \label{tab:phases}
	}
\end{table}

\subsubsection*{Telescopes}
Each telescope within our proposed intensity interferometer array is essentially a stellar spectrograph with moderately high spectral resolution, an ``internal amplitude interferometer'' to create the path extension, and relatively loose optical tolerances (see \Sec{sec:opt_tolerances} below). Due to the high photon statistics required to tease out the weak intensity correlation, each telescope should have a large aperture and be equipped with ultra-fast photodetection electronics. The telescope design is illustrated in \Fig{fig:telescope}. The fiducial photodetection efficiency is taken to be $\eta = 0.5$, while other reference specifications---the aperture diameter $D$, timing resolution $\sigma_t$, spectral resolution $\mathcal{R}$, and number $n_\mathrm{arr}$ of telescopes per array site---are shown in table~\ref{tab:phases}. No gains are expected from a spectral resolution higher than $\sqrt{2} c \overline{k} \sigma_t \approx 5 \times 10^4$ at $\overline{k} = 2\pi / (500 \, \mathrm{nm})$ and $\sigma_t = 10\,\mathrm{ps}$, cfr.~\Eq{eq:C1pointsource}, which corresponds to a timing precision sufficiently fast to fully resolve the coherence time of the light in the spectral channel centered at a wavelength of $500\,\mathrm{nm}$. Larger apertures are not inconceivable, especially given the low to moderate tolerances required on the most of the optical surfaces, in particular the primary mirror. Large steerable telescopes with aperture diameters of $10$--$30\,\mathrm{m}$ are already in operation or construction for Cherenkov cosmic ray imaging~\cite{weekes2002veritas,hinton2004status,baixeras2004commissioning,anderhub2013design,cta2011design}.

\subsubsection*{Variable path extension}
The main conceptual difference between an ordinary spectrograph and a constituent telescope of an EPIC array is that the latter requires an internal beamsplitter and recombiner (with unequal path lengths). In light of \Eqs{eq:ellopt1} and~\ref{eq:ellopt2}, to cover all possible source separations $\theta_{ba}$ up to the isoplanatic patch angle $\theta_0$ of \Eq{eq:isoplanatic}, the internal interferometer of each telescope $p$ should have the capability to create a differential optical path length of
\begin{align}
	 \ell_p \sim \frac{\theta_{ba} d}{2} \lesssim \frac{\theta_0 d}{2} \sim 5\,\mathrm{cm} \left( \frac{\theta_0}{2 \,\mathrm{arcsec}}\right) \left( \frac{d}{10\,\mathrm{km}}\right).
\end{align}
This can be achieved by a corresponding tuning of the width $w_p$ of the internal interferometer (at fixed $\gamma$ and $\gamma'$), cfr.~\Fig{fig:telescope} and \Eq{eq:ell}. The precision to which $\ell_p$ is \emph{adjustable} should be on the order of a wavelength, so that the intensity interferometer can operate near the main (ghost) fringe, to adjust $\vect{\theta}^\mathrm{ref}_{ba}$ such that $\vect{\delta \theta}_{ba} \cdot \hat{\vect{d}} \sim \sigma_{\theta_\mathrm{res}}$, and to correct for aperture smearing, i.e.~the slightly different angle of incidence between the wavefronts from each source (the $\vect{b}$-dependence in \Eqs{eq:ellopt1} and~\ref{eq:ellopt2}). (In fact, as long as the aperture smearing effects are corrected to sub-wavelength precision, $|\vect{\delta \theta}_{ba} \cdot \vect{d}| \lesssim \sigma_{\Delta \theta}$ is also acceptable.) Aperture smearing is negated by taking $\gamma' - \gamma$ equal to $\theta_{ba} f / 2f'$ and with a relative tilt in the same direction as $\hat{\vect{\theta}}_{ba}$ (see figure~\ref{fig:telescope} and appendix~\ref{sec:geometricdelay}).

The positioning and control of the path extension require sub-wavelength tolerances over tens of centimeters, which are mild compared to what has been demonstrated in laboratory and observatory settings~\cite{gao2015measurement,2021MeScT..32d2003S}. For example, in the CHARA interferometer, a laser metrology unit controls the mirrors to a root-mean-squared error of $10$--$20\,\mathrm{nm}$ over distances of tens of meters~\cite{2005ApJ...628..453T}, while mirrors in adaptive optics systems achieve similar or better control over smaller distances and angles~\cite{2014SPIE.9148E..03B,10.1117/12.2560017}.

\subsubsection*{Array configuration}
In this work, we primarily consider the case of a single pair of telescopes separated by a distance $\vect{d}$. Interesting science can be obtained from a single pair of telescopes (\Sec{sec:applications}), but the potential of an intensity interferometry array can grow quadratically with the number of telescopes in the array. This quadratic scaling can be made precise: the signal-to-noise ratio on the excess intensity correlation scales linearly with the number of telescope pairs $n^\mathrm{arr}$ sharing the same baseline in the array (\Eq{eq:SNR}), so for the same light-centroiding precision $\sigma_{\delta \theta}$ (\Eq{eq:sigmaCentroid1}), the required integration time $t_\mathrm{obs}$ scales as $1/(n^\mathrm{arr})^2$. Equivalently, the light-centroiding precision can thus improve as $\sigma_{\delta \theta} \propto 1/n_\mathrm{arr}$ for fixed integration time.

For a fixed number of telescopes, the optimal layout of an EPIC array is determined by the interplay of several considerations~\cite{holdaway1999interferometric}. First, the optimal projected baseline separation falls in the $1\,\mathrm{km}$--$100\,\mathrm{km}$ range, primarily depending on the angular size of the source (and more weakly on source temperature). Second, the array should ideally have dense coverage in the $(u,v)$ plane for faithful ``image'' reconstruction of the emission region(s). Third, the array should have sensitivity to many (pairs of) sources of varying angular sizes and locations on the sky. Fourth, the SNR should be maximized.

The first three considerations dictate that, for a source at zenith, the telescopes should be distributed along the boundary of (a slightly perturbed) Reuleaux triangle~\cite{keto1997shapes}, a curved triangle of constant width (and the constant-width curve with the least rotational symmetry), with an overall size determined by the maximum desired baseline. This configuration creates the most uniform sampling (exact uniformity being impossible~\cite{golay1971point}) in a circular disk in Fourier space, suitable for image reconstruction~\cite{kogan1997optimization} and for detecting sources of varying angular sizes. Because such a telescope arrangement has a near-isotropic response, it is a good compromise for a source not exactly at zenith, and for Earth rotation synthesis and lower-elevation observations (third consideration). Slight, random perturbations from the perfectly symmetric Reuleaux triangle shape further break rotational symmetry, though ring- or ellipse-like configurations may be more desirable for low-declination observations~\cite{boone2002interferometric}. 

However, the fourth consideration favors some built-in degeneracy for an EPIC array layout (e.g.~in a square lattice or Y-shaped configuration) due to the aforementioned $n_\mathrm{arr}$ scaling of the signal-to-noise ratio for degenerate baselines. Furthermore, studies for optimal interferometric configurations have focused largely on the case of narrow-bandwidth radio interferometers; an EPIC array with good spectroscopic capabilities would have a broadband response, i.e.~any \emph{single} baseline would densely cover a ``radial spoke'' in Fourier space over an octave or more, which may affect the optimal array layout, and is an interesting topic for future study.

\subsection{Observational Procedure}
\label{sec:observation}
In sections~\ref{sec:snr} and~\ref{sec:applications}, we implicitly assume that the source separation vector $\vect{\theta}_{ba}$ is already roughly known, and then study the ultimate light-centroiding capabilities and applications of a pair of EPIC telescopes. In practice, however, $\vect{\theta}_{ba}$ may only be predetermined to some precision $\sigma_{\delta \theta}^{(0)}$ from prior observations. If this precision is of order the intrinsic angular resolution (or better), i.e.~$\sigma_{\delta \theta}^{(0)} \lesssim \sigma_{\theta_\mathrm{res}}$, then the EPIC telescopes can adjust their path extensions of \Eqs{eq:ellopt1} and~\ref{eq:ellopt2} to have the reference angle $\vect{\theta}_{ba}^\mathrm{ref}$ within about one fringe (or a few fringes) of the true separation $\vect{\theta}_{ba}$. In this case, the light-centroiding sensitivity estimates of \Sec{sec:snr} apply directly. 

The majority of stars and quasars on which EPIC is applicable should have good global astrometry in the \textit{Gaia} catalog~\cite{prusti2016gaia}, which is essentially complete up to the apparent magnitudes $m_V \lesssim 15$ of interest. \textit{Gaia} DR4 and DR5 will provide time-averaged sky localization expected at the $\mathcal{O}(20\,\mathrm{\mu as})$ level for these bright sources~\cite{lindegren2021gaia}. Relative astrometry for two sources at this precision will typically be better than Phase I's resolution, and not much worse than the optimal resolutions of Phase II and III. This prior knowledge should thus be sufficient to unambiguously perform extended-path intensity correlation, i.e.~determine $\vect{\theta}_{ba} \cdot \vect{d}$ from the set of excess fractional intensity correlations $C$ in each spectral channel~\cite{shortpaper}.

To understand how an EPIC measurement would work in practice, let us study the time dependence of the correlator over the course of an observing night. We choose variable path extensions $\ell_1^\mathrm{opt}(t)$ and $\ell_2^\mathrm{opt}(t)$ according to \Eqs{eq:ellopt1} and~\ref{eq:ellopt2} so that the $\vect{\theta}_{ba}^\mathrm{ref}(t)$ is consistent with the prior astrometry on the source from \textit{Gaia}. Due to the latter's finite precision, we will generally not satisfy \Eq{eq:path_extension} \emph{exactly}, and instead have a time-dependent difference:
\begin{alignat}{2}
	\vect{\theta}_{ba}(t) \cdot \vect{d}(t) - \left[\ell_1^\mathrm{opt}(t) + \ell_2^\mathrm{opt}(t)\right] 
	&= \delta\vect{\theta}_{ba}(t) \cdot \vect{d}(t),
\end{alignat}
with $\delta\vect{\theta}_{ba}(t) = \vect{\theta}_{ba}(t) - \vect{\theta}^\mathrm{ref}_{ba}(t)$ of characteristic size $\sigma_{\delta \theta}^{(0)}$.
The differential astrometric information in an EPIC observation enters through the fringes in \Eq{eq:C2pointsource2opt}, which are proportional to:
\begin{alignat}{2}
	\cos\left[ k \delta\vect{\theta}_{ba}(t) \cdot \vect{d}(t) \right]
	&= \cos\left( k \left[ \delta \theta_{ba}^\delta d_z \cos(\mathrm{dec}) + \mathcal{A} \sin\left(\frac{2 \pi t}{\mathrm{day}} + \Phi \right) \right] \right), \label{eq:arg_correlator}
\end{alignat}
where $\mathcal{A}$ is defined as $\mathcal{A}^2 \equiv \left(d_x^2 + d_y^2\right) \left[\left(\delta\theta_{ba}^\alpha\right)^2 \cos^2(\mathrm{dec})  + \left(\delta\theta_{ba}^\delta\right)^2 \cos^2(\mathrm{ra}) \right]$, and $\Phi$ as  $\tan \Phi \equiv \left[-\delta \theta_{ba}^\alpha d_y \cos (\mathrm{dec})+ \delta \theta_{ba}^\delta d_x \sin (\mathrm{dec})\right]/\left[{\delta \theta_{ba}^\alpha d_x \cos (\mathrm{dec})+\delta \theta_{ba}^\delta d_y \sin (\mathrm{dec})} \right]$. The component(s) $d_z$ ($d_x$, $d_y$) is the baseline projection towards the celestial North pole (lie in celestial equator plane), and ``ra'' and ``dec'' refer to the right ascension and declination of the source pair, while superscripts $\alpha$ and $\delta$ denote components along ra and dec.

\begin{figure}[t!]
	\centering
	\includegraphics[width=\textwidth]{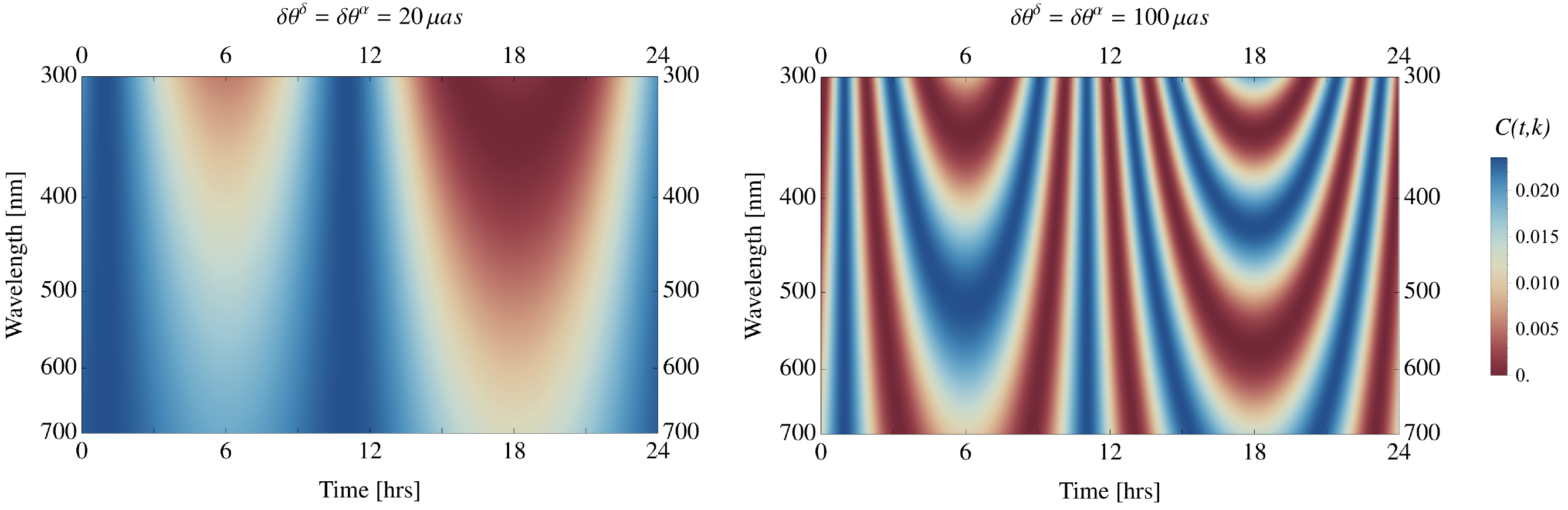}
	\caption{The expected excess fractional intensity correlation between two sources $a,b$ of equal brightness, as a function of time and wavelength. We assume a mismatch of the reference position $\vect{\theta}_{ba}^\mathrm{ref}$ and the true position $\vect{\theta}^{ba}$ of $\delta \theta^\alpha_{ba} = \delta \theta^\delta_{ba} = 20 \, \mathrm{\mu as}$ (left) and $100 \, \mathrm{\mu as}$ (right), baseline vector components $(d_x,d_y,d_z) = (1 \, \mathrm{km}, 0.5 \, \mathrm{km}, 0.5 \, \mathrm{km})$, spectral resolution $\mathcal{R} = 5{,}000$, and $\sigma_t = 10 \, \mathrm{ps}$. }
	\label{fig:epicobs}
\end{figure}

As \Eq{eq:arg_correlator} makes clear, the time variation in the argument of the correlator is coming almost exclusively from the term with amplitude $\mathcal{A}$ (proportional to the magnitude of the baseline in the equatorial plane) and sidereal frequency. The phase contribution from this change over an observation time $t_\mathrm{obs}$ is approximately:
\begin{alignat}{2}
	\mathcal{O}(1) \times \left( \frac{\sqrt{d_x^2 + d_y^2}}{1\,\mathrm{km}}\right) \left( \frac{500 \, \mathrm{nm}}{\lambda} \right) \left(\frac{\sqrt{(\delta \theta^\alpha)^2 \cos^2 \theta_{a}^\delta + (\delta \theta^\delta)^2 \sin^2\theta_{a}^\delta}}{20  \, \mathrm{\mu as}}\right)\left(\frac{ t_\mathrm{obs}}{10 \, \mathrm{hours}}\right).
\end{alignat}
In other words, for baselines at non-vanishing latitudes and with reference positions with \textit{Gaia} uncertainties, one can expect order-unity variation in the intensity correlation throughout an observing night, with smaller wavelengths experiencing faster ``fringe scanning''. We depict this effect in \Fig{fig:epicobs}, where over the course of 24 hours, the excess fractional intensity correlation in each spectral channel would sweep from left to right. In practice, one would chop up the observation period into shorter intervals, such that the correlation is constant in each interval separately. The source separation $\delta \vect{\theta}_{ba}$ can be extracted from fitting to the correlation in these time intervals across all spectral channels.

In some cases, the prior knowledge on the source separation may be significantly worse than the angular resolution of the EPIC pair of telescopes, but practically always better than their dynamic range: $\sigma_{\theta_\mathrm{res}}\ll \sigma_{\delta \theta}^{(0)} < \sigma_{\Delta \theta}$. 
Poorly known separations can occur when sources are within $0.3\,\mathrm{arcsec}$ of each other (e.g.~photometric or spectroscopic binaries, double stars, or unresolved multiple images of a strongly lensed quasar) so that they are typically not resolved in standard imaging telescopes, either due to diffraction or seeing limitations. 
Because the map $C \mapsto \overline{k} \vect{d} \cdot \vect{\theta}_{ba}$ is multivalued with roughly $\mathcal{R}$ fringes, a measurement on the correlation $C$ in each spectral channel can correspond to $\mathcal{O}(\sigma_{\delta \theta}^{(0)} / \sigma_{\theta_\mathrm{res}})$ physically viable solutions, no matter how high the signal-to-noise ratio. This (discrete) degeneracy in parameters can, in principle be broken by combining the information from all spectral channels, but this procedure can be computationally demanding given the large data volume and the motion of Earth and the source, both of which change the dot product angle $\hat{\vect{d}} \cdot \vect{\theta}_{ba}$.

In a large \emph{array} of EPIC telescopes, one can perform a logarithmic search for the separation vector $\vect{\theta}_{ba}$. One could use a pair of nearby EPIC telescopes (with small $d$ or serendipitously small $\vect{d} \cdot \hat{\vect{\theta}}_{ba}$) whose angular resolution $\sigma_{\theta_\mathrm{res}}^{(1)}$ is of order the prior uncertainty $\sigma_{\delta \theta}^{(0)}$ on the source separation; this first pair of EPIC telescopes would then improve the relative light-centroiding precision to $\sigma_{\delta \theta}^{(1)}$, an improvement over $\sigma_{\theta_\mathrm{res}}^{(1)} \sim \sigma_{\delta \theta}^{(0)}$ by a factor roughly equal to the SNR (see \Eq{eq:sigmaCentroid1}). A second pair of telescopes with a larger projected baseline $\vect{d} \cdot \hat{\vect{\theta}}_{ba}$ and corresponding resolution $\sigma_{\theta_\mathrm{res}}^{(2)} \sim \sigma_{\delta \theta}^{(1)}$ could then improve the light-centroiding precision to $\sigma_{\delta \theta}^{(2)}$, and so on. An EPIC array could thus ``zoom in'' on the source separation quite efficiently, reaching its ultimate operating point near the main fringe with only a logarithmically large number of steps, since the relative centroiding precision improves as $\mathcal{O}(\mathrm{SNR}^n)$ after $n$ iterations. Fortuitously, this should be possible in Reuleaux triangle, circular, and Y-shaped array configurations, which have centrally condensed response functions, i.e.~dense Fourier coverage at small spatial frequencies~\cite{keto1997shapes,boone2002interferometric}. This protocol would be robust but bring some overhead to the observation time of any new pair of unresolved sources.

\subsection{Photodetectors}
\label{sec:photodetectors}

The foremost requirement for (extended-path) intensity interferometry is a very fast photon counter. The fringe contrast scales as $C \propto 1/\sigma_t$ and the overall SNR as $\propto 1/\sqrt{\sigma_t}$, where $\sigma_t$ is the standard deviation for the differential photon arrival time, which includes the timing resolution of both detectors, readout electronics, and synchronization errors, cfr.~\Eqs{eq:sigmat}. The timing precision of photodetectors is conventionally quantified by their time jitter $t_\mathrm{jitter}$, the full width at half maximum (FWHM) of the signal curve as a function of time. For ease of comparison, we assume the signal curve is Gaussian, and take the standard deviation for a single detector's timing resolution to be $t_\mathrm{res} = t_\mathrm{jitter} / (2 \sqrt{2 \ln 2})$. Assuming the readout electronics are not the bottleneck and both telescopes share the same type of photodetector, then the \emph{relative} timing precision from \Eq{eq:sigmat} will be $\sigma_t \simeq t_\mathrm{jitter} / (2 \sqrt{\ln 2}) \approx 0.72 \, t_\mathrm{jitter}$.

\begin{figure}
	\centering
	\includegraphics[width=0.7 \textwidth]{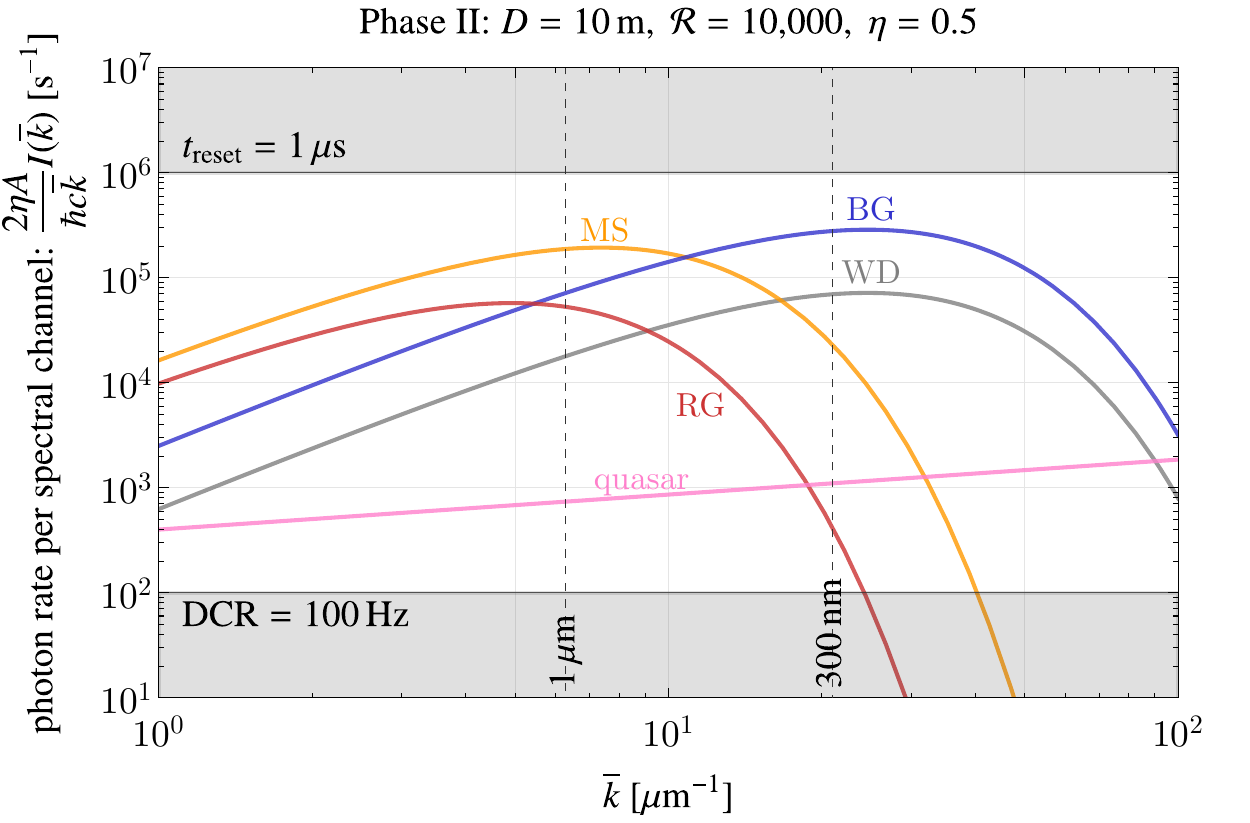}
	\caption{Rate of recorded photons per spectral channel incident on the aperture of one telescope. We use the same parameters as in figure~\ref{fig:sigmaC} for the source pairs and Phase II detectors (also in table~\ref{tab:phases}).
	The upper shaded region corresponds to photon rates in excess of $10^6\,\mathrm{s}^{-1}$, where significant sensitivity loss would occur (due to missed photon counts) for a photodetector reset time of $t_\mathrm{reset} = 1\,\mathrm{\mu s}$.
	Likewise, the lower shaded region corresponds to source photon rates below the dark count rate (DCR) of the photodetector, where loss of fringe contrast and thus sensitivity would occur because dark counts from electronic noise are uncorrelated.
	If $N_\mathrm{subap}$ subapertures are used, then photon rates are reduced by a factor of $N_\mathrm{subap}^{-1}$, weakening the reset time constraint but tightening the DCR requirement.
	}
	\label{fig:cps}
\end{figure}

In addition, we require that the photodetectors have a sufficiently small reset time $t_\mathrm{reset}$, i.e.~the ``dead time'' the detector needs to revert to its original state after a single-photon absorption event. The classes of photodetectors with the best timing jitters typically cannot distinguish between single-photon or multiphoton events (in contrast to photon-number-resolving devices). A reset time (much) smaller than the typical time separation between photon events (in each spectral channel) is thus necessary to have a high duty cycle and not wash out intensity fringe correlations. For most sources under consideration and our assumed telescope parameters (table~\ref{tab:phases}), sub-microsecond reset times $t_\mathrm{reset} \lesssim 1 \, \mu\mathrm{s}$ should be sufficient, see \Fig{fig:cps} for example.

Thirdly, we desire a high quantum efficiency $\eta_p$ (cfr.~\Eq{eq:Ihat1}) to maximize the SNR (\Eq{eq:SNR}) and light-centroiding precision (\Eq{eq:sigmaCentroid2}). The fact that each photodetector pixel is only responsible for reading out a narrow bandwidth is advantageous in this regard, as each pixel can be tuned to maximize the photodetection efficiency in that particular spectral channel.

Finally, photodetectors in general will have spurious events at rate known as the dark count rate (DCR), which may arise intrinsically (e.g.~from electronic noise) or environmentally (from cosmic rays, blackbody radiation, stray light, etc.). Some environmental backgrounds can be partially shielded by an appropriate spectral filter on the pixels, as each pixel of the photodetector only needs to read out a narrow band. However, the tolerances on the DCR are somewhat loose, as the relatively large photon flux from the sources of interest constitutes our main source of statistical noise: in the sea of incoming photons, we are searching for those very rare photon pairs that simultaneously (after correcting for propagation differences) impinge on both telescopes. The DCR is not the limiting factor when below the sources' combined photon flux. If larger, it would reduce the intensity correlation fringe contrast, as the DCR would give false positives that are uncorrelated between the two telescopes.

The above four requirements point towards to two possible technologies for single-photon detectors, namely superconducting nanowire single photon detectors (SNSPDs) and single photon avalanche diodes (SPADs), also known as Geiger mode avalanche photodiodes (GmAPD). The present state of the art for these two classes of devices is summarized below. 

Given the demonstrated technology and the real-word applications of both SNSPDs and SPADs outside pristine laboratory conditions, we are optimistic that arrays of photodetectors tailored to the needs of multichannel intensity interferometry are feasible. In figure~\ref{fig:cps}, we plot the photon rate in a single spectral channel of an EPIC Phase II telescope, for the same benchmark source pairs as in figure~\ref{fig:sigmaC}, showing that these rates fall between achievable DCRs and inverse reset times $t_\mathrm{reset}^{-1}$. Ultrafast photon detection over a large array of pixels (each optimized at different wavelengths) has not been a priority of most current research efforts, but it is clear from our summaries below that the requisite technology exists, and that there is no obstacle preventing its further development. Our proposal offers unique prospects for using this cutting-edge technology for a novel use case in astronomy, and we hope the exciting scientific applications (\Sec{sec:applications}) will further intensify research and development in this area.

\subsubsection*{Superconducting Nanowire Single Photon Detectors}
\label{sec:snspds}

Single photon detection with superconducting nanowires, first developed about two decades ago, is a technique with extraordinary promise for quantum optics and many other applications, owing to picosecond-level timing jitter, sub-millihertz DCR, near-unity quantum efficiency, sub-nanosecond recovery times~\cite{SteinhauerReview,Hadfield2020}, and ability to operate from the infrared (IR) to optical range~\cite{ShawBerggren2020} and even in the ultraviolet (UV)~\cite{WollmanUV}. 
The physical principle upon which SNSPDs operate is simple: after absorption of a single photon by the superconducting nanowire, a large number of Cooper pairs are broken, inducing a local hotspot in the normal, non-superconducting phase. A bias current is applied so that this hotspot subsequently expands due to Joule heating, further increasing the resistance, generating a voltage pulse that can be amplified and registered with room-temperature electronics. The energy of the hotspot is made to dissipate quickly thereafter to return the nanowires back to their superconducting state. 

The system timing jitter for single SNSPDs has steadily improved over the last decade, from $18\,\mathrm{ps}$ ($15\,\mathrm{ps}$ intrinsic)~\cite{You2013} to as low as $2.6\,\mathrm{ps}$ for $400\,\mathrm{nm}$ and $532\,\mathrm{nm}$ visible light with a niobium nitride (NbN) nanowire at $0.9\,\mathrm{K}$ operating temperatures \cite{ShawBerggren2020}, with possible further improvements. 
Recently, SNSPDs have made further advances, from the UV to the IR. A UV SNSPD with an active area of $56\,\mathrm{\mu m}$ in diameter was demonstrated to have a timing jitter down to $62\,\mathrm{ps}$, a system efficiency of up to 84\% for narrowband designs, a DCR of $0.25\,\mathrm{counts/hr}$, and detector dead time less than $250\,\mathrm{ns}$~\cite{WollmanUV}. On the other side of the spectrum, a free-space coupled NbN SNSPD was shown to have a $14\,\mathrm{ps}$ time jitter, a DCR below $0.1\,\mathrm{Hz}$, a system efficiency of up to $84\%$, with an active area of $22\times 15$~$\mu$m$^2$ at $1550\,\mathrm{nm}$ wavelengths~\cite{SpiropuluShaw2021}. Optimization of the system's design and fabrication can push photodetection efficiencies as high as 98\%~\cite{NamVerma2020}, and with tungsten silicide (WSi) nanowires, the energy threshold can be lowered to perform single-photon detection at $10\,\mathrm{\mu m}$ wavelengths.

Multiplexing SNSPDs into arrays with many pixels---while simultaneously retaining their exquisite performance---presents new challenges. To date, the largest realized SNSPD array consists of $32\times32$ pixels with a row-column multiplexing architecture~\cite{Namkilo}, optimized at $1550\,\mathrm{nm}$ with an area of $1.6 \times 1.6 \, \mathrm{mm^2}$ ($0.96\times 0.96 \, \mathrm{mm}^2$ active area) and operated at $0.73\,\mathrm{K}$, which exhibited a timing jitter of at least $250\,\mathrm{ps}$. Due to the small filling factor and low pixel efficiency, the combined system efficiency was only 8\%, but there are clear paths to improve these factors~\cite{Namkilo}, and other thermal-coupling array designs may be scaled into the megapixel range~\cite{NamThermal}. The array cannot distinguish between photons that impinge on the detector with a relative arrival time less than the timing jitter. Saturation of the photodetector array restricts the maximum apparent brightness of sources, as we illustrate in figure~\ref{fig:cps}. We should note that lab measurements using this array were done over second-long timescales, so scaling up to hours of observation is yet to be demonstrated. Finally, we note that a \emph{single} nanowire alone can serve \emph{multiple} pixels using time-of-flight information: ref.~\cite{berggren2017} reports a $50\,\mathrm{ps}$ timing jitter at $405\,\mathrm{nm}$ and MHz photon counts, with one wire resolving 590 effective pixels. Large SNSPD arrays with active areas up to $3.1\times 3.1\,\mathrm{mm^2}$ have been considered for use in exoplanet transit spectroscopy~\cite{ShawExo2021} (e.g.~on the Origins space telescope~\cite{origins}), for which exceptional photometric stability is required in the IR range of $2.8$--$20\,\mathrm{\mu m}$.

\subsubsection*{Single Photon Avalanche Photodiodes}

A SPAD is a diode operated well above its reverse-bias breakdown voltage. Because of the very large internal electric field, a single photoelectron is accelerated to energies sufficiently high to ionize other electrons in its path, thus creating a self-sustaining avalanche, which is registered as a ``click'' (hence the Geiger mode monicker). The detector's dead time is set by the timescale for quenching the avalanche current and restoring it back to the operating bias. 

Already in 2008, single SPADs $20\,\mathrm{\mu m}$ in diameter could reach efficiencies up to 42\% with a timing jitter of $39\,\mathrm{ps}$, kHz-level DCR, and reset times of about $600\,\mathrm{ns}$~\cite{Zappa2008}; ten years later, these performance numbers were improved with a $10$--$80\,\mathrm{\mu m}$ diameter bipolar-CMOS-DMOS SPAD to $t_\mathrm{jitter} \approx 28\,\mathrm{ps}$, $\eta = 70\%$, $t_\mathrm{reset} = 50\,\mathrm{ns}$ at $820\,\mathrm{nm}$ and operating temperature of $-20^\circ\,\mathrm{C}$~\cite{Zappa2018}. The envelope continues to be pushed, with reports of $8.7\,\mathrm{ps}$ timing jitters up to MHz count rates~\cite{becker2022}, and further improvements on the horizon.

SPADs have been assembled into arrays as large as $256 \times 256$, with demonstrated timing jitters of $300\,\mathrm{ps}$, reset times of $100\,\mathrm{ns}$, $70\%$ efficiency, and DCR of $10\,\mathrm{Hz}$~\cite{LargeFormatAull}. These arrays can be manufactured to support Gbit/s data rates (relevant for telecommunications), which is on the order of the photon fluxes relevant even for Phase 3 dishes of EPIC. \cite{LargeFormatAull}.
The Europa Lander astrobiology mission is to be equipped (as part of its LIDAR system) with a $2048 \times 32$ array of Si SPADs with a DCR of $4.5\,\mathrm{kHz}$, peak detection efficiency of 70\% at $532\,\mathrm{nm}$~\cite{EuropaLidar}, and timing jitter target of $250\,\mathrm{ps}$~\cite{EuropaSlides}; a smaller $1024 \times 32$ version for the Europa Lander Sensor Array brought the latter figure down to $33\,\mathrm{ps}$~\cite{SOIspad}.

\subsection{Dispersive element}
\label{sec:grating}
As discussed in \Sec{sec:photodetectors}, despite major leaps in timing resolution of modern single-photon counters, the relative timing uncertainty $\sigma_t$ is still much larger than the period of electromagnetic waves in the optical band.  The fringe contrast in intensity interferometry increases with decreasing spectral width $\sigma_k$ (\Eq{eq:C1pointsource}). For a (point-like) source with a broad spectrum, we thus have $C \sim 1/c \sigma_k \sigma_t$, but the flux is reduced as $I \propto \sigma_k$. For a \emph{single} spectral channel, this implies that the SNR is independent of $\sigma_k$ (\Eq{eq:sigmaC}). However, by combining \emph{multiple} spectral channels of width $\sigma_k$ across an $e$-fold around wavenumber $\overline{k}$, the signal-to-noise ratio can be enhanced as $\mathrm{SNR} \propto (\overline{k}/\sigma_k)^{1/2} = \mathcal{R}^{1/2}$, as derived in \Eq{eq:SNR}.
The most natural way to spectrally split the incoming light is via a dispersive prism or a diffraction grating, as in modern spectrographs. In the following, we perform a simplified analysis of this spectroscopic procedure and note some associated challenges unique to intensity interferometry.

\begin{figure}[h!]
\centering
\includegraphics[width=0.7\textwidth]{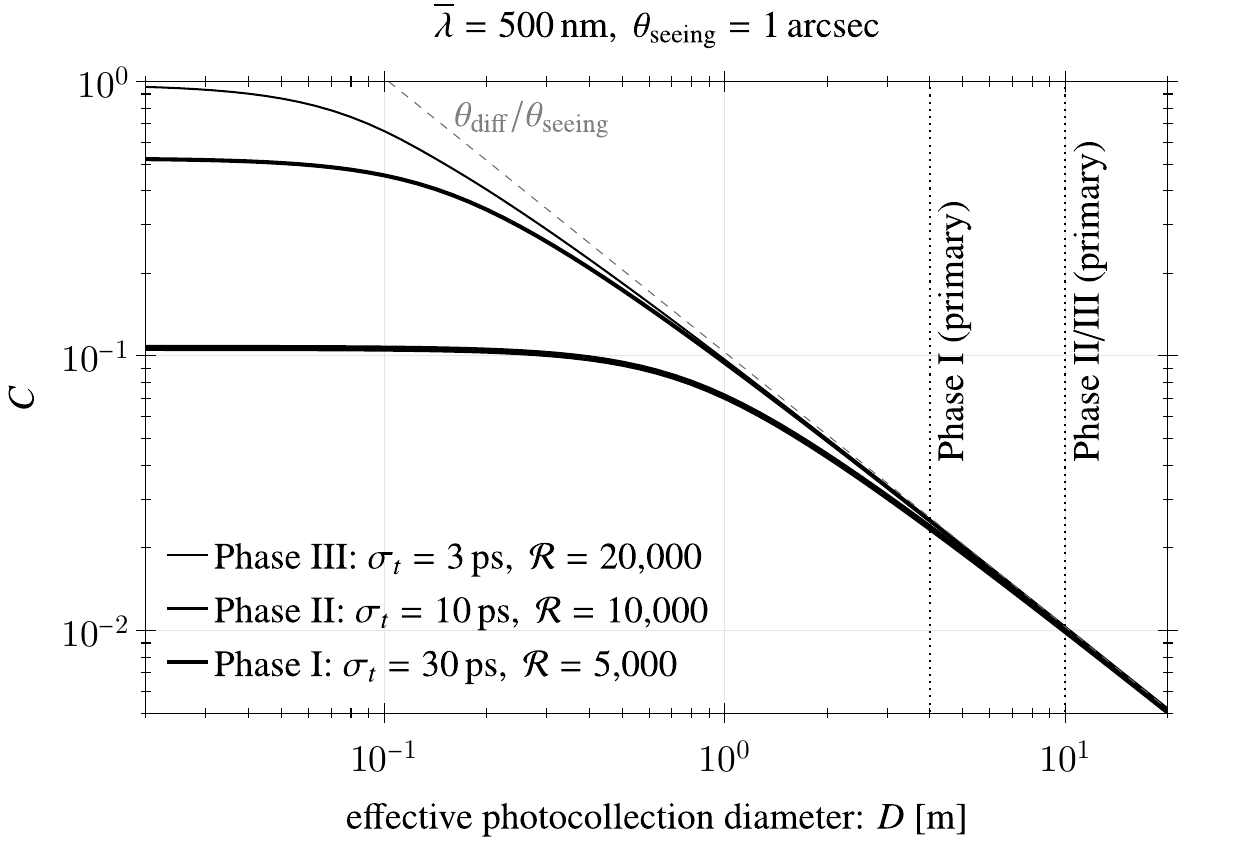}
\caption{
	Intensity correlation of \Eq{eq:Grating_correlation} for a single pixel whose central wavelength is $\overline{\lambda} = 500\,\mathrm{nm}$, as a function of the effective linear size of the photocollection element, and assuming a seeing angle of $\theta_\mathrm{seeing} = 1\,\mathrm{arcsec}$. We plot this correlation for the timing and spectral resolutions $\sigma_t$ and $\mathcal{R}$ assumed for Phases~I,~II,~and~III of EPIC. 
	The horizontal axis can be read either as the diameter of the primary mirror (vertical black dashed lines) for a setup with only one photodetector array, or as the linear subaperture size in the spectroscopic direction (see ``Potential Solutions'' below). The spectroscopic suppression of $C\sim\theta_\mathrm{diff}/\theta_\text{seeing}$ (gray dashed line) is the limiting factor for $D \gtrsim {2 \pi c\sigma_t}/{\mathcal{R} \theta_\mathrm{seeing}}$. The ideal subaperture size is roughly this threshold value: $1\,\mathrm{m}$, $0.2\,\mathrm{m}$, and $0.1\,\mathrm{m}$ for EPIC Phases~I,~II,~and~III, corresponding to a number of subapertures of $N_\mathrm{subap} \sim 4, 50, 100$.}
\label{fig:grating_suppression}
\end{figure}

The main (and perhaps surprising) result of this section is that the intensity correlation for a multichannel intensity interferometer can receive an additional suppression for large-aperture telescopes and excellent spectral and timing resolution. Without mitigating solutions discussed below, the excess fractional intensity correlation in any one spectral channel parametrically scales as:
\begin{alignat}{1}
	C \sim \min \left \lbrace 1, \frac{1}{c \sigma_k \sigma_t}, \frac{\theta_\mathrm{diff}}{\theta_\mathrm{seeing}} \right \rbrace ,  \label{eq:C_ceiling}
\end{alignat}
which is also depicted in figure~\ref{fig:grating_suppression}.
Comparing to e.g.~\Eq{eq:C1pointsource}, the fractional correlation is limited by the ratio of the diffraction-limited angle $\theta_\mathrm{diff} \equiv {2\pi}/{\overline{k} D}$ of a single telescope to the seeing angle $\theta_\mathrm{seeing}$ (\Eq{eq:thetaseeing}). This correlation ceiling  would be the limiting factor for the large-aperture telescopes with high spectral resolution and exquisite timing resolution in all three EPIC Phases of table~\ref{tab:phases}. We conclude this section with possible solutions that can recover the sensitivity assumed throughout section~\ref{sec:theory}, i.e.~remove the ${\theta_\mathrm{diff}}/{\theta_\mathrm{seeing}}$ ceiling in \Eq{eq:C_ceiling}.

In order to illustrate this potential suppression as lucidly as possible, we study the following simplified setup: a plane wave incident on a grating of size $N d_\mathrm{g}$, with $d_\mathrm{g}$ the periodicity of the grating and $N$ the number of (illuminated) slits. We concern ourselves only with the first diffraction peak focused onto a photodetector array, of which a single pixel records the wavenumbers
$k\in\parea{\bar{k}-\sigma_k/2,\bar{k}+\sigma_k/2}$, with $\sigma_k\equiv \bar{k}/\mathcal{R}$. In appendix~\ref{app:waveoptics}, we carry out the full calculation within scalar diffraction theory in great detail, from the point of incidence on the primary mirror to the photodetector array.

The electric field at wavenumber $k$ far away from a diffraction grating is given by the standard result~\cite{BornWolf}:
\begin{equation}
E_k(\theta_\text{F},t)\propto A_k\frac{\sin\pare{k\theta_\text{F}{Nd_\mathrm{g}}/{2}}}{\sin\pare{k\theta_\text{F}{d_\mathrm{g}}/{2}}}e^{-ikt}, 
\label{eq:Ephotodetector}
\end{equation}
with $\theta_F\equiv x_\text{F}/f_c$, where $x_\text{F}$ is the physical linear coordinate on the photodetector, $f_\text{c}$ the focal length of the camera which focuses the collimated light from the grating onto the photodetector, and $A_k$ the field amplitude at wavenumber $k$. The first diffraction peak at wavenumber $\overline{k}$ occurs at $\theta_\mathrm{F} \simeq \overline{\theta}_F \equiv 2\pi/(\overline{k} d)$ (in the small-angle approximation). We are free to redefine $\theta_\mathrm{F} \to \overline{\theta}_\mathrm{F} + \theta_\mathrm{F}$ so that $\theta_\mathrm{F} = 0$ corresponds to the center of the pixel corresponding to the spectral channel centered at $\overline{k}$. This pixel subtends a small angular width equal to $\overline{\theta}_F / \mathcal{R}$. With this redefinition of $\theta_\mathrm{F}$ and expanding $\sin x \simeq x$ at $x \ll 1$, the electric field has the form:\begin{equation}
E_k(\theta_\text{F},t)\propto A_k\frac{\sin\parea{k\theta_\text{F}{Nd_\mathrm{g}}/{2}+ \pi N\pare{ {k}/{\overline{k}}-1}}}{k\theta_\text{F}{Nd_\mathrm{g}}/{2}+ \pi N\pare{ {k}/{\overline{k}}-1}}e^{-i
k t}.
\label{eq:Ephotodetector2}
\end{equation}
Writing the intensity correlation as $C=\frac{\avg{I_1(0) I_2(t)}}{\avg{I_1}\avg{I_2}}-1\equiv\frac{\mathcal{N}}{\mathcal{D}}$, we find for the numerator $\mathcal{N}$ and denominator $\mathcal{D}$:
\begin{align}
\mathcal{N}&\propto\int\frac{\di t}{\sigma_t}\iint\di\theta_\text{F}^{(1)}\di\theta_\text{F}^{(2)}\iint\di k\,\di k' E_k(\theta_\text{F}^{(1)},0)E_{k'}^*(\theta_\text{F}^{(1)},0)E_k^*(\theta_\text{F}^{(2)},t) E_{k'}(\theta_\text{F}^{(2)},t),\label{eq:gratingN}\\
\mathcal{D}&\propto\left|\int\di\theta_\text{F}\int\di k \,\left|E_k(\theta_\text{F},0)\right|^2\right|^2,
\label{eq:gratingD}
\end{align}
where the superscript $(p)$ labels the telescope $p=1,2$ location of the pixel. For simplicity, we incorporate the relative timing uncertainty via a uniform distribution of width $\sigma_t$. Both expressions contain four factors of the electric field in \Eq{eq:Ephotodetector2}. The numerator involves products of electric fields with {different} wavenumbers $k$ and $k'$ at each telescope. In contrast, the denominator is simply the product of the average intensities at each telescope, involving products of electric fields with the \emph{same} wavenumber at each telescope. A heuristic understanding of the suppression effect is thus that large pixels---larger than the diffraction limit---contain several spatially non-overlapping electromagnetic modes with \emph{independent} temporal fluctuations, so that $C = \mathcal{N}/\mathcal{D}$ does \emph{not} attain the limit $C \to 1$ as $\sigma_t \to 0$.

We can gain some (analytic) intuition by taking the size of the pixels to be large (formally infinite) while simultaneously endowing $A_k$ with (uniform) support only over a wavenumber range of width $\sigma_k$.\footnote{In practice, the spectral width $\sigma_k$ of the pixel is linearly proportional to its angular size, so this limit is not physical---we only provide it for illustrative purposes.} In this limit, which turns out to be a reasonable approximation, we find (see appendix~\ref{app:grating}):
\begin{equation}
C = \frac{\mathcal{N}}{\mathcal{D}}\simeq \frac{2}{\sigma_k^2}\int_0^{\sigma_k} \di\sigma \, (\sigma_k-\sigma) e^{-\frac{1}{2}\sigma^2\sigma_t^2} \left[ \frac{\sin \sigma / \sigma_d}{\sigma / \sigma_d} \right]^2.
\label{eq:Grating_correlation}
\end{equation}
This integral consists of a convolution of two effective ``filters'': one Gaussian filter with a bandwidth $\sigma_t^{-1}$, and one sinc-squared filter with bandwidth $\sigma_d\equiv\bar{k}/(\pi N)$. In the limit of poor timing resolution, $(c\sigma_t)^{-1}\ll\sigma_d$, the latter filter is nearly unity over the integration range, and we recover the standard result $C\sim1/{c \sigma_k \sigma_t}$ (from e.g.~\Eq{eq:C1pointsource}). However, in the opposite limit $(c \sigma_t)^{-1}\gg\sigma_d$ of good timing resolution, it is the Gaussian filter that is unity, yielding $C \sim \pi \sigma_d/\sigma_k \ll 1/c \sigma_k \sigma_t$. 

The pixel corresponding to the spectral channel centered on $\overline{k}$ subtends an angle $\overline{\theta}_\mathrm{F}/\mathcal{R}$ to capture wavenumbers in the range $k\in\parea{\bar{k}-\sigma_k/2,\bar{k}+\sigma_k/2}$, but it also needs to be sufficiently large to contain the entire seeing disk of the source(s). Atmospheric aberrations cause a spatial spread (which furthermore fluctuates on millisecond time scales) on the order of the magnified seeing angle $\theta_\mathrm{seeing} f / f'$. The illuminated area of the grating has a linear size $N d_\mathrm{g}$ which is related to the aperture diameter $D$ of the telescope by the same (inverse) magnification factor (the focal length ratio of primary and secondary mirrors): $N d_\mathrm{g} = D f'/f$. These relations imply that containment of the seeing disk is equivalent to:
\begin{alignat}{1}
\frac{\overline{\theta}_\mathrm{F}}{\mathcal{R}} \gtrsim \frac{f}{f'} \theta_\mathrm{seeing} \quad \Leftrightarrow \quad \frac{N}{\mathcal{R}} \gtrsim \frac{\theta_\mathrm{seeing}}{\theta_\mathrm{diff}} \approx 40 \left(\frac{D}{4\,\mathrm{m}}\right) \left( \frac{500\,\mathrm{nm}}{\overline{\lambda}} \right) \left( \frac{\theta_\text{seeing}}{1\,\mathrm{arcsec}} \right), \label{eq:grating_oversize}
\end{alignat}
which is nothing more than the usual requirement in spectroscopy that gratings have to be oversized in practice to account for seeing, since atmospheric fluctuations reduce the grating's nominal resolution $N$ to the much smaller effective resolution $\mathcal{R}$. When this inequality is saturated---$N/\mathcal{R} \simeq \theta_\mathrm{seeing}/\theta_\mathrm{diff}$---then the suppression found in \Eq{eq:Grating_correlation} is equal to:
\begin{alignat}{1}
	C \leq \frac{\pi \sigma_d}{\sigma_k} \simeq \frac{\theta_\mathrm{diff}}{\theta_\mathrm{seeing}}, \label{eq:C_ceiling2}
\end{alignat}
i.e.~the parametric upper bound stated in \Eq{eq:C_ceiling} in the beginning of the section. In figure~\ref{fig:grating_suppression}, we show that Phase I of EPIC would already be limited by (and saturate) \Eq{eq:C_ceiling2}, and thus not achieve its theoretical SNR without modification; the suppression would be even worse for Phases II and III.

\subsubsection*{Potential solutions}
There are at least two (and possibly more) straightforward solutions to mitigate this spectroscopic suppression, albeit at the expense of additional complexity and cost.

The first and simplest strategy is to use multiple $N_\mathrm{subap}$ subapertures in each EPIC telescope, since the correlation ceiling from \Eq{eq:C_ceiling2} scales as $C \propto \theta_\mathrm{diff}$ and thus with the inverse linear size of photocollection. One could take the configuration from figure~\ref{fig:telescope_2}, and after the path-extension stage, geometrically split the light with $N_\mathrm{subap}$ separate gratings, focusing mirrors, and photodetector arrays. Each photodetector array would thus have a smaller effective photocollection size (in the spectroscopic direction), and operate near its nominal fringe contrast of $C \sim 1/c \sigma_k \sigma_t$ with the appropriate subaperture number $N_{\mathrm{subap}} \sim  4, 50, 100$ for Phases~I,~II,~III, respectively (see figure~\ref{fig:grating_suppression}).\footnote{A related technique involves correlation of sub-pupil intensities to perform aperture synthesis~\cite{2021MNRAS.505.2328G}.} The advantages of this approach are that tolerances on the primary and secondary mirrors are not severe, and that the photodetector reset time constraints are relaxed by a factor $N_\mathrm{subap}^{-1}$. The disadvantage (besides cost and complexity) is that the photodetector DCR requirements are more stringent by the same factor $N_\mathrm{subap}^{-1}$.

A second potential solution is the employment of adaptive optics (AO), a well-established technology to recover the diffraction limit of the telescope by using wavefront sensors and thin deformable mirrors to cancel atmospheric aberrations. AO is already in use on many spectrographs (see ref.~\cite{2016PASP..128l1001J} for a review). In an EPIC context, this would allow for pixel sizes $\overline{\theta}_F/\mathcal{R} \simeq \theta_\mathrm{diff}$, for which the spectroscopic suppression from \Eq{eq:C_ceiling2} does not occur. The advantage of this approach is that it can allow for even higher spectral resolutions ($\mathcal{R} \gtrsim 100{,}000$) and correspondingly larger fringe contrast $C$. This elegant approach would however require tight tolerances on the polishing of \emph{all} optically active surfaces, and come at considerable cost especially for large apertures.

Besides those brute-force approaches, there may exist a method to correct for these intra-pixel dispersive phases with clever optical configurations, possibly consisting of arrays of microlenses and/or microprisms, which have demonstrated lenslet pitch on the order of tens of microns to fit SNSPD or SPAD pixel sizes~\cite{paschotta2008encyclopedia}. Any such optical elements would have to cancel the relative propagation phases of a fractional wavenumber range of order $1/\mathcal{R}$, i.e.~``undo'' the spectral splitting within the pixel itself so that $\mathcal{N} \simeq \mathcal{D}$ in the limit $\sigma_t \to 0$ in \Eqs{eq:gratingN} and~\ref{eq:gratingD}.\footnote{A minimal configuration would be a diverging lens to recollimate the beam, a corrective dispersive element such as a microprism, and potentially a final refocusing lens, all directly above each pixel. A simple calculation of the requirements of the microprism dictates resolving power $b\frac{\di n}{\di \lambda}=N$, where $b$ is the base of the prism and $n$ its index of refraction, which should be accurate to about one part in $N/N_\text{eff}$, where $\sigma_t\equiv \bar{k}/{\pi N_\text{eff}}$, i.e.~$N_\text{eff}\sim 15{,}000$ for a $10 \, \mathrm{ps}$ timing resolution. The feasibility of achieving this on the scale of individual pixels is left to future study.} We have checked that simple modifications---defocusing, recollimating the light onto the pixel, irregular gratings, or a non-co-phased segmented mirror---by themselves do not resolve the spectroscopic suppression. We leave investigations of possible clever optical configurations to future work.

\subsection{Tolerances}
\label{sec:opt_tolerances}

\subsubsection*{Optical tolerances}

One of the attractive features of intensity interferometry is the mildness of the required tolerances on the telescope optics, in stark contrast to the tight tolerances for imaging telescopes and (especially optical) amplitude interferometers. 
In the simplest realization of intensity interferometry, without spectral splitting nor path extension, the primary requirement is that the wavefronts in either telescope do not acquire relative path differences (or equivalently, time delays) larger than the relative timing resolution $\sigma_t$ of the photodetectors. The phototubes used at the NSII in the 1970s had electronic bandwidths on the order of $60\,\mathrm{MHz}$~\cite{1967MNRAS.137..375H} or timing resolutions of $\sigma_t \sim 10 \, \mathrm{ns}$. This poor timing translated to a tolerance on the order of a meter on the location of the receivers, and was therefore clearly not the limiting factor of the observatory. Even for our Phase III parameter of $\sigma_t = 3\,\mathrm{ps}$, significant loss of contrast can avoided as long as there are no relative path differences greater than a millimeter, about 4--5 orders of magnitude above the sub-wavelength rms error needed for conventional diffraction-limited astronomical observations. For the same reason, intensity interferometry is also quite immune to atmospheric distortions and delays (\Sec{sec:atm}).

The elements with the most stringent tolerances of (multichannel) EPIC fall into two categories: those that may cause differential phases to the light from both sources, and those with common effects. The path extension belongs to the former, as it is designed to (spatially) split the light from both sources: the two-photon amplitude depicted in the left panel of figure~\ref{fig:C1} corresponds to the fringe useable for wide-angle differential astrometry (cfr.~\Eqs{eq:ellopt1} and~\ref{eq:ellopt2}). Therefore, the relative path length $\ell_p$ needs to be \emph{known} in each telescope $p = 1,2$ to a fraction of a wavelength. In principle, it needs to be \emph{adjustable} to within a range of order $\sigma_k^{-1}$ from the optimal values of \Eqs{eq:ellopt1} and~\ref{eq:ellopt2} to maintain fringe contrast, though it is advisable in practice to keep it well within that range to avoid fringe confusion. These considerations apply to the two beamsplitters and two reflecting flat mirrors of the path extension module. We provide a simple analysis of the beamsplitters in appendix~\ref{sec:beamsplitters}, where we also elaborate on additional phases from the splitters' finite width and index of refraction, which also need to be characterized to achieve optimal operation of an EPIC interferometer.

Optical elements that affect the light of both sources simultaneously---the primary and secondary mirrors, diffraction grating, and camera---should be positioned and polished well enough for the resulting image to fit within the pixel corresponding to each spectral channel, so that the desired spectral resolution $\mathcal{R}$ is achieved. This tends to be a significantly stricter requirement than the aforementioned polishing to within $\sigma_t$; numerically, it amounts to upper bounds on primary aberrations of $\mathcal{O}(\text{few}\,\mathrm{\mu m})$ for the parameters under consideration. This is a much looser requirement than on conventional imaging telescopes aiming to achieve diffraction-limited resolution: tolerances on an EPIC telescope are looser by a factor of about $f'D / f\overline{\lambda}  \mathcal{R} \sim 200$ for $D = 10\,\mathrm{m}$, $\overline{\lambda} = 500\,\mathrm{nm}$, $f/f'=10$ and $\mathcal{R} = 10{,}000$, the fiducial parameters for EPIC Phase~II. We provide a more detailed account on effects from aberrations in appendix~\ref{app:aberrations}. These less restrictive tolerances imply that some of the expensive and time-consuming aspects of conventional large-mirror polishing could be skipped \cite{bely2003design}, helping to make the cost-effective construction of large EPIC arrays a realistic goal. 

As we explained in \Sec{sec:grating}, our assumed spectral resolution in combination with exquisite timing resolution may lead to a suppression in fringe contrast that has to be mitigated. The tolerance requirements mentioned above are for optical configurations than can correct for this intra-pixel dispersion. If one decides to solve the issue raised in \Sec{sec:grating} via multiple sub-apertures, then the requirement is that the aberrations yield an image for a point source that is no larger than the seeing disk. If one employs adaptive optics instead, then the aberration requirements tighten to those of a conventional imaging telescope operated at the diffraction limit.

\subsubsection*{Geodety, synchronization, and calibration}

Extracting astrometrically useful information (such as the angle $\vect{\theta}_{ba}$ between two sources) from the intensity correlation $C$ requires excellent geodety and synchronization---a precise knowledge of the relative positions (i.e.~baseline $\vect{d} = \vect{r}_2 - \vect{r}_1$) and times (i.e.~timing offset $\tau = t_2 - t_1$) at each telescope. These techniques have been developed in support of VLBI programs over decades and continue to improve~\cite{1999ASPC..180..463F,thompson2017interferometry}.

Geological shifts, tides, and Earth's precession and nutation can cause drifts in the relative positions of the telescopes to each other as well as to the celestial reference frame, typically on the order of centimeters per year~\cite{thompson2017interferometry}. Accurate geodetic data are thus key for accurate measurements, since the baseline $\vect{d}$ is to be known to a precision of $c \sigma_t$ (about $1\,\mathrm{cm}$ for Phase I, and $1\,\mathrm{mm}$ for Phase III, cfr.~table~\ref{tab:phases}). 
On the $1\,\mathrm{km}$--$100\,\mathrm{km}$ scale of the baselines considered here, the requirements are somewhat less stringent compared to VLBI. In particular, for sources observed near zenith, the vertical motion of Earth's surface, which is less precisely known, contributes subdominantly to the differential astrometric phase of interest ($\overline{k} \vect{\theta}_{ba} \cdot \vect{d}$). The VLBI global geodetic measurements are planned to improve by an order of magnitude compared to current precision as part of the next-generation  VLBI Global Observing System (VGOS)~\cite{hase2012emerging,behrend2019roll}, which is currently being implemented. A geodetic analysis of the data from a recent demonstration of a subset of the VGOS system yielded a root-mean-square deviation of the baseline length residuals about the weighted mean of $1.6\,\mathrm{mm}$~\cite{niell2018demonstration}. The expected improvements will result in residual path length errors of order a millimeter, or timing precision of a few picoseconds across the globe~\cite{rioja2020precise}, sufficient for the anticipated tolerances for EPIC.

The fast photon detection and short intensity fluctuation times set stringent requirements on the timing stability of the telescopes. In particular, the timing offset between the two telescopes needs to be known to a precision better than the quadrature sum of the photodetectors' timing resolutions (cfr.~\Eq{eq:sigmat}) in order to be a subleading cause for fringe contrast loss. A relative synchronization of $\sigma_t \lesssim 10\,\mathrm{ps}$ over an integration time of $t_\text{obs} = 10^4\,\mathrm{s}$ corresponds to a required Allan deviation of $10^{-14}$. For shorter baselines, it may be possible to achieve this by sharing a common frequency reference, but for longer baselines, each telescope (site) needs to be equipped with its own high-precision atomic clock. Allan deviations at this level are achievable with commercially available hydrogen masers, both active~\cite{mhm} and passive~\cite{CH1-76A}, and are employed and tested by e.g.~the Event Horizon Telescope~\cite{EventHorizonTelescope:2019uob}. 

If $\mathrm{SNR} > 1$ can be achieved on the intensity correlation fringe on shorter integration times with large apertures and/or bright sources, the requirements on the timing stability can be relaxed, making possible the use of cheaper rubidium~\cite{8040C,3352A} or cesium~\cite{5071A} frequency standards, which in commercial packages can have Allan deviations down to $\mathcal{O}(10^{-13})$. We note that in addition to the time-delayed (ghost) intensity correlation fringe at e.g.~$\ell = \ell_1^\mathrm{opt}$ and $\ell' = -\ell_2^\mathrm{opt}$ most useful for relative astrometry, one can also employ the ``main'' intensity correlation fringe in the first line of \Eq{eq:C2pointsource2}, which  has about \emph{four} times the contrast of the ghost fringe and can thus be detected \emph{sixteen} times faster (at the same SNR). This main fringe can be used to self-calibrate and tease out any drifts in baseline and time difference, and can therefore ensure the telescopes' clocks remain synchronized over the course of the observation. 

Finally, each telescope's optical path and path extension needs to be calibrated and regularly monitored. This can be done experimentally with a known light source (e.g.~laser) at the other ``input ports'' on the right of the beamsplitters in \Fig{fig:telescope_2}. A calibration should measure (frequency-dependent) phase shifts from propagation \emph{within} the beamsplitter, the distance $w_p$, and angles $\gamma$ and $\gamma'$. Likewise, the mapping of pixels to wavelengths needs to be established, but this is a standard procedure in spectroscopy, and can be performed by observation of a bright source with a known spectrum.

\section{Applications}
\label{sec:applications} 

Historically, stellar intensity interferometry has mainly been used to investigate the morphology of bright emission regions. In particular, one can determine stellar radii and shapes (projected on the celestial sphere) to high precision, and study star spots, stellar winds, and limb darkening, which are all quantified by the form factors $\mathcal{F}(\overline{k},\mathbf{d})$ in \Eq{eq:C2pointsource2opt_finite}. For more information, we refer the reader to the classic intensity interferometry papers on this subject cited in \Sec{sec:intro}. Emission region morphology studies of \emph{single sources} are also possible in a extended path intensity correlator, \emph{at the same time as differential astrometric measurements between two sources}. For example, one can collect data for differential astrometry on two sources $a$ and $b$, and afterwards select the appropriate fringe by adjusting $\tau$ for the desired application. For differential astrometry, one would use $\tau$ equal to that of \Eq{eq:tauopt}, while the emission morphology of source $j = a,b$ could be investigated by taking e.g.~$\tau = -\hat{\vect{\theta}}_j \cdot \vect{d}/c$ (cfr.~\Eq{eq:C2pointsource2}), as well as potentially other fringe combinations.

In this section, we present a (partial) science case for the differential astrometry capabilities of an extended-path intensity correlator (see ref.~\cite{2021ARA&A..59...59B} for a general overview). The path-extension modification of an intensity interferometer increases its effective field of view up to a few arcseconds, limited by atmospheric fluctuations, cfr.~figure~\ref{fig:thetaIso}. This (literally and figuratively) increases the scope of stellar intensity interferometry, enabling \emph{extreme-precision differential astrometry between two widely separated sources}, with the hierarchy:
$\sigma_{\delta \theta} \ll \sigma_\text{res} \ll \theta_{ba} \lesssim \theta_0$. In other words, the light-centroiding precision $\sigma_{\delta \theta}$ from \Eq{eq:sigmaCentroid2} is much better than the angular resolution $\sigma_{\theta_\text{res}}$ from \Eq{eq:sigmatheta}, which can be orders of magnitudes smaller than the source separation $\theta_{ba}$. The fringe is still detected at maximum contrast as long as the sources are separated by less than the isoplanatic angle $\theta_0$ from \Eq{eq:isoplanatic}. 

We identify five use cases where differential astrometry between two bright sources with EPIC has the potential to provide unprecedented accuracy and precision, opening up new prospects for astrophysical and cosmological measurements. The five uses cases are: binary-orbit characterization (\Sec{sec:binaries}), exoplanet detection (\Sec{sec:exoplanets}), stellar acceleration measurements (\Sec{sec:acceleration}), stellar microlensing (\Sec{sec:stellar_microlensing}), and quasar microlensing by low-mass dark matter halos (\Sec{sec:quasar_microlensing}). This last signature is entirely new, and will be explored in more detail in a follow-up paper~\cite{companionquasar2023}. This list of applications is almost certainly not exhaustive, but hopefully sufficient to motivate research and development in support for EPIC Phase I and subsequent Phases with larger apertures and arrays and/or better spectroscopy and timing.

\subsection{Binaries}
\label{sec:binaries}

Differential astrometry over a narrow field of view is ideally suited to characterization of binary orbits, since the two sources are automatically nearby, and orbital measurements only demand good precision on the \emph{relative} separation of the sources.  Close binary characterization and distance measurement was already achieved in the early days of intensity interferometry~\cite{1971MNRAS.151..161H}. Our EPIC setup allows for observation of binary pairs that are much more widely separated and fainter.

With data only from an intensity interferometer, the period $P$ and all angles (including inclination) of the binary orbit can be determined. With additional spectroscopic radial-velocity measurements of each of the binary constituents, the full Keplerian orbit can be fixed without degeneracies, including the masses $M_a$, $M_b$, and the line-of-sight distance $D_\text{s}$ to the binary. We first review some basic notation of binary orbits, and then present mock analyses on potential binary targets.

Let $\vect{r} \equiv \vect{r}_b - \vect{r}_a \equiv \lbrace x, y, z \rbrace$ be the relative separation between the two sources in the observer's frame, with $z$ denoting the line-of-sight direction. In this frame, we can describe any Keplerian orbit of period $P$ with the six orbital elements: semi-major axis $a$, eccentricity $e$, inclination $i$, longitude of ascending node $\Omega$, argument of periapsis $\omega$, and true anomaly $\nu$:
\begin{align}
x &= r\left[\cos \Omega \cos(\omega + \nu) - \sin \Omega \sin(\omega+\nu) \cos i\right]; \\
y &= r\left[\sin \Omega \cos(\omega + \nu) + \cos \Omega \sin(\omega+\nu) \cos i\right];\\
z &= r\left[ \sin(\omega+\nu) \sin i\right].
\end{align}
The separation $r$ evolves as $a(1-e^2)/(1+e\cos \nu)$. With that substitution, all time dependence is encapsulated though the variation of the true anomaly $\nu$, which is related to the eccentric anomaly $E$, and to the mean anomaly $\mathcal{M}$ evolving linearly in time $t$ as:
\begin{align}
\nu = 2 \arctan \left\lbrace \sqrt{\frac{1+e}{1-e}} \tan \frac{E}{2} \right\rbrace; \quad E - e \sin E = \mathcal{M} = \frac{2 \pi}{P} (t-t_0),
\end{align}
where $t_0$ is the reference epoch.

With 7 or more purely astrometric measurements (ideally spaced out roughly evenly over the orbit) obtained by an intensity interferometer, which can only measure the projected angles on the sky, namely $\lbrace x/D_\text{s}, y/D_\text{s}\rbrace$, one can fit for the six ``dimensionless'' orbital elements $\lbrace a/D_\text{s}, e, i, \Omega, \omega, t_0/P \rbrace$ and one dimensionful quantity, the orbital period $P = 2 \pi a^{3/2} (G M)^{-1/2}$ with $M = M_a + M_b$ the total mass. However, there would always remain one size-distance-mass degeneracy along the ``direction'' $a \propto D_\text{s} \propto (G M)^{1/3}$. This final degeneracy can be broken with radial velocity measurements of the constituents: $\dot{z}_a$ and $\dot{z}_b$. Their difference $\dot{z}_b - \dot{z}_a = \dot{z}$ constrains the ratio $\sqrt{a / G M}$ independent of $D_\text{s}$, while their ratio $\dot{z}_b / \dot{z}_a$ (after subtracting out the time-averaged mean radial velocity of the binary's center of mass) constrains the mass ratio $M_a/M_b$ in direct proportionality.

As a simplified example, we perform an Markov Chain Monte Carlo (MCMC) analysis on a mock data set reminiscent of the spectroscopic binary W0135-052 composed of an unresolved but detached pair of DA white dwarfs~\cite{saffer1988discovery,bergeron1989determination}. The parameters of this system are: binary period $P = 1.556 \, \mathrm{day}$, effective temperatures $T_a = 7450 \pm 500 \, \mathrm{K}$, $T_b= 6920 \pm 500 \, \mathrm{K}$, component masses $M_a = 0.47 \pm 0.05 \, M_\odot$ and $M_b = 0.52 \pm 0.05 \, M_\odot$, velocity semi-amplitudes $K_a = 76.5\,\mathrm{km/s}$ and $K_b = 66.3 \, \mathrm{km/s}$, and a parallax-based distance $D_\text{s} = 12.35 \pm 0.43 \, \mathrm{pc}$. The masses are only estimated after measurements of the surface gravity and assumptions about the stellar composition and mass-radius relation~\cite{bergeron1989determination}.

\begin{figure}
\centering
\includegraphics[width=1\textwidth, trim = 0 0 0 0]{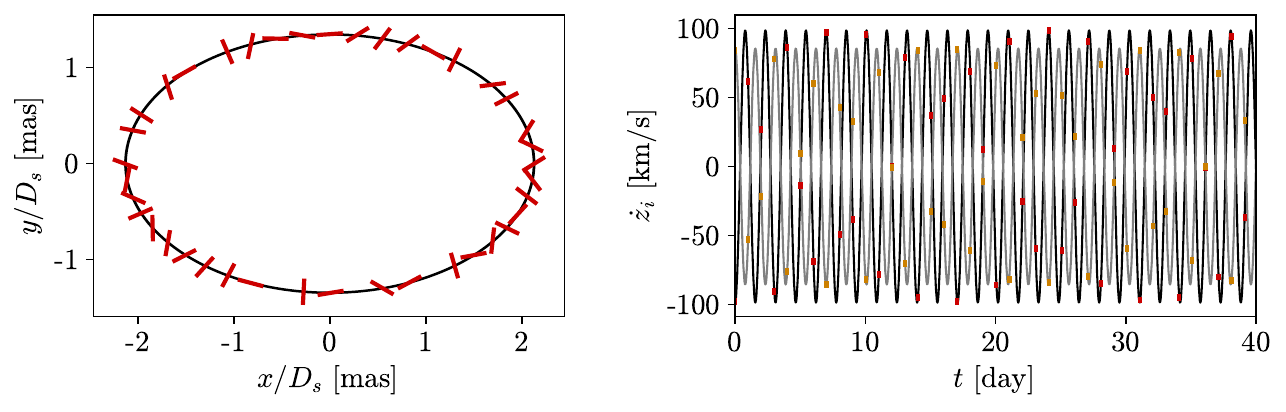}
\caption{Mock data (red/orange points with error bars) for astrometric (left panel) and spectroscopic (right panel) measurements of the doubly-degenerate binary WD0135-052A/B. The EPIC Phase I astrometric uncertainties $\sigma_{\delta \theta} = 5.6 \times 10^{-11}\,\mathrm{rad}$ are magnified by a factor of 10 for clarity.}\label{fig:binary_WD_mock}
\end{figure}

For simplicity, we assumed ``truth parameters'' for the mock data set of $P = 1.556\,\mathrm{day}$, $D_\text{s} = 12.35 \,\mathrm{pc}$, velocity semi-amplitudes of $K_a = 76.5\,\mathrm{km/s}$ and $K_b = 66.3 \, \mathrm{km/s}$, primary mass $M_a = 0.47\,M_\odot$, secondary mass $M_b = M_a K_a / K_b = 0.5423 \, M_\odot$, ignored the systemic line-of-sight radial velocity, and assumed a circular orbit ($e = 0$, $\omega$ undefined) with inclination $i = 0.886 \, \mathrm{rad} = 50.8^\circ$ and longitude of ascending node $\Omega = 0$. We set $t_0 = 0$, and simulated 40 measurements taken over 40 days (1 day between each epoch), with astrometric precision of $\sigma_{\delta \theta} = 5.6 \times 10^{-11} \, \mathrm{rad}$ along a random axis in $(x/D_\text{s},y/D_\text{s})$ space, and radial velocity precision of $\sigma_{\dot{z}_i} = 1\, \mathrm{km/s}$, both per epoch. This astrometric precision is in line with the expectation for this binary for EPIC Phase I, and the spectroscopic precision to that of refs.~\cite{napiwotzki2020eso,maxted2000radial}. We show mock data for these truth parameters in figure~\ref{fig:binary_WD_mock}.

\begin{figure}
\centering
\includegraphics[width=1\textwidth, trim = 0 0 0 0]{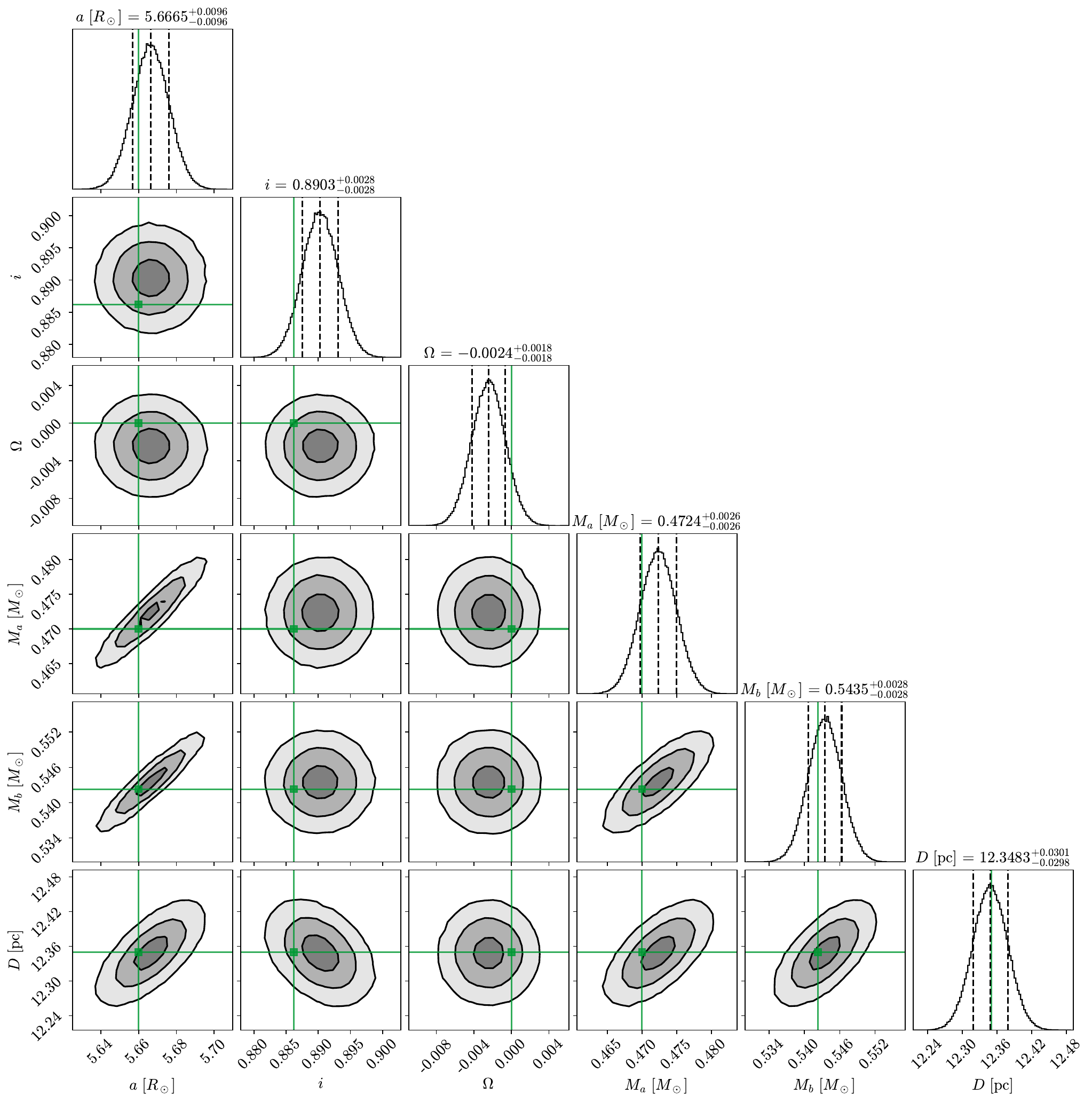}
\caption{MCMC posteriors on the parameters $\lbrace a, i, \Omega, M_a, M_b, D_\text{s} \rbrace$ for a (restricted) binary orbit of WD0135-052A/B, based on the mock data of figure~\ref{fig:binary_WD_mock}. The truth parameters are indicated by green lines (and squares). The contours are $n=1,2,3$ sigma contours (levels $1-e^{-n^2/2}$) for the 2D slices, and the vertical black dashed lines indicate the 16\%, 50\%, and 84\% quantiles (also quoted numerically in the titles) in the 1D histograms.}\label{fig:binary_WD_params}
\end{figure}

In figure~\ref{fig:binary_WD_params}, we show the results of a MCMC parameter estimation of $\lbrace a, i, \Omega, M_a, M_b, D_\text{s} \rbrace$ based on the mock data from figure~\ref{fig:binary_WD_mock}. For this system, the ``astrometric angles'' such as $a/D_\text{s}$, $i$, and $\Omega$ can be determined at $10^{-3}\,\mathrm{rad}$ level of precision, whereas the dimensionful quantities $M_i$, $a$, and $D_\text{s}$ separately are slightly less tightly constrained (but still at a precision of about $0.5\%$ or better fractionally) due to the limited spectroscopic precision.  Nevertheless, it is clear that even with these conservative assumptions, binary mass and distance measurements should be possible at unprecedented precision. For example, observations of WD0135-052A/B (or other similar nearby systems~\cite{holberg201625}) with results as depicted in figure~\ref{fig:binary_WD_params} would yield valuable constraints on the mass-radius relation of white dwarfs~\cite{bergeron2019measurement, chandra2020gravitational, panei2000mass}, as the same intensity interferometry observations would also yield both of the white dwarfs' radii. With EPIC Phase II, differential astrometric precision of about $\sigma_{\delta \theta} \approx 5.6 \times 10^{-12} \, \mathrm{rad}$ should be possible (cfr.~figure~\ref{fig:thetaCentroid3}), while the spectroscopic precision of $\sigma_{\dot{z}_i} \sim 10\, \mathrm{m/s}$ should be possible with dedicated observations by modern instruments such as ESPRESSO at the VLT~\cite{pepe2021espresso}. A similar observation run with such capabilities should be able to characterize all orbital parameters at a fractional precision of $10^{-4}$.

This type of analysis can give even more striking results on very bright stars. For example, the spectroscopic binary $\gamma^2$~Velorum, comprising a O-type blue supergiant and a Wolf-Rayet (WR) star at $D_\text{s} = 379 \pm 4 \, \mathrm{pc}$. An astrometric precision per epoch of $\sigma_{\delta \theta} \sim 10^{-13}\,\mathrm{rad}$ should already be achievable with EPIC Phase I, and radial velocity precision of $\sigma_{\dot{z}_i} \sim 20\, \mathrm{m/s}$ per epoch should also be possible on such a bright binary, yielding component masses with a fractional precision at or below the $10^{-4}$ level. At that level (or better), one could conceivably start to measure the mass loss rate of the WR star at $10^{-4} \, M_\odot / \mathrm{yr}$, and obtain even more precise line-of-sight distance and semi-major axis. These ``spectro-geometric'' distance determinations on binaries involving red giants and/or Cepheid variables may also help calibrate the cosmic distance ladder independently, though not to very large line-of-sight distances.

\subsection{Exoplanets}
\label{sec:exoplanets}

In the previous section, we have seen that EPIC can be a powerful tool for characterizing binary orbits, or orbits of more complicated triple/quadruple systems, comprising \emph{visible} constituents. However, as long as there are two visible sources---either a visible binary or a double star from an accidental chance alignment---one can search for anomalous astrometric deviations due to invisible dark companions. This includes exoplanet searches, as outlined in~\cite{shortpaper}. Depending on the eccentricity and inclination of the exoplanet orbit, the gravitational attraction from the exoplanet will introduce an additional astrometric wobble on its host star, which will trace out an ellipse projected onto the celestial sphere (on top its unperturbed motion).
For the special (but most motivated) case of a circular orbit with mass $M_\text{p}$, semi-major axis $a_\text{p}$, inclination $i_\text{p}$, period $P_\text{p}$, and reference epoch $t_0$, the magnitude of the host star's astrometric wobble is:
\begin{align}
\Delta \theta_a \simeq \frac{M_\text{p}}{M_a} \frac{a_\text{p}}{D_a} \sqrt{1 - \sin^2 i_\text{p} \, \sin^2\left( 2 \pi  \frac{t - t_0}{P_\text{p}} \right) }.
\label{eq:wobble}
\end{align}
The host star's mass is $M_a$ and its line-of-sight distance $D_a$.

\begin{figure}
\centering
\includegraphics[width=1\textwidth, trim = 0 0 0 0]{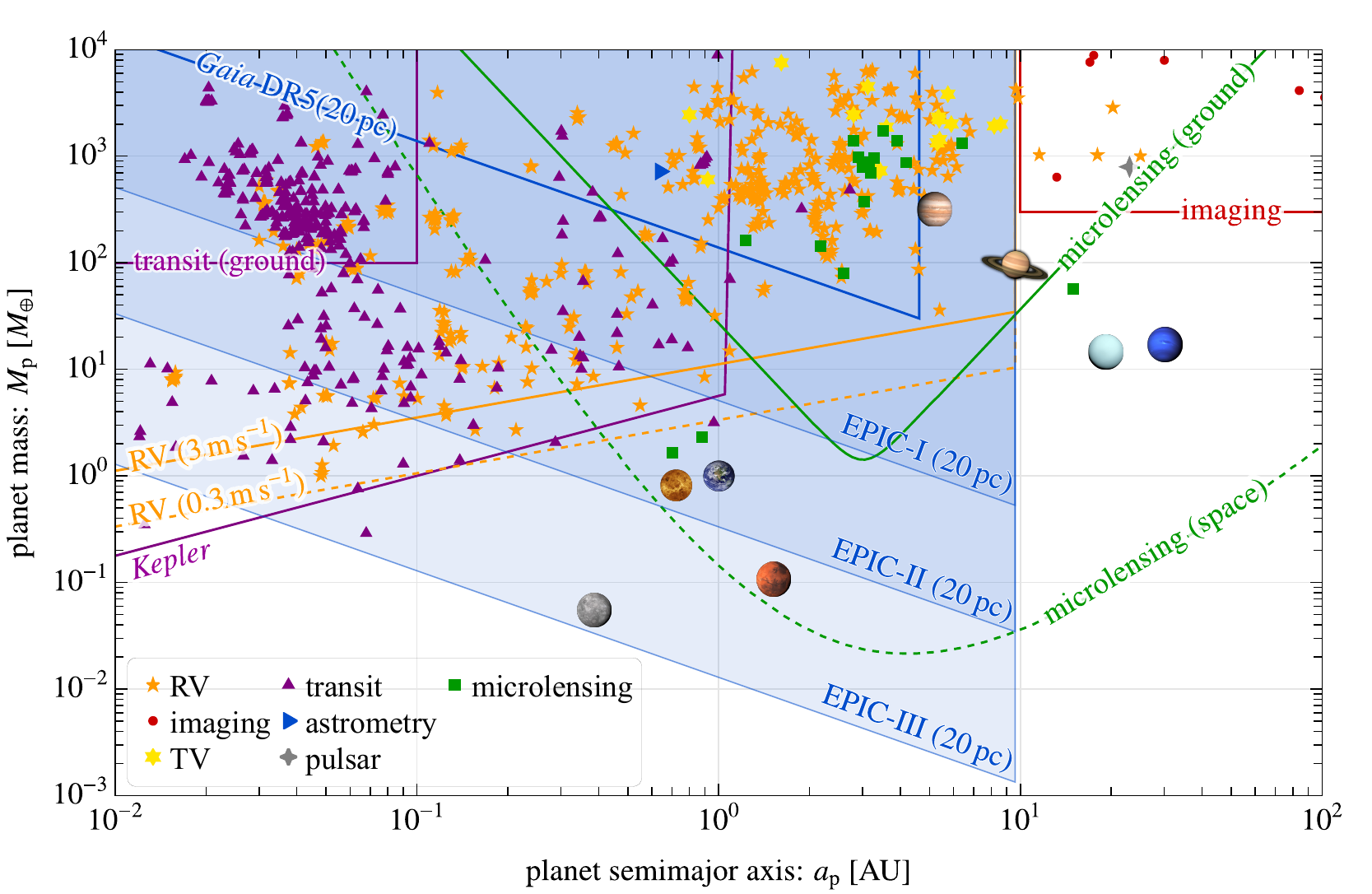}
\caption{Sensitivity of exoplanet detection techniques as a function of semimajor axis $a_\text{p}$ and mass $M_\text{p}$. The astrometric 3-$\sigma$ detection sensitivity of EPIC Phases I-III is shown by the blue regions for a Sun-like star at $D =  20\,\mathrm{pc}$. The markers depict confirmed exoplanets in multiple-star systems (those where EPIC is definitely applicable), classified by discovery method from the NASA Exoplanet Archive~\cite[acquired May 2023]{akeson2013nasa}. The astrometric sensitivity of \textit{Gaia} DR5 is outlined by the dashed blue line, that of radial velocity (RV) surveys by the orange lines, transit methods by the purple lines, direct imaging by the red line, and microlensing by the green lines (adapted from ref.~\cite{zhu2021exoplanet}). The eight Solar System planets and planets discovered eclipse timing variations (TV) are also overlaid (image credit: NASA). Figure from companion paper~\cite{shortpaper}.}\label{fig:exoplanets}
\end{figure}

The method of looking for anomalous astrometric deviations of stellar light centroids has not been the most prolific exoplanet discovery technique historically, but the \textit{Gaia} satellite has already found dozens of candidates in early data releases~\cite{2022arXiv220605595G}, and is expected to find thousands~\cite{perryman2014astrometric} of high-mass planets, preferentially at large semimajor axis $a_\text{p}$ around nearby stars, due to the scalings of \Eq{eq:wobble}. In figure~\ref{fig:exoplanets}, we plot the expected sensitivity of \textit{Gaia} at its fifth data release (DR5) containing ten years of observations, for a Sun-like star at $D_a = 20\,\text{pc}$ in dashed blue. The sensitivity contour corresponds to a final-mission sensitivity of $\sigma_{\Delta \theta_a} \approx 7\,\mathrm{\mu as}$, and a 3-$\sigma$ detection threshold, for orbits shorter than the mission duration, i.e.~$a_\text{p} \lesssim 4.6\,\mathrm{AU}$.

An EPIC setup can surpass this sensitivity by a couple of orders of magnitude for nearby stars, or alternatively, extend the volume over which the technique is applicable. For a pair of Sun-like stars at $D_a = 20\,\text{pc}$, the per-epoch light centroiding precision is $\sigma_{\delta \theta} \approx \lbrace 4.5, 0.29, 0.011\rbrace  \, \mathrm{\mu as}$ for EPIC Phases $\lbrace \mathrm{I, II, III} \rbrace$. With $N_\text{obs} = 300$ observations (e.g.~a cadence of about one per month over 30 years), the final wobble sensitivity is of order $\sigma_{\Delta \theta_a} \approx \lbrace 0.26, 0.017, 0.00065 \rbrace \,\mathrm{\mu as}$, indeed better than \textit{Gaia} by a large factor for this particular application (for nearby stars, with a companion in close proximity). We show contours of 3-$\sigma$ sensitivity as blue regions in figure~\ref{fig:exoplanets} for $D_a =  20  \,\mathrm{pc}$. Earth-Sun-like systems ($M_a = M_\odot$, $M_\text{p} = M_\oplus$, and $a_\text{p} = 1\,\mathrm{AU}$) should be observable by EPIC Phase II at distances of $D_a \lesssim 20\,\mathrm{pc}$ ($D_a \lesssim 400\,\mathrm{pc}$ by Phase III), as long as a sufficiently bright reference star is nearby.

The exoplanet parameter space accessible to an EPIC is complementary to that of other methods. Exoplanet transits causing repeating, temporary dips in the host star's brightness have been detected from the ground~\cite{2006PASP..118.1407P,2004PASP..116..266B} and especially with the \textit{Kepler} satellite~\cite{borucki2010kepler}, which has produced the largest exoplanet catalog to date; the approximate parameter space is depicted by purple lines in figure~\ref{fig:exoplanets}. These surveys have a moderate bias toward small semimajor axes due to the higher transit frequency. Likewise, many exoplanets are detected via the radial velocity (RV) they impart on their host star~\cite{2016PASP..128f6001F}, but this method also favors small semimajor axes, as the radial velocity scales as $\dot{z}_a \sim \sqrt{G M_\text{p}^2 / M_a a_\text{p}}$. We plot the RV detection sensitivity for precisions of $3\,\mathrm{m \, s^{-1}}$ in solid orange, and $0.3\,\mathrm{m \, s^{-1}}$ in dashed orange. Direct imaging also favors large planets far away from their host star (red)~\cite{2016PASP..128j2001B}. Finally, chance alignments of exoplanetary systems with background stars can lead to detection of even very low-mass systems; the approximate sensitivity with ground-based observations is depicted in solid green, and with space-based observations (e.g.~\textit{Nancy Grace Roman Space Telescope}) in dashed green, following ref.~\cite{penny2019predictions}.

\subsection{Accelerations}
\label{sec:acceleration}

The exquisite astrometric light-centroiding of EPIC enables precise measurements of the relative transverse angular accelerations between two stars, and thus of the difference in their gravitational potential gradient (or at least two components thereof). This is a \emph{direct, local} probe of the gravitational potential and the underlying matter distribution, unlike techniques based on the MW's rotation curve~\cite{eilers2019circular} (not local, since they are based on the enclosed mass within the orbit) or on velocity dispersions~\cite{kuijken1989mass,holmberg2000local,bovy2012local,schutz2018constraining,widmark2019measuring} (not direct, since they rely on assumptions of equilibrium and symmetry, which are furthermore violated~\cite{bennett2019vertical,antoja2018dynamically,helmi2018merger}). 

The line-of-sight component of the MW acceleration (relative to that of the Solar barycenter) can be measured by next-generation spectrographs through the resulting changes in radial velocity~\cite{ravi2019probing,silverwood2019stellar,chakrabarti2020toward} and by pulsar timing arrays through changes in the pulsar's spin period~\cite{phillips2021milky} or in the orbital period of pulsar binaries~\cite{phillips2021milky,chakrabarti2021measurement,bovy2020purely}. The latter orbital period drift is already constraining the vertical acceleration towards the Galactic disk at moderate signal-to-noise ratio, with further improvements expected over time.
Ref.~\cite{buschmann2021galactic} has proposed using \textit{Gaia} and future astrometric observatories (potentially in tandem) to measure the transverse angular acceleration of MW stars, forecasting that \textit{Gaia} alone should have significant evidence for the MW's transverse angular acceleration by the end of its mission, if large-angle systematics can be kept under control. 

In this section, we revisit the idea from ref.~\cite{buschmann2021galactic} and show that EPIC can significantly surpass the sensitivity of \textit{Gaia} by a large factor, albeit on a much smaller number of stars due to its limitations on apparent brightness. For a \emph{single} pair of stars, it may approach a (relative) angular acceleration sensitivity of $\sigma_{\delta \alpha} \sim 10^{-3}\,\mathrm{\mu as \, yr^{-2}}$ over a 30-year observing campaign, close to the expected size of Galactic accelerations.

\begin{figure}[t!]
    \centering
    \includegraphics[width=1.00\textwidth]{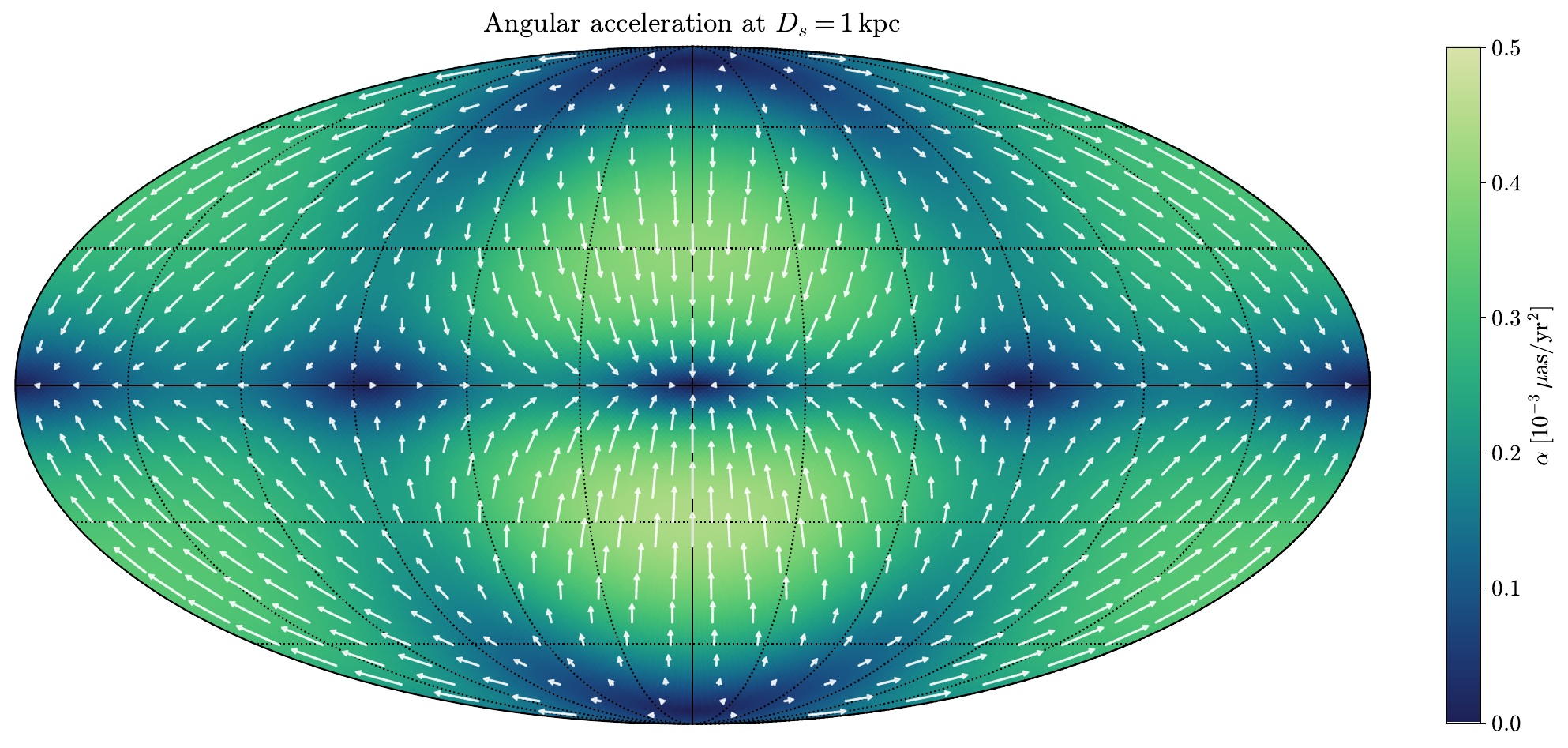}\\
    \includegraphics[width=0.49\textwidth]{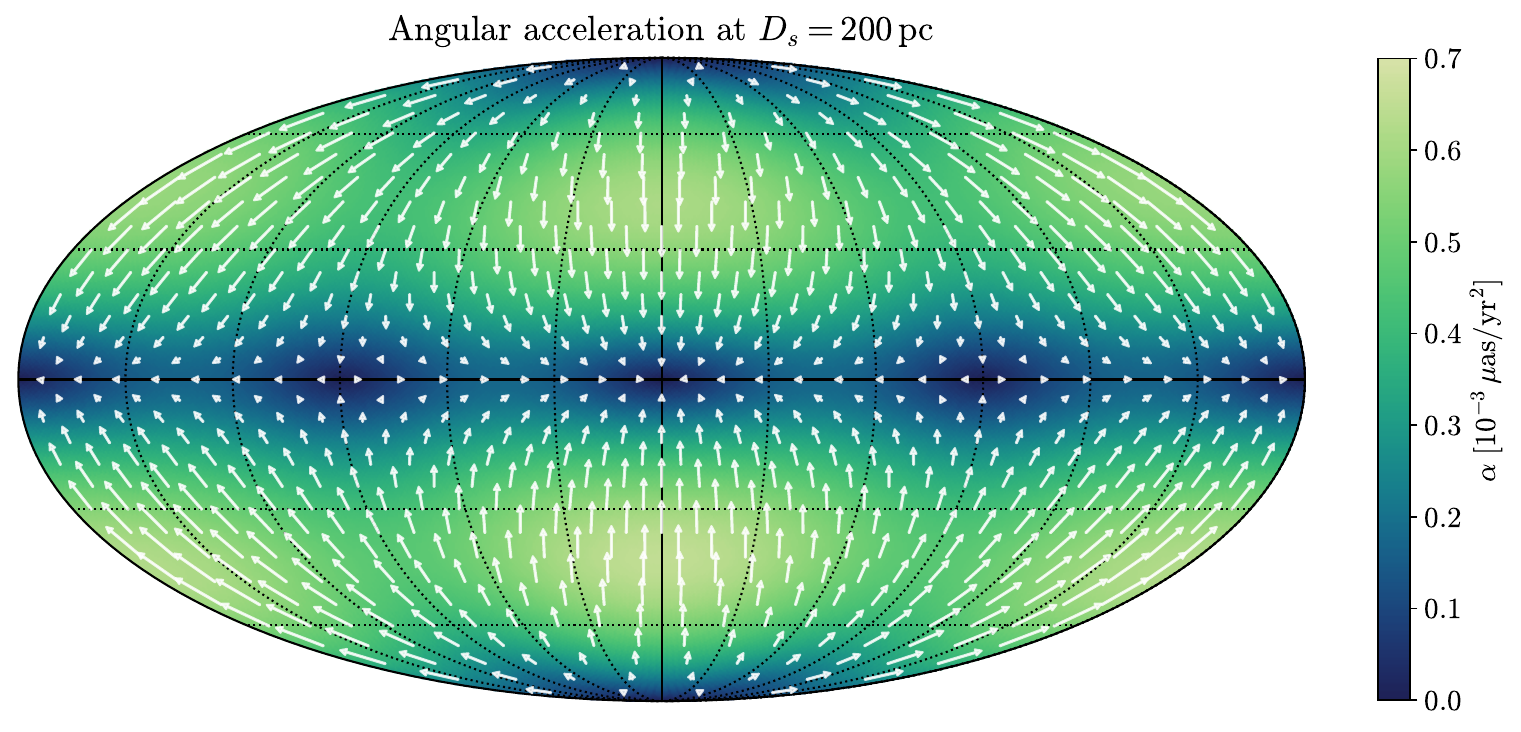}
    \includegraphics[width=0.49\textwidth]{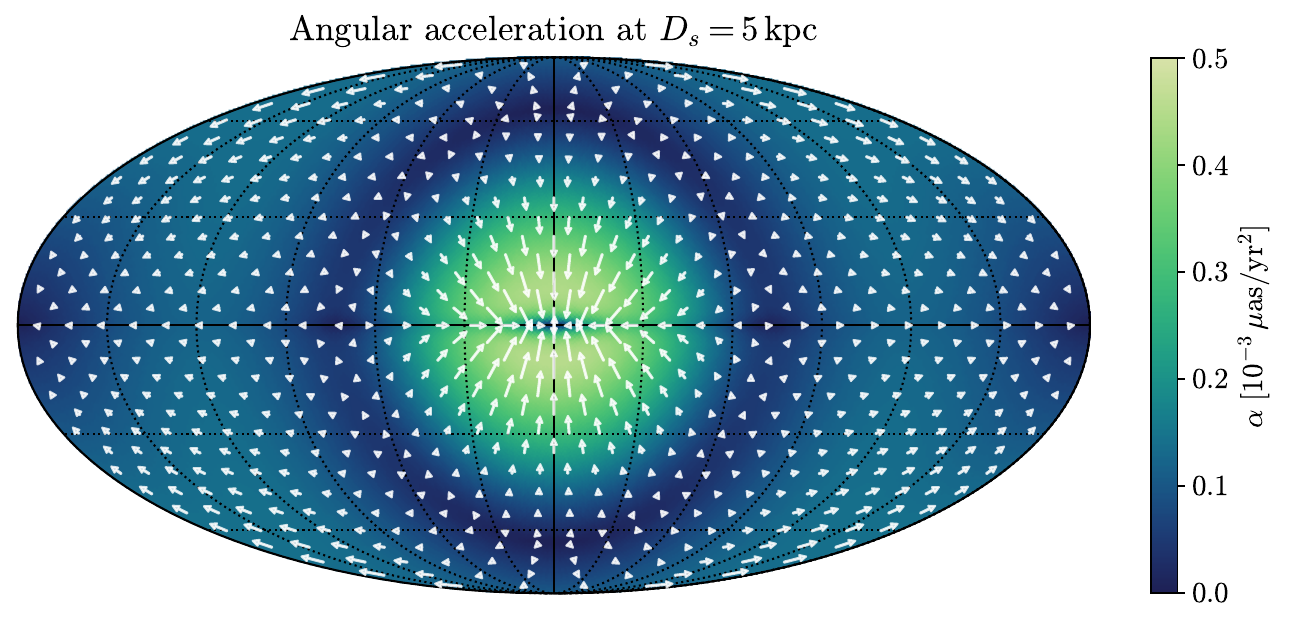}
    \caption{Apparent angular acceleration vector field in the Milky Way in galactic coordinates at three line-of-sight distances $D_\text{s} = 1\,\mathrm{kpc}$ (top), $D_\text{s} = 200\,\mathrm{pc}$ (bottom left), and $D_\text{s} = 5\,\mathrm{kpc}$ (bottom right). The angular accelerations towards the Galactic disk are dominant (but no larger than $10^{-3} \, \mathrm{\mu as} \, \mathrm{yr}^{-2}$, as indicated by the color bars) for small distances, while for $D_\text{s} \gtrsim 1\,\mathrm{kpc}$ the bulge and DM halo also give important contributions. The nodes at $(l,b) \approx \lbrace (0^\circ,0^\circ), (90^\circ,0^\circ),  (180^\circ,0^\circ),  (270^\circ,0^\circ),  (0^\circ,90^\circ),  (0^\circ,-90^\circ)\rbrace$ for small $D_\text{s}$ are dictated by symmetry. } \label{fig:acc_MW}
    \end{figure}

To estimate the size of the Galactic accelerations, we follow ref.~\cite{buschmann2021galactic} and use a simple model of the MW matter components, which is a combination of a bulge, disk, and spherical DM halo. We use the the Hernquist bulge model~\cite{hernquist1990} with total mass $M_\text{bulge} = 5 \times 10^9 \, M_\odot$ and scale radius $r_\text{bulge} = 0.6\,\mathrm{kpc}$, a Miyamoto-Nagai disk model~\cite{miyamoto1975three} with mass $M_\text{disk} =  6.8 \times 10^{10} \, M_\odot$, scale length $a_\text{disk} = 3 \, \mathrm{kpc}$, and scale height $b_\mathrm{disk} = 280\,\mathrm{pc}$, and the Navarro-Frenk-White (NFW) profile~\cite{nfw1996} for the DM halo with scale radius $r_s = 16 \, \mathrm{kpc}$ and local density (at $r_\odot = 8\,\mathrm{kpc}$ from the Galactic Center) of $\rho_\odot = 0.3\,\mathrm{GeV/cm^3} = 0.008 \, M_\odot/\mathrm{pc}^3$. Within this model, it is then straightforward to calculate the gravitational potential $\Phi$ and the resulting acceleration $\vect{a}(\vect{r}) = - \vect{\nabla} \Phi$ at any position $\vect{r}$ in the MW. The apparent angular acceleration is then given by
\begin{align}
\vect{\alpha}(l,b,D_\text{s}) = \frac{\delta a}{D_\text{s}} \left[\hat{\vect{\delta a}} - \hat{\vect{D}}_s (\hat{\vect{D}}_s \cdot \hat{\vect{\delta a}}) \right], \label{eq:alpha_gal}
\end{align}
in terms of the acceleration relative to that of the solar barycenter $\vect{\delta a} \equiv \vect{a}(\vect{r}) - \vect{a}(\vect{r}_\odot)$, the distance vector $\vect{D}_s \equiv \vect{r} - \vect{r}_\odot$ to the star, and their unit vectors $\hat{\vect{\delta a}}$ and $\hat{\vect{D}}_s$, respectively. We plot the fiducial angular acceleration of \Eq{eq:alpha_gal} as a function of of galactic longitude $l$ and latitude $b$ in figure~\ref{fig:acc_MW}, for three distances $D_\text{s} = \lbrace 200\,\mathrm{pc}, 1\,\mathrm{kpc}, 5\,\mathrm{kpc} \rbrace$. Typical Galactic acceleration magnitudes are at the level of $0.3 \times 10^{-3} \, \mathrm{\mu as} \, \mathrm{yr}^{-2}$.

An EPIC telescope can track the relative angular separation $\vect{\theta}_{ba}$ between two sources $a$ and $b$, and thus the relative angular acceleration between two sources $\vect{\alpha}_{ba} \equiv \vect{\alpha}_b - \vect{\alpha}_a$ according to \Eq{eq:alpha_gal}. In practice, both of the components of $\vect{\theta}_{ba}$ would be fit to the model:
\begin{align}
\theta_{ba}^j(t) \simeq \varpi_0^j \cos \left( 2 \pi \frac{t}{\mathrm{yr}} + \gamma^j \right) +  \theta_0^j +  \mu_0^j (t - t_0) + \alpha_0^j \frac{(t-t_0)^2}{2}.
\end{align}
where $j = 1,2$ is a component superscript (e.g.~$(\text{ra},\text{dec})$), and $\varpi_0^j$, $\theta_0^j$, $\mu_0^j$, and $\alpha_0^j$ are the fit parameters for the relative parallax, angular position, velocity, and acceleration, respectively. The phases $\gamma^j$ are known based on the position of the source. Assuming measurements are taken at a fixed repetition frequency $f_\text{rep}$ over a survey time $\mathcal{T}$ (so a total number of measurement epochs $f_\text{rep} \mathcal{T}$), the covariance matrix on the estimated fit parameters for one component $p$ from a least-squares analysis would then be:
\begin{align}
\begingroup
\renewcommand*{\arraystretch}{1.5}
\mathrm{cov} \left[ \widehat{\theta}_i, \widehat{\theta}_j \right] =
\frac{\sigma_{\delta \theta}^2}{f_\mathrm{rep}} \begin{pmatrix}
\frac{2}{\mathcal{T}} & \frac{15}{\pi^2 \mathcal{T}^3} & 0 & \frac{-360}{\pi^2 \mathcal{T}^5} \\
\frac{15}{\pi^2 \mathcal{T}^3} & \frac{9}{4 \mathcal{T}} & 0 & \frac{-30}{\mathcal{T}^3} \\
0 & 0 & \frac{12}{\mathcal{T}^3} & 0 \\
\frac{-360}{\pi^2 \mathcal{T}^5} & \frac{-30}{\mathcal{T}^3} & 0 & \frac{720}{\mathcal{T}^5}
\end{pmatrix};
\endgroup
\qquad \widehat{\theta}_i = \left\lbrace \widehat{\varpi}_0^j, \widehat{\theta}_0^j, \widehat{\mu}_0^j, \widehat{\alpha}_0^j \right\rbrace, \label{eq:covalpha}
\end{align}
where $\sigma_{\delta \theta}$ is the single-epoch light-centroiding precision of \Eq{eq:sigmaCentroid1} as before. 

The Galactic acceleration is prohibitively small to be measured by EPIC Phase I, and difficult to detect on single star pairs even for EPIC Phase II, though perhaps positive evidence for the effect can be obtained on aggregate. EPIC Phase III, however, could map out the Galactic acceleration field on single star pairs. Its relative light-centroiding precision for two Sun-like stars at a distance of $200\,\mathrm{pc}$ would be $\sigma_{\delta \theta} \approx 0.11\,\mathrm{\mu as}$, and the resulting error on the relative acceleration would be $\sigma_{\delta \alpha} \simeq \sqrt{720/f_\text{rep}\mathcal{T}^5} \sigma_{\delta \theta} \approx 0.18 \times 10^{-3}\,\mathrm{\mu as} \, \mathrm{yr}^{-2}$ for $f_\text{rep} = 1/\text{month}$ and $\mathcal{T} = 30\,\mathrm{yr}$, smaller or comparable to the expected accelerations of stars in the Milky Way shown in figure~\ref{fig:acc_MW}. Brighter stars would yield higher accuracy and sensitivity to accelerations at larger distances.

We leave a detailed projection on how a EPIC could be used to infer the structure of the Milky Way (e.g.~as done in ref.~\cite{buschmann2021galactic}) to future work. Such an analysis would entail finding suitable bright source pairs, where the two constituents of the pair would have significantly different line-of-sight distances, otherwise their predicted relative accelerations would be too similar, with little signal remaining in their relative angular accelerations. An ideal configuration would consist of bright nearby stars close to distant sources, such as red/blue giants or quasars.  

\subsection{Stellar microlensing}
\label{sec:stellar_microlensing}

Gravitational microlensing is the phenomenon whereby a compact foreground object, either luminous (e.g.~a star) or dark (e.g.~a black hole, neutron star, or compact dark matter object), temporarily passes nearly in front of a background source, bending its light and splitting its image into two (often unresolved) images~\cite{1986ApJ...304....1P}. Such an event produces transient signals in both photometric and astrometric observables~\cite{1986ApJ...304....1P,1995A&A...294..287H,1995AJ....110.1427M,1995ApJ...453...37W}.
Due to experimental realities, the photometric signal---a characteristic non-repeating magnification event in the light curve of a background source---has historically been the primary driver in the search for compact lenses~\cite{1991ApJ...371L..63P,1991ApJ...372L..79G,1991ApJ...374L..37M}, even excluding primordial black holes as the totality of dark matter over a wide mass range. 
Ground-based microlensing surveys such as Optical Gravitational Lensing
Experiment (OGLE)~\cite{1994AcA....44..227U}, Microlensing
Observations in Astrophysics (MOA)~\cite{2001MNRAS.327..868B}, and the Korea Microlensing Telescope Network (KMTNet)~\cite{2016JKAS...49...37K} continue to detect thousands of photometric microlensing events every year in dense stellar fields, and will soon be joined by a similar survey~\cite{penny2019predictions} with the \textit{Nancy Grace Roman Space Telescope}~\cite{2015arXiv150303757S}.
Astrometric microlensing has been observationally challenging~\cite{2016ApJ...830...41L,2017ApJ...843..145K,2018MNRAS.476.2013R} but has recently become a useful tool, yielding positive stellar microlensing detections~\cite{2017Sci...356.1046S,2018MNRAS.480..236Z}, confirming the discovery of the first isolated black hole~\cite{2022ApJ...933L..23L,2022ApJS..260...55L,2022ApJ...933...83S,2022ApJ...937L..24M}, and spurring forecasts for astrometry-only analyses of luminous lenses~\cite{MassesRD2} and ``blind'' searches for dark compact lenses~\cite{van2018halometry,2023arXiv230100822C}.

Since the optical depth for Galactic microlensing is tiny and building a vast catalog with EPIC is infeasible, blind searches for dark compact lenses with EPIC are not practical. However, EPIC's excellent astrometric capabilities enable precise parameter estimation of two classes of microlensing events. Firstly, weak microlensing events between two luminous sources as predicted by e.g.~\textit{Gaia} may pinpoint the lens mass along the lines of ref.~\cite{MassesRD2} but with higher precision. Secondly, for photometrically-flagged strong microlensing events of dark compact lenses, the relative separation of the two source images can be measured (as proposed in ref.~\cite{2001A&A...375..701D}) by EPIC, again with great potential for parameter estimation. In this section, we will explore EPIC's potential for these two use cases.

The characteristic angular scale of a microlensing event by a point-like lens of mass $M_\mathrm{L}$ at a line-of-sight distance $D_\mathrm{L}$ of a source at distance $D_\mathrm{S}$ is set by the (angular) Einstein radius:
\begin{alignat}{1}
\theta_\mathrm{E} = \sqrt{\frac{4 G M_\mathrm{L}}{c^2 D_\mathrm{L}} \frac{D_\mathrm{S} - D_\mathrm{L}}{D_\mathrm{S}} } \approx 2.85 \, \mathrm{mas} \sqrt{\frac{M_\mathrm{L}}{M_\odot} \frac{1\,\mathrm{kpc}}{D_\mathrm{L}}\frac{D_\mathrm{S} - D_\mathrm{L}}{D_\mathrm{S}} } \label{eq:theta_E}.
\end{alignat}
At an angular impact parameter $\vect{\beta}$ from the lens to the (unlensed) source location, the source image is split into two (denoted by $+$ and $-$ subscripts), with astrometric deflection angles $\vect{\Delta \theta}_{\pm}$ and photometric magnifications $\mu_\pm$:
\begin{alignat}{2}
\vect{\Delta \theta}_{\pm} &= \frac{\pm \sqrt{\beta^2 + 4 \theta_\mathrm{E}^2} - \beta}{2} \hat{\vect{\beta}} \label{eq:point_deflection}; \\
\mu_\pm &= \frac{\beta^2 + 2 \theta_\mathrm{E}^2}{2 \beta \sqrt{\beta^2 + 4 \theta_\mathrm{E}^2}} \pm \frac{1}{2} \label{eq:point_mag}; \\
\beta^j &\equiv \left(\hat{\vect{\theta}}_\mathrm{S} - \hat{\vect{\theta}}_\mathrm{L} \right)^j_\mathrm{unlensed} \equiv \varpi_0^j \cos \left(2 \pi \frac{t}{\mathrm{yr}} + \gamma^j \right) + \theta_0^j + \mu_0^j (t-t_0). \label{eq:beta}
\end{alignat}
On the RHS of the last line, we have parametrized the unlensed trajectory by 5 parameters: the relative parallax $\varpi_0$ (whose direction and phases $\gamma_j$ are known), and the relative separation $\vect{\theta}_0$ and proper motion $\vect{\mu}_0$ (two components each, not to be confused with the magnifications $\mu_\pm$) at some reference time $t_0$ (taken to be the midpoint of the observation period). 

In this notation, the two aforementioned signatures for EPIC are as follows. Analogous to the \textit{Gaia} forecasts of ref.~\cite{MassesRD2}, EPIC can measure the relative separation of a luminous lens and (the primary ``$+$'' image of) a source
\begin{alignat}{2}
    \vect{\theta}_{\mathrm{SL}} \equiv \hat{\vect{\theta}}_{\mathrm{S}+}  - \hat{\vect{\theta}}_\mathrm{L} = \vect{\beta} + \vect{\Delta \theta}_+ \quad \text{(weak lensing, luminous lens)} \label{eq:microlensing_sig_1}.
\end{alignat}
Likewise, EPIC can also be employed to execute the proposal from ref.~\cite{2001A&A...375..701D} to measure the relative separation $\vect{\theta}_{+-}$ between the two images of the source in a microlensing event by a non-luminous lens:
\begin{alignat}{2}
    \vect{\theta}_{+-} \equiv \hat{\vect{\theta}}_{\mathrm{S}+} - \hat{\vect{\theta}}_{\mathrm{S}-} = \sqrt{\beta^2 + 4 \theta_\mathrm{E}^2} \, \hat{\vect{\beta}} \quad \text{(strong lensing, dark lens)} \label{eq:microlensing_sig_2}.
\end{alignat}
The observables in \Eqs{eq:microlensing_sig_1} and~\ref{eq:microlensing_sig_2} can both be fit by 6 parameters: the 5 parameters in \Eq{eq:beta}, and the lens mass $M_\mathrm{L}$, using the well-known fact that one can write $\theta_\mathrm{E}^2 = 4 G M_\mathrm{L} \varpi_0/ \mathrm{AU}$ in terms of the lens mass $M_\mathrm{L}$ and the relative parallax $\varpi_0$. In both cases, the lens mass can be extracted directly, unlike in photometry-only microlensing measurements with only the total magnification $\mu_+ + \mu_-$ as an observable.

\begin{figure}[t]
    \centering
    \includegraphics[width=1\textwidth]{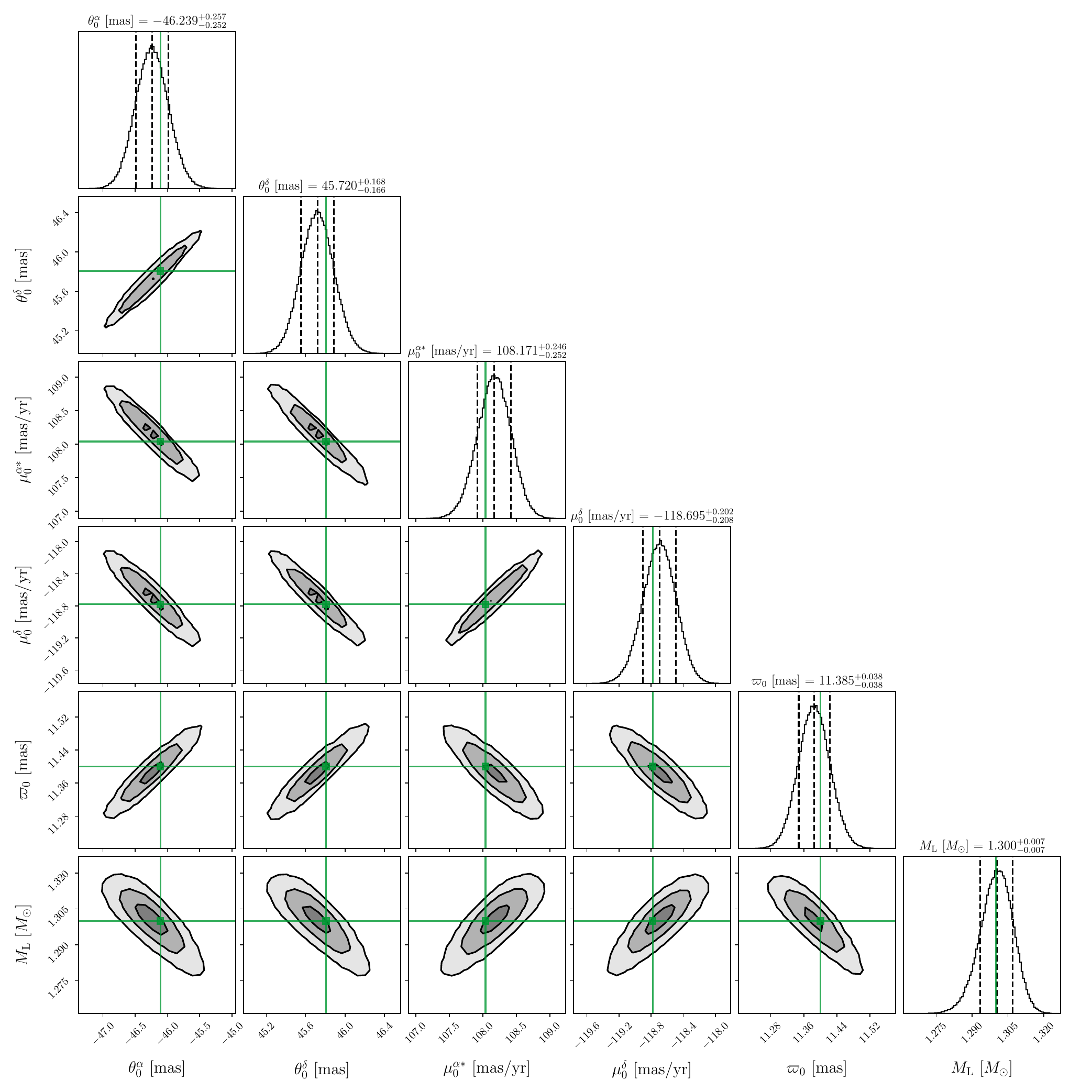}
    \caption{MCMC posteriors of the full relative astrometric solution $\{\theta_0^\alpha,\,\theta_0^\delta,\,\mu_0^{\alpha*},\,\mu^\delta_0,\,\varpi_0,\,M_\text{L}\}$ from a mock data set of six relative separation measurements of the microlensing event 4410342766254199168-35742 predicted in ref.~\cite{gaiamicrolensing}. The measurements span a period of one year around the time of closest approach of the two stars, each having a precision $\sigma_\theta = 11 \, \mathrm{\mu as}$ Remarkably, EPIC Phase II can determine the mass of the stellar lens at the $0.46\%$ level, even with just enough measurements to break the parameter degeneracy.}
    \label{fig:microlensing_mass}
\end{figure}

To assess EPIC's astrometric microlensing capabilities, we generate two mock data sets. The first is for the weak microlensing event 4410342766254199168-35742 between two luminous stars predicted from \textit{Gaia} data in ref.~\cite{gaiamicrolensing}. We assume ``truth values'' for the system of 
$\lbrace -46.1 \, \text{mas}$, $45.8 \, \text{mas}$, $108.04\,\text{mas/yr}$, $-118.78 \, \text{mas/yr}$, $11.4 \, \text{mas}$, $1.3 \, M_\odot \rbrace$ 
for  
$\lbrace \theta_0^\alpha$, $\theta_0^\delta$, $\mu_0^{\alpha *}$, $\mu_0^{\delta}$, $\varpi_0$, $M_\text{L} \rbrace$, the two components each (right ascension $\alpha$ and declination $\delta$, with $\mu_0^{\alpha *} \equiv \mu_0^\alpha \cos \delta$) of relative separation $\vect{\theta}_0$ and proper motion $\vect{\mu}_0$, and for the relative parallax $\varpi_0$ and lens mass $M_\mathrm{L}$. 
We set the parallax phases $\gamma_j$ to zero and $3\pi/2$ for the right ascension and declination components, respectively.  
We generate a mock data set according to eq.~\ref{eq:microlensing_sig_1} comprised of six relative angular separation measurements, spanning a period of a year around the closest approach. We assume an astrometric light-centroiding precision of $\sigma_{\delta\theta}= \, 11 \mathrm{\mu as}$ per epoch, which is the expectation per observing night given the quoted apparent magnitude of the two stars in ref.~\cite{gaiamicrolensing} and Phase II instrumental parameters of an EPIC interferometer. We perform an MCMC parameter estimation of $\lbrace \theta_0^\alpha$, $\theta_0^\delta$, $\mu_0^{\alpha*}$, ${\mu_0^\delta}$, $\varpi_0$, $M_\text{L} \rbrace$, whose posteriors we show in figure~\ref{fig:microlensing_mass}.

The second mock data set is for the black-hole microlensing event MOA-2011-BLG-191/OGLE-2011-BLG-0462 from refs.~\cite{2022ApJ...933L..23L,2022ApJS..260...55L,2022ApJ...933...83S,2022ApJ...937L..24M}. We create a (counterfactual) simulation of the dark lens' mass estimation had EPIC taken data around the time that OGLE detected the photometric amplification. For our mock data generation, we assume a relative separation of $\lbrace\theta_0^\alpha,\theta_0^\delta\rbrace = \lbrace - 0.23, -0.17 \rbrace \, \mathrm{mas}$ at the time of closest (unlensed) approach (taken to be the reference epoch $t_0$), commensurate with the reported closest approach of $\mathrm{min} \, \beta \approx 0.05 \, \theta_\mathrm{E}$ and Einstein angle $\theta_\mathrm{E} \approx 5.81 \, \mathrm{mas}$ from ref.~\cite{2022ApJ...933...83S}. 
We take the mean values of their inferred velocities as our ``truth parameters'', namely $\{\mu^{\alpha*}_\text{L},\mu^\delta_\text{L}\}=\{4.36,3.06\} \, \mathrm{mas/yr}$ and $\{\mu^{\alpha*}_\text{S},\mu^\delta_\text{S}\}=\{2.263,-3.597\} \, \mathrm{mas/yr}$, with the relative proper motion $\vect{\mu}_0 \equiv \vect{\mu}_\mathrm{S}- \vect{\mu}_\mathrm{L}$. Likewise, we take the mass of the dark lens to be $M_\mathrm{L} = 7.1 \, M_\odot$ and a relative parallax magnitude of $\varpi_0 = 0.6 \, \mathrm{mas}$.  We also do not study the effects of the unequal brightness of the two images, nor their lensing-induced elongation (primarily along the axis perpendicular to $\hat{\vect{\theta}}_{+-}$), which could slightly affect the form factor. 

\begin{figure}[t]
    \centering
    \includegraphics[width=1\textwidth]{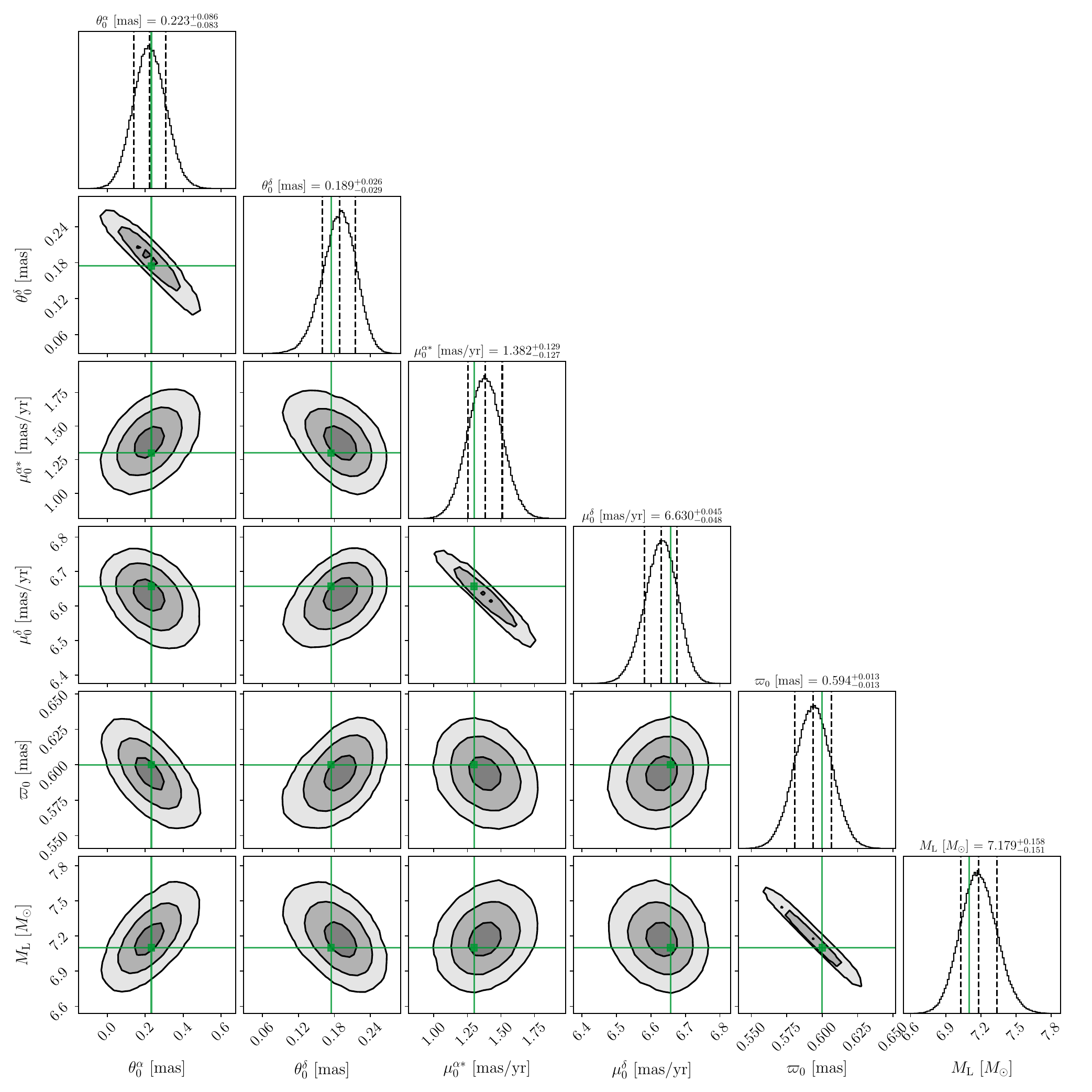}
    \caption{MCMC posteriors of the full relative astrometric solution $\{\theta_0^\alpha,\,\theta_0^\delta,\,\mu_0^{\alpha*},\,\mu_0^\delta,\,\varpi_0,\,M_\text{L}\}$ from a mock data set of six relative separation measurements of the two images of MOA-2011-BLG-191/OGLE-2011-BLG-0462~\cite{2022ApJ...933L..23L,2022ApJS..260...55L,2022ApJ...933...83S,2022ApJ...937L..24M}. The measurements span a period of two years around the time of closest approach of the star to the dark lens, each having a precision $\sigma_\theta= 6 \, \mathrm{\mu as}$ assuming Phase II instrumental parameters. The achieved determination of the mass of the dark lens---in this case an isolated black hole---is at the $1.7\%$ level.}
    \label{fig:dark_lens}
\end{figure}

Given the faintness of this particular source star, whose I-band magnitude varies in the range $13 \lesssim m_\mathrm{I} \lesssim 16$ during the microlensing event, Phase II of EPIC would yield a low SNR, so Phase III parameters might be needed for high significance. On the other hand, the source is far enough that larger baselines can be employed: a Sun at the $1.5 \, \mathrm{kpc}$ distance to the lens, would have an optimal baseline of $4 \, \mathrm{km}$ baseline, yielding an angular resolution of $\sigma_{\theta_\mathrm{res}} \approx 6 \, \mathrm{\mu as}$ in the I band, and even better light centroiding. Assuming a conservative light-centroiding precision of $\sigma_{\delta \theta} \simeq \sigma_{\theta_\mathrm{res}}$, we generate a mock data set which comprises six relative-angle measurements between the two images of the source, according to eq.~\ref{eq:microlensing_sig_2}, interspersed within two years around the closest approach. We perform an MCMC parameter estimation of $\{\theta_0^\alpha,\theta_0^\delta,\mu_0^{\alpha*},\mu_0^{\delta},\varpi_0,M_\text{L}\}$, whose posteriors we show in figure~\ref{fig:dark_lens}.

The capability of \textit{Gaia} in determining stellar masses from star-star microlensing is projected to be at best a few percent: according to ref.~\cite{MassesRD2}, the mission can achieve stellar mass determination of about 5 to 10\% for about seven events and below 15\% for thirteen events total. As seen from the marginalized posterior for the lens mass in figure~\ref{fig:microlensing_mass}, even with the conservative assumptions of a Phase II interferometer and just six observing epochs, EPIC can greatly surpass \textit{Gaia}'s capability per microlensing event by at least an order of magnitude, thereby achieving a 0.46\% determination of the lens mass.

The microlensing event MOA-2011-BLG-191/OGLE-2011-BLG-0462 from an isolated black hole yielded a determination of the dark lens' mass to about 18\% using \textit{HST} astrometry~\cite{2022ApJ...933L..23L,2022ApJS..260...55L,2022ApJ...933...83S,2022ApJ...937L..24M}. The observable in this case was the luminosity-weighted light centroid separation of \emph{both} images of the source star, since \emph{HST} cannot resolve them from each other.
In contrast, an EPIC interferometer, much like the proposal of ref.~\cite{2001A&A...375..701D} to use the VLT, can directly resolve the angular separation of the two images, thereby allowing for an order of magnitude better measurement of the dark lens' mass, down to a precision level of 1.7\% with Phase II specifications, as seen from  figure~\ref{fig:dark_lens}. Phase III could improve the significance and light centroiding by more than an order of magnitude, far surpassing the capabilities of optical interferometers, whose baselines can be at most a few hundred meters and light-centroiding precision of about $10 \, \mathrm{\mu as}$~\cite{2001A&A...375..701D}. It would be interesting to pursue a more detailed analysis of dark lens microlensing events, including the modification of the stellar sizes due to elongation, which could give a form factor suppression akin to that of eq.~\ref{eq:C2pointsource2opt_finite} if resolved, but could also provide information about the lens structure.

\subsection{Strongly lensed quasars}
\label{sec:quasar_microlensing}

Intensity interferometry, including the extended-path variation detailed in this work, is only feasible on compact sources with a high surface brightness. The most obvious sources \emph{at cosmological distances} are quasars, extremely luminous emission regions powered by the accretion of matter onto a supermassive black hole. (Other potential targets at large distances include dense stellar clusters and supernovae, but intensity interferometry applications for these sources are left to future work.) Their overall apparent brightness is not too small for advanced Phases of EPIC. The morphology of quasar emission regions is not well understood, and only an intensity interferometer may have the requisite effective angular resolution to directly resolve the nano-arcsecond structure of quasar central engines. Complementary indirect methods have also been proposed~\cite{moustakas2019astro2020}.

About 1 in 500 quasars are known to be strongly lensed into two or more images~\cite{treyer2004astrometric}. These images are often magnified by about an order of magnitude in apparent brightness, and are typically separated by about an arcsecond. (Smaller image separations are also possible but are not easily resolved, so some ``single quasars'' may actually consist of two or more duplicate images.) The first such system was $\mathrm{Q0957+561}$~\cite{walsh19790957}, while the most striking and best-studied system is $\mathrm{Q2237+0305}$~\cite{huchra19852237,kochanek2004quantitative}. A few hundred multiply-imaged quasars are known at present~\cite{inada2008sloan,mosquera2011microlensing}, with an explosion of this sample (approaching $10^4$) expected from upcoming surveys~\cite{oguri2010gravitationally}, most notably LSST but also the Dark Energy Survey, DESI, and Euclid.

EPIC is ideally suited to perform relative astrometry on any pair of strongly-lensed quasar images. The relative locations of these images  are not so interesting per se, but with its exquisite light-centroiding precision, EPIC can track \emph{changes} in their relative separation. Such changes can occur due to the (relative) peculiar velocities of the quasar-lens system, which lead to differential proper motion between the various quasar images because of the relative distortions caused by the lensing potential. The apparent image proper motions also sweep through the fine-grained structure of the gravitational potential, most notably the micro-caustic network of stars in the lens galaxy. Near any micro-caustic crossing, one expects (correlated) photometric~\cite{kochanek2004quantitative,jaroszynski1992microlensed,rauch1991microlensing,wambsganss1990interpretation,webster1991interpreting} and astrometric~\cite{treyer2004astrometric,williams1995,lewis1998quasar} variability. A companion paper~\cite{companionquasar2023} to this work fleshes out an entirely new signature---stochastic astrometric weak lensing fluctuations from low-mass dark matter halos.

In this work, we will content ourselves with making rough projections on the relative astrometric light-centroiding precision between two images, and how this information can be used to measure the effective proper motion and macro-lensing Jacobians. These measurements would allow, for the first time, highly significant determinations of transverse proper motions of objects at cosmological distances and strong-lensing geometries for multiply-imaged quasars in their respective rest frames. The lensing map from an image-plane location $\vect{\theta}$ to a source-plane location $\vect{\beta}$ is given by the lensing map $\vect{\varphi}$,
\begin{align}
\vect{\beta} = \vect{\varphi}(\vect{\theta}) = \vect{\theta} - \vect{\alpha}(\vect{\theta}),
\end{align}
with $\vect{\alpha}(\vect{\theta})$ the well-known (reduced) deflection angle, which can be written as:
\begin{align}
\vect{\alpha}(\vect{\theta}) = 2 \int_0^{D_\mathrm{q}} \dd D' \frac{D_\mathrm{q} - D'}{D_\mathrm{q}} \vect{\nabla}_\perp \Phi(D'\vect{\theta}, D').
\end{align}
In the above, $D_\mathrm{q}$ is the angular diameter distance to the quasar, and $\Phi$ is the 3D gravitational potential, which in almost all cases of relevance is dominated by the lens galaxy at some distance $D_\mathrm{L}$. The macro-lensing Jacobian is the determinant of the lensing map, and is given by:
\begin{align}
A_{ij} \equiv \frac{\partial \beta_i}{\partial \theta_j}.
\end{align}
Let $B$ be the inverse of the $2\times2$ matrix $A$, i.e.~$B_{ij} = (A^{-1})_{ij}$, which is a function of $\vect{\theta}$. The total image magnification is given by $\det B$.
The inverse map $\vect{\varphi}^{-1} : \vect{\beta} \to \vect{\theta}$ is, in general, multi-valued (leading to multiple images). Let us denote the multiple quasar image locations as $\vect{\theta}^\mathrm{I} = \vect{\varphi}^{-1}(\vect{\beta}_\mathrm{q})|_\mathrm{I}$ with $\mathrm{I} = \mathrm{A, B, \ldots}$ a discrete label for the images, and $\vect{\beta}_\mathrm{q}$ the quasar's location on the source plane. 

The effective apparent angular motion between the quasar and the lens galaxy (in the source plane) is 
\begin{align}
    \vect{\mu} \equiv \frac{\dd \vect{\beta}}{\dd t} = \frac{\vect{v}_\mathrm{q}}{1+z_\mathrm{q}} \frac{1}{D_\mathrm{q}} - \frac{\vect{v}_\mathrm{L}}{1+z_\mathrm{L}}\frac{1}{D_\mathrm{L}} + \frac{\vect{v}_\mathrm{o}}{1+z_\mathrm{L}} \frac{D_\mathrm{q} - D_\mathrm{L}}{D_\mathrm{L}D_\mathrm{q}}
\end{align}
where $\vect{v}_\mathrm{q}$, $\vect{v}_\mathrm{L}$, and $\vect{v}_\mathrm{o}$ are the proper velocities of the quasar, lens galaxy, and observer, respectively~\cite{kayser1986astrophysical}. 
With these definitions, the trajectory for each of the images is:
\begin{align}
\theta^\mathrm{I}_i(t) = \theta^\mathrm{I}_i(t_0) + \sum_j B^\mathrm{I}_{ij} \left[ \mu_j (t - t_0) + \varpi_j \cos\left(2 \pi \frac{t}{\mathrm{yr}} + \beta_j \right) \right] \label{eq:thetaI}
\end{align}
with $B^\mathrm{I}_{ij} \equiv B_{ij}[\vect{\theta}^\mathrm{I}(t_0)]$. For example, for Huchra's lens at $z_\mathrm{L} = 0.0394$ which quadruply lenses the quasar $\mathrm{Q2237{+}0305}$, we have 
\begin{align}
    \mu \simeq \frac{v_\mathrm{L}}{(1+z_\mathrm{L})D_\mathrm{L}} \approx 0.252 \, \mathrm{\mu as/yr} \left(\frac{v_\mathrm{L}}{200\,\mathrm{km/s}} \right); \qquad \varpi \simeq \frac{\mathrm{AU}}{(1+z_\mathrm{L})D_\mathrm{L}} \approx 0.0060 \, \mathrm{\mu as}. \label{eq:muHuchra}
\end{align}
These astrometric deviations are then further magnified by the strong lens, parametrized by the (inverse) Jacobian $B^\mathrm{I}_{ij}$, which can further amplify astrometric perturbations by up to a factor of 10 (or even higher~\cite{fan2019discovery}) along some direction(s), depending on the image and strong lens under consideration.

\begin{figure}
    \centering
    \includegraphics[width=0.7\textwidth, trim = 0 0 0 0]{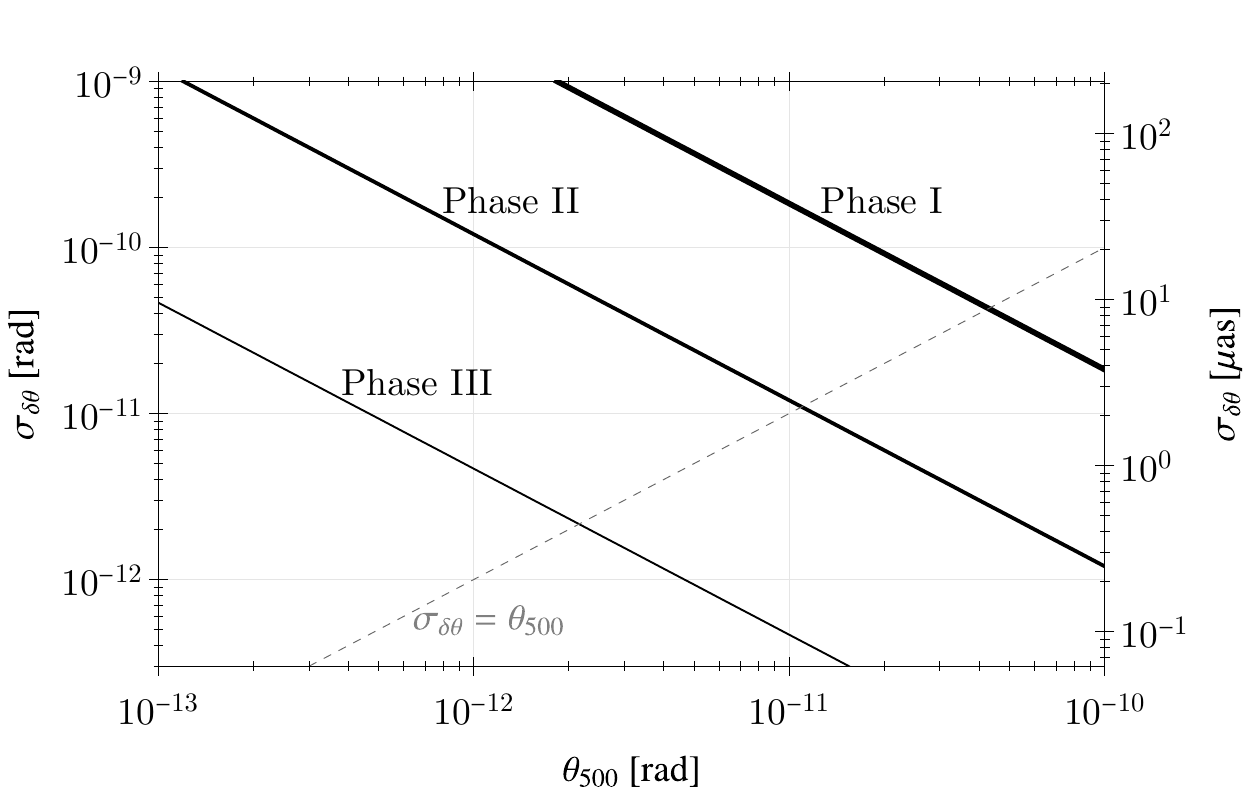}
    \caption{Light-centroiding precision $\sigma_{\delta \theta}$ from \Eq{eq:sigmaCentroid2} (at the optimal baseline distance $d^\text{opt}$ of \Eq{eq:doptq}) between two identical quasar images with apparent angular optical radius $\theta_{500}$. For EPIC Phase III, we project that $\sigma_{\delta \theta,\mathrm{q}}^\mathrm{opt} \approx 4.64 \times 10^{-13} \,\mathrm{rad} \, (10^{-11}\,\mathrm{rad}/\theta_{500})$ in our simple quasar emission model.
    \label{fig:thetaCentroid500}}
    \end{figure}

An error estimate analogous to the one in \Eq{eq:covalpha} for this simpler model without acceleration gives uncertainties on the estimators of the parallax, relative separation, and proper motion (in each direction $j = 1,2$) of:
\begin{align}
\begingroup
\renewcommand*{\arraystretch}{1.5}
\begin{pmatrix}
\sigma_{\widehat{\varpi}_0^j} \\
\sigma_{\widehat{\theta}_0^j} \\
\sigma_{\widehat{\mu}_0^j} 
\end{pmatrix} 
= 
\frac{\sigma_{\delta \theta}}{\sqrt{f_\mathrm{rep}\mathcal{T}}} 
\begin{pmatrix}
\sqrt{2} \\
1 \\
\frac{2\sqrt{3}}{\mathcal{T}} 
\end{pmatrix}
\approx 
\begin{pmatrix}
    0.168 \, \mathrm{\mu as}  \\
    0.119 \,  \mathrm{\mu as} \\
    0.0138 \, \mathrm{\mu as}/\mathrm{yr}
\end{pmatrix}
\left(\frac{\sigma_{\delta \theta}}{10^{-11}\,\mathrm{rad}}\right) \left(\frac{10/\mathrm{yr}}{f_\mathrm{rep}}\right)^{1/2}
\endgroup
\label{eq:covq}
\end{align}
with the latter estimate at $\mathcal{T} = 30\,\mathrm{yr}$. The quoted uncertainties are for the apparent parallax and proper motion in the image plane; the magnification by the $B^\mathrm{I}_{ij}$ factors for each image in the pair will further reduce the uncertainties for the quantities in the source plane.
We plot the expected light-centroiding precision $\sigma_{\delta \theta}$ from \Eq{eq:sigmaCentroid2} (at the optimal baseline distance $d^\text{opt}$ of \Eq{eq:doptq}) between two identical quasars in figure~\ref{fig:thetaCentroid500} for the three Phases of EPIC.
Comparison of \Eqs{eq:muHuchra} and~\ref{eq:covq} shows that the relative motions between any image pair should be measurable at high fractional precision over a long time period with EPIC Phase II and III, while the parallax is extremely challenging to measure, even for this especially close lens. 

In the simple model of \Eq{eq:thetaI}, we have ignored the effects of fine-grained structures in the lens galaxy, both from the stellar population and from dark matter substructures. Astrometric deflections from stellar substructure have been proposed as potentially observed signatures in ref.~\cite{treyer2004astrometric,lewis1998quasar} and from medium-scale dark matter subhalos in ref.~\cite{williams1995}; these are all interesting effects that can be explored with an intensity interferometer. A companion paper~\cite{companionquasar2023} will explore the astrometric noise caused by dark matter substructures at extremely small scales (substellar masses).

\section{Conclusions}
\label{sec:conclusions}

We have proposed the \emph{Extended-Path Intensity Correlator} (EPIC)~\cite{shortpaper}, an optical modification of intensity interferometers that introduces an adjustable,  differential geometric extension in the light paths from widely-separated sources. EPIC expands the field of view of traditional intensity interferometry---by up to six orders of magnitude beyond the angular resolution---while preserving the light-centroiding precision. 
In detail, the EPIC setup consists of a large primary mirror and a collimating secondary mirror that guides the light into the path-extension system (figure~\ref{fig:telescope}). The latter has the form of a Mach-Zehnder interferometer of unequal path lengths, and serves a dual role: it introduces the geometric path extension and corrects for the relative tilt of the sources' wavefronts. The wavefront tilt correction avoids the fringe suppression present when a constituent telescope resolves the source separation. The path extension allows for angular distances between sources at arcsecond separations to be determined with microarcsecond precision.

Our proposal arrives at a time when technological advances can be employed together to increase the signal-to-noise ratio and extend the intensity interferometry technique to a broader range of targets.  We develop a multichannel observation scheme over a broad frequency range, combined with large telescopes, fast photodetectors, and synchronization over large baseline distances. We outline the current capabilities in spectroscopy and picosecond-level timing, and identify the specific advances needed from the recently-demonstrated kilopixel single-photon detector arrays for the EPIC proposal.  Modern development of atomic clocks and infrastructure for large baseline radio interferometry  make possible picosecond-level synchronization and millimeter-level baseline  characterization  over the hours-long observation times and kilometer-plus distances needed for intensity interferometry. 

We find that EPIC will be able to target stars as faint as magnitude 15, as well as bright quasars, with telescope apertures no larger than $10\,\mathrm{m}$ in diameter. The maximum angular source separation for EPIC is limited by the atmosphere's isoplanatic angle---a few arcseconds in the optical---within which atmospheric fluctuations are correlated.

This work improves on differential astrometry strategies with intensity interferometry in several ways. The feasibility of measuring the relative separation of sources with traditional interferometry was first demonstrated on the close spectroscopic binary $\alpha$ Vir at the original NSII~\cite{1971MNRAS.151..161H}, which determined all six Keplerian orbital parameters. Recently, ref.~\cite{Stankus:2020hbc} has proposed to target widely separated sources (by more than the isoplanatic angle) with a \emph{fixed} path extension using a single beam splitter and a GHz photon bandwidth per spectral channel. Such a setup  imposes exacting (and potentially infeasible)  requirements on the adaptive optics and spectroscopic systems.  Using the \emph{variable} path extension proposed here, we can track the separation of sources at or near the main path-extended fringe (center of figure~\ref{fig:C2b}) and employ the differential wavefront tilt correction to avoid fringe suppression for sources separated by more than a few milliarcseconds. 
Our approach yields a per-epoch light-centroiding precision that scales as $t_\mathrm{obs}^{-1/2}$ (\Eq{eq:sigmaCentroid2}), compared to the fringe-scanning method of ref.~\cite{Stankus:2020hbc} at extremely high fringe orders, which could lead to fringe confusion and scales as $t_\mathrm{obs}^{-3/2}$.

In addition, while multi-band observation schemes have been proposed in recent years~\cite{2014JKAS...47..235T,Stankus:2020hbc,Chen:2022ccn}, this work is the first to study the optical tolerances induced by such a scheme.   The few picosecond time-resolution assumed in this work allows for errors in optical paths as long as a millimeter. The desired spectral resolution and the small pixel sizes of ultra-fast photon counters are the limiting factors, and can tolerate primary-mirror imperfections as large as tens of wavelengths, in contrast to imaging telescopes and amplitude interferometers that require sub-wavelength polishing~\cite{bely2003design}.

We also provide the first quantitative study of a fringe contrast suppression for very fast photodetectors and large pixels---unique to ultra-fast and multi-channel intensity interferometry and an entirely new effect unveiled in this work. This perhaps surprising effect originates for setups with photodetector arrays and spectroscopic splitting:  atmospheric and instrumental aberrations require the angle subtended by each pixel to be larger than the diffraction limit of each telescope, resulting in uncorrelated fluctuations within a pixel.  We identify several possible solutions to this problem, and focus on the conceptually simplest: splitting the beam exiting the EPIC optical delay system into several sub-beams, each with its own dispersion element and photodetector array. 

 Finally,  we calculate effects from atmospheric aberrations and their limitation on the maximum source separation of order the isoplanatic angle for a ground-based intensity interferometer; correcting for atmospheric turbulence over angular separations significantly larger than the isoplanatic patch is beyond the capabilities of present adaptive optics systems~\cite{bely2003design}. In the future, should a lunar observatory be established, an EPIC observatory would be an clear candidate that would benefit from the absence of an atmosphere;  study of a large opening angle design without atmospheric limitations is left to future work.
 
We present several scientific applications that would experience major leaps forward by leveraging the differential astrometric capabilities of EPIC. 
The target sources of interest in the Milky Way are binary systems, accidental double stars, and strong microlensing events.
EPIC could detect exoplanets astrometrically through the wobble they impart on their host stars with a reach orders of magnitude better than \textit{Gaia}'s projected DR5~\cite{shortpaper}. An EPIC interferometer targeting a binary system could determine all Keplerian angles to a sub-degree level and, combined with radial velocity measurements, determine the masses of the constituent stars and the line-of-sight distance to the per-mil level or better. This  level of precision would open up possibilities of measurements of stellar mass loss and cosmic distance ladder calibration. Observations of accidental double stars can be used to detect astrometric microlensing and determine stellar masses to the sub-percent level and, with later Phases of the setup, map the potential of the Milky Way by directly measuring the induced stellar accelerations at the $\mathrm{nas / yr^2}$ level, significantly surpassing \textit{Gaia} sensitivity~\cite{buschmann2021galactic}. Finally, EPIC could target \emph{extragalactic} sources such as quasars (and possibly supernovae). We study the specific case of measuring the astrometric evolution of the images in a multiply-imaged quasar to characterize the strong-lensing system and determine peculiar velocities at cosmological distances.

Directions for future work are numerous. At the level of  design, identifying a simpler solution to the aforementioned spectroscopic suppression is important to reducing both complexity and cost, potentially allowing the construction of several baselines and many telescopes per observation site. Potential sources of observational systematic uncertainties, such as contamination from ambient light or other sources in a crowded field, should be characterized in more detail. On the experimental side, it is crucial to demonstrate kilo-pixel arrays of single-photon detection devices tailored to the requirements of EPIC, including broadband sensitivity to a wide range of wavelengths. Naturally, a laboratory demonstration of the EPIC's path-extension system  would be the first step to large-scale implementation.

Further study of the full scientific potential of the applications is also of interest. While we use real example systems to illustrate the power of our technique, we have not performed an exhaustive survey of potential sources. For instance, thousands of microlensing events are predicted to occur in the next few decades based on the \textit{Gaia} catalog alone~\cite{gaiamicrolensing}, hundreds of which could be targeted by EPIC. \textit{Gaia} data can also be used to construct a larger double-star catalog to  map out the Milky Way's potential via the relative transverse angular accelerations it imparts on source pairs. Finding sufficiently bright and distant binary systems of Cepheids and other giant stars would aid in refining the first rung of the cosmic distance ladder.

We have also foretold the possibility to probe sub-solar-mass dark matter substructure using astrometry on the multiple images of strongly lensed quasars, which will be studied in detail in a future paper~\cite{companionquasar2023}. In addition, we expect  EPIC may test general relativity in the vicinity of Sagittarius A*~\cite{Ghez_1998,2008ApJ...689.1044G,2009ApJ...692.1075G}, with potentially sufficient sensitivity to measure the spin of the central black hole and surrounding mass distribution~\cite{companionSagA}. Finally, it would be interesting to generalize our analysis of multi-channel intensity interferometry (with and without path extension) to other extended, high surface brightness sources, such as the various types of novae and the cores of globular clusters and galaxies, and investigate corresponding applications.

Extended-Path Intensity Correlation will enable new sets of astronomical measurements at unprecedented precision. It has the potential to transform our knowledge about stellar systems and the Milky Way, and may ultimately provide a new window into the dark universe and physics Beyond the Standard Model.

\acknowledgments{We thank Gordon Baym, Megan Bedell, Karl Berggren, Michael Blanton, Matteo Cantiello, Cyril Creque-Sarbinowski, Liang Dai, Neal Dalal, Julianne Dalcanton, Peter Graham, David Hogg, Miguel Morales, Oren Slone, and David Spergel for valuable conversations and input, and Calvin Chen, Jacob Crawford, Cyril Creque-Sarbinowski, David Dunsky, Marius Kongsore, and Jessie Yang for comments on the manuscript.
KVT thanks Jason Aufdenberg, Matthew Brown, James Buckley, Dainis Dravins, David Kieda, Michael Lisa, Nolan Matthews, Andrei Nomerotski, Ue-Li Pen, Naomi Vogel, Shiang-Yu Wang, and Luca Zampieri for fruitful conversations, and Sebastian Karl for pointing out ref.~\cite{franson1989bell},  during the 2023 Workshop on Stellar Intensity Interferometry at The Ohio State University.
KVT is supported by the National Science Foundation under Grant PHY-2210551. 
NW is supported by NSF under award PHY-2210498, by the BSF under grant 2018140, and by the Simons Foundation. 
MB is supported by the DOE Office of Science under Award Number DE- SC0022348, University of Washington Royal Research Fund and through the Department of Physics and College of Arts and Science at the University of Washington. 

The Center for Computational Astrophysics at the Flatiron Institute is supported by the Simons Foundation.
Research at Perimeter Institute is supported in part by the Government of Canada through the Department of Innovation, Science and Economic Development and by the Province of Ontario through the Ministry of Colleges and Universities.

Part of this work was performed at the Aspen Center for Physics, which is supported by National Science Foundation grant PHY-1607611 and PHY-2210452. The participation of MB at the Aspen Center for Physics was supported by the Simons Foundation.
MB, MG, and KVT thank the Institute for Nuclear Theory at the University of Washington for its kind hospitality and stimulating research environment. The  INT is supported in part by the U.S. Department of Energy grant No.~DE-FG02-00ER41132.

This work has made use of data from the European Space Agency (ESA) mission {\it Gaia} (\url{https://www.cosmos.esa.int/gaia}), processed by the {\it Gaia} Data Processing and Analysis Consortium (DPAC, \url{https://www.cosmos.esa.int/web/gaia/dpac/consortium}). Funding for the DPAC has been provided by national institutions, in particular the institutions participating in the {\it Gaia} Multilateral Agreement. This research has made use of the NASA Exoplanet Archive, which is operated by the California Institute of Technology, under contract with the National Aeronautics and Space Administration under the Exoplanet Exploration Program.
}

\appendix

\appendix

\section{Notation}
\label{app:notation}

In tables~\ref{tab:notationE}--\ref{tab:notationKDII}, we relist all variables used in this work in the first column, along with their definitions in the second column.

\begin{table}[h!]
\centering
\small
\begin{tabular}{l | l}
\hline
\hline
Variable & Definition\\
\hline 
$E_p$ & Total electric field at telescope $p$\\
$E_{i\alpha}$ & Electric field amplitude of $i$ and of energy $k_\alpha$\\
$I$, $\widetilde{I}_j$ & Intensity, fractional intensity of emitter $j$ \\
$k = 2\pi / \lambda$, $\overline{k}$ & Wavenumber (in terms of wavelength), mean wavenumber \\
$\hat{\vect{\theta}}_j$ & Unit vector to $j$\\
$\vect{\theta}_{ij} = \hat{\vect{\theta}}_i - \hat{\vect{\theta}}_j$ & Relative angular distance of $i$ and $j$.\\
$r_{ip}$ & Distance of $i$ from telescope $p$\\
$\phi^\text{em}_{i\alpha}$ & Intrinsic phase of emitter $i$ and energy $k_\alpha$\\
$\widetilde{\phi}_{i\alpha}^{(p)}$ & Aberration phase at telescope $p$\\
${\di I}/({\di k \di \Omega})$ & Intensity (power per area) per unit wavenumber and unit solid angle\\
\hline
\hline
\end{tabular}
\caption{Electric field variables.}
\label{tab:notationE}
\end{table}

\begin{table}[h!]
\centering
\small
\begin{tabular}{l | l}
\hline
\hline
Variable & Definition\\
\hline 
$\bold{d}$ & Baseline vector\\
$\bold{b}$ & Aperture vector\\
$A_i$ & Area of aperture of telescope $i$\\
$D$ & Diameter of primary mirror\\
$f$, $f'$, $f_\text{c}$ & Focal lengths of primary, secondary, and camera\\
$n_p^\text{arr}$ & Number of telescopes at array site $p$\\
$\alpha_{nlm}$ & Optical aberration amplitude\\
\hline
$\ell_p$ & Geometric path extension at telescope $p$\\
$\gamma$ & Angle of first beam-splitter\\
$\gamma',\delta'$ & Angles of wavefront correction mirror\\
$\epsilon$ & Transmission coefficient of beam splitter\\
\hline
$N$ & Number of illuminated lines on diffraction grating \\
$d_\text{g}$ & Periodicity of grating\\
$s$ & Slit size of grating\\
$\theta_\text{F}$ & Angle from camera to the pixel plane\\
$\mathcal{R}$ & Spectroscopic resolution\\
$\sigma_k$ & Range of wavenumbers on incident on pixel\\
\hline
$\tau$ & Timing offset\\
$\sigma_t/\sqrt{2}$ & Timing precision per photodetector \\
$t_\text{obs}$ & Observation time per epoch\\
$t_\text{jitter}$ & Time jitter of photodetector (FWHM)\\
$t_\text{reset}$ & Photodetector reset time\\
DCR & Dark count rate of photodetector\\
$N_p$ & Total number of photons incident at $p$ after time $t_\text{obs}$\\
$\eta_p$ & Photodetection efficiency of $p$\\
\hline
$C(\bold{d},\tau)$ & Intensity correlation\\
$\hat{X}$ & Estimator of $X$ \\
$\avg{X}$ & Expectation value of $X$ (averaged over mode phases) \\
$\overline{X(t)}$ & Time average of $X(t)$ over observation time \\
$\sigma_{\hat{C}}$ & Standard deviation of $\hat{C}$\\
$\sigma_{\theta_\text{res}}$ & Angular resolution of the interferometer\\
$\sigma_{\Delta \theta}$ & Angular dynamic range of the interferometer \\
$\sigma_{\hat{\boldsymbol{\theta}}}$ & Global astrometric precision \\
\hline
\hline
\end{tabular}
\label{tab:notationI}
\caption{Instrumental variables.}
\end{table}

\begin{table}[h!]
\centering
\small
\begin{tabular}{l | l}
\hline
\hline
Variable & Definition\\
\hline 
$n[\bold{r}]$ & Atmospheric index at 3D position $\bold{r}$\\
$D_n(r)$ & Structure function, $r$-dependent\\
$h$ & Altitude in the atmosphere\\
$C_n(h)$ & Structure function, dependence on altitude $h$\\
$l_0$ & Inner turbulence scale\\
$L_0$ & Outer turbulence scale\\
$B_n(r)$ & Correlation function of atmospheric index of refraction\\
$\theta_0(\overline{k},\gamma)$ & Isoplanatic patch angle at an angle $\gamma$ away from zenith\\
$\sigma_\phi^\text{atm}$ & Standard deviation of differential phase fluctuation due to the atmosphere\\
$\theta_\text{seeing}$ & Seeing angle\\
$r_0$ & Fried parameter\\
\hline
\hline
\end{tabular}
\label{tab:notationA}
\caption{Atmospheric fluctuation variables from section~\ref{sec:atm}, appendix~\ref{app:atm}, and appendix~\ref{app:atm_seeing}.}
\end{table}

\begin{table}[h!]
\centering
\small
\begin{tabular}{l | l}
\hline
\hline
Variable & Definition\\
\hline
$R_\text{s}$ & Stellar radius\\
$D_\text{s}$ & Distance to star/binary\\
$\theta_\text{s}$ & Angular radius of star\\
$M_j$ & Mass of star $j$\\
$T_j$ & Temperature of $j$\\
$\mathcal{F}_j$ & Finite-size form factor of $j$\\
\hline
$T_{500} = 2 \pi \hbar c / 500 \, \mathrm{nm}$ & Black body temperature corresponding to $500\,\mathrm{nm}$ \\
$R_\text{500}$ & Physical radius in quasar's local frame with temperature $T_\text{500}$\\
$z_\mathrm{q}$ & Quasar redshift\\
$R_\text{isco}$ &  Innermost stable circular orbit radius of quasar\\
$T_{\mathrm{q},0}$ & Quasar reference temperature\\
\hline
\hline
\end{tabular}
\label{tab:notationS}
\caption{Single source parameters.}
\end{table}

\begin{table}[h!]
\centering
\small
\begin{tabular}{l | l}
\hline
\hline
Variable & Definition\\
\hline
$f_\text{rep}$ & Observational repetition frequency \\
\hline
$G$ & Gravitational constant\\
$M$ & Total binary mass\\
$a$ & Semi-major axis\\
$e$ & Eccentricity\\
$i$ & Inclination\\
$\Omega$ & Longitude of ascending node\\
$\omega$ & Argument of periapsis\\
$\nu$ & True anomaly\\
$t_0$ & Reference epoch\\
$E$ & Eccentric anomaly\\
$\mathcal{M}$ & Mean anomaly\\
$P$ & Period of binary\\
$\dot{z}_i$ & Radial velocity of star $i$\\
$K_i$ & Radial velocity semi-amplitude of star $i$\\
\hline
\hline
\end{tabular}
\label{tab:notationKDI}
\caption{Binary system parameters from sections~\ref{sec:binaries} and~\ref{sec:exoplanets}.}
\end{table}

\begin{table}[h!]
\centering
\small
\begin{tabular}{l | l}
\hline
\hline
Variable & Definition\\
\hline 
$\mu_+$,  $\mu_-$ & Magnification factor of $+$ and $-$ images \\
$\vect{\mu}$ & Angular velocity\\
$\vect{\alpha}$ & Angular acceleration\\
$\varpi$ & Parallax\\
$\theta_\text{E}$ & Einstein angle\\
\hline
$\Phi$ & Gravitational potential\\
$\boldsymbol{\beta}$ & Source-plane location\\
$\boldsymbol{\varphi}$ & Lensing map\\
$A_{ij}$, $B_{ij}$ & Macro-lensing Jacobian and its inverse\\
\hline
\hline
\end{tabular}
\caption{Microlensing parameters from sections~\ref{sec:acceleration} and~\ref{sec:quasar_microlensing}.}
\label{tab:notationKDII}
\end{table}

\clearpage
\newpage

\section{Derivations}
\label{app:derivations}

\subsection{Intensity correlation noise}
\label{app:snr}
In what follows, we derive expressions for the expectation values of various moments of the intensity estimator $\hat{I}_p(t_p)$ from eq.~\ref{eq:Ihat1}. Firstly, its time average
\begin{alignat}{1}
    \overline{\hat{I}_p(t_p)} \equiv \frac{1}{t_\mathrm{obs}}\int_{t_p - t_\mathrm{obs}/2}^{t_p + t_\mathrm{obs}/2} \dd t' \, \hat{I}_p(t')
\end{alignat}
is expected to equal the true intensity $\langle I_p\rangle$, since it is an unbiased estimator:
\begin{align}
\avg{ \overline{\hat{I}_p(t_p)}}= \frac{\hbar c \overline{k}}{\eta_p A_p} \int_{t_p-t_\text{obs}/2}^{t_p+t_\text{obs}/2}\frac{\di t'}{t_\text{obs}} \sum_{i=1}^{N_p} \int \frac{\di t_i}{t_\text{obs}} \frac{e^{\frac{-(t'-t_i)^2}{\sigma_t^2}}}{\sqrt{\pi} \sigma_t} = \frac{\hbar c \overline{k}}{\eta_p A_p} \frac{N_p}{t_\text{obs}} \equiv \langle I_p \rangle.
\end{align}
Again, because $\hat{I}_p(t_p)$ from eq.~\ref{eq:Ihat1} is an unbiased estimator of the local intensity field, it also faithfully reflects the intensity \emph{correlations} at detectors $p=1$ and $p=2$:
\begin{align}
\avg{ \overline{ \hat{I}_1(t_1)\hat{I}_2(t_2) }} &= \avg{\overline{ \hat{I}_1(t_1)}} \avg{\overline{\hat{I}_2(t_2)}} \left[1 + \underbrace{C(\vect{d},\ell,t_2 - t_1)}_{\text{correlations} \ll 1}\right] \\
&\simeq \frac{\hbar^2 c^2 \overline{k}^2}{\eta_1 \eta_2 A_1 A_2}  \frac{N_1 N_2}{t_\text{obs}^2} = \langle I_1 \rangle \langle I_2 \rangle \label{eq:corrIest}.
\end{align}
In the second line, we have approximated the excess intensity correlations to be small ($C \ll 1$), which is usually a good approximation. Equation~\ref{eq:corrIest} implies that $\hat{C}$ from eq.~\ref{eq:C2estimate} is also an unbiased estimator of $C$.

The statistical error on the time-averaged intensities is negligibly small, because:
\begin{align}
& \avg{\bigg[\overline{\hat{I}_p(t_p)}\bigg]^2}  \simeq \left(\frac{\hbar c \overline{k}}{\eta_p A_p}\right)^2 \left( \prod_{i=1}^{N_p}\int\frac{\di t_i}{t_\text{obs}} \right) \int\frac{\di t'}{t_\text{obs}} \left(\sum_{i=1}^{N_p}\frac{e^{\frac{-(t'-t_i)^2}{\sigma_t^2}}}{\sqrt{\pi} \sigma_t} \right)  \int\frac{\di t''}{t_\text{obs}}  \left(\sum_{i=1}^{N_p}\frac{e^{\frac{-(t''-t_i)^2}{\sigma_t^2}}}{\sqrt{\pi} \sigma_t} \right) \nonumber \\
&\simeq \left(\frac{\hbar c \overline{k}}{\eta_p A_p}\right)^2 \left(\frac{N_p}{t_\mathrm{obs}}\right)^2 =  \langle I_p \rangle^2.
\end{align}
In other words, the fractional variance of $\overline{\hat{I}_p(t_p)}$ is of order $\mathcal{O}(1/N_p)$, exceedingly small for very bright sources with $N_p \gg 1$ needed to measure $C$ with high SNR. The (fractional) statistical error on the intensity correlations is significantly larger, since it receives contributions only from near-simultaneous photon arrivals. We find:
\begin{align}
&\avg{\bigg[\overline{\hat{I}_1(t)\hat{I}_2(t+\tau)}\bigg]^2} = \frac{(\hbar c \overline{k})^4}{(\eta_1 \eta_2 A_1 A_2)^2}\left( \prod_{i=1}^{N_1}\int\frac{\di t_i^{(1)}}{t_\text{obs}} \right) \left( \prod_{m=1}^{N_2}\int\frac{\di t_m^{(2)}}{t_\text{obs}} \right) \int\frac{\di t'}{t_\text{obs}}\int\frac{\di t''}{t_\text{obs}} \nonumber \\
&\phantom{=}\times \left[\sum_{i=1}^{N_1}\frac{e^{\frac{-\left(t'-t_i^{(1)}\right)^2}{\sigma_t^2}}}{\sqrt{\pi} \sigma_t} \right]
\left[\sum_{m=1}^{N_2}\frac{e^{\frac{-\left(t'-t_m^{(2)}\right)^2}{\sigma_t^2}}}{\sqrt{\pi} \sigma_t} \right]
\left[\sum_{j=1}^{N_1}\frac{e^{\frac{-\left(t''-t_j^{(1)}\right)^2}{\sigma_t^2}}}{\sqrt{\pi} \sigma_t} \right]
\left[\sum_{n=1}^{N_2}\frac{e^{\frac{-\left(t''-t_n^{(2)}\right)^2}{\sigma_t^2}}}{\sqrt{\pi} \sigma_t} \right] \nonumber \\
&=\frac{(\hbar c \overline{k})^4}{(\eta_1 \eta_2 A_1 A_2)^2}\frac{1}{t_\text{obs}^2}\left( \prod_{i=1}^{N_1}\int\frac{\di t_i^{(1)}}{t_\text{obs}} \right) \left( \prod_{m=1}^{N_2}\int\frac{\di t_m^{(2)}}{t_\text{obs}} \right) \sum_{\substack{i,j\\m,n}} \frac{e^{ \frac{-1}{2\sigma_t^2}\left[\left(t_i^{(1)} - t_m^{(2)}   \right)^2 + \left(t_j^{(1)} - t_n^{(2)}  \right)^2\right] }}{2 \pi \sigma_t^2}\\
&\simeq \frac{(\hbar c \overline{k})^4}{(\eta_1 \eta_2 A_1 A_2)^2}\frac{1}{t_\text{obs}^2} \left\lbrace \left[ \sum_{i,m} \int\frac{\di t_i^{(1)}}{t_\text{obs}} \int\frac{\di t_m^{(2)}}{t_\text{obs}}  \frac{e^{ \frac{-1}{2\sigma_t^2}\left(t_i^{(1)} - t_m^{(2)}   \right)^2  }}{\sqrt{2 \pi}\sigma_t }  \right]^2  \right. \nonumber\\
&\phantom{\simeq \frac{(\hbar c \overline{k})^4}{(\eta_1 \eta_2 A_1 A_2)^2}\frac{1}{t_\text{obs}^2} \Bigg \lbrace \,} 
+ \left. \sum_{i,m}  \int\frac{\di t_i^{(1)}}{t_\text{obs}} \int\frac{\di t_m^{(2)}}{t_\text{obs}}  \frac{e^{ \frac{-1}{\sigma_t^2}\left(t_i^{(1)} - t_m^{(2)}   \right)^2  }}{2 \pi \sigma_t^2 } \right\rbrace \nonumber \\
&=\frac{(\hbar c \overline{k})^4}{(\eta_1 \eta_2 A_1 A_2)^2} \left[\left(\frac{N_1 N_2}{t_\text{obs}^2}\right)^2 + \frac{N_1 N_2}{t_\text{obs}^3 \sqrt{4\pi} \sigma_t} \right] \nonumber
\end{align}
The latter term in the last equality is responsible for the statistical variance on $\hat{C}$ in the low-SNR limit for each spectral channel. Putting everything together, we find:
\begin{align}
\sigma_{\hat{C}}^2 &\equiv \big\langle{ \hat{C}^2 } \big\rangle- \big\langle{ \hat{C} }\big\rangle^2 \simeq  \frac{\avg{\bigg[\overline{\hat{I}_1(t)\hat{I}_2(t+\tau)}\bigg]^2} -  \bigg[ \avg{\overline{\hat{I}_1(t)\hat{I}_2(t+\tau)}}\bigg]^2}{\langle I_1 \rangle^2 \langle I_2 \rangle^2} = \frac{t_\text{obs}}{\sqrt{4 \pi} \sigma_t} \frac{1}{N_1 N_2}.
\end{align}
This corresponds to eq.~\ref{eq:sigmaC} from the main text.

\subsection{Atmospheric aberration}\label{app:atm}

\begin{figure}
\centering
\includegraphics[width = 0.7\textwidth]{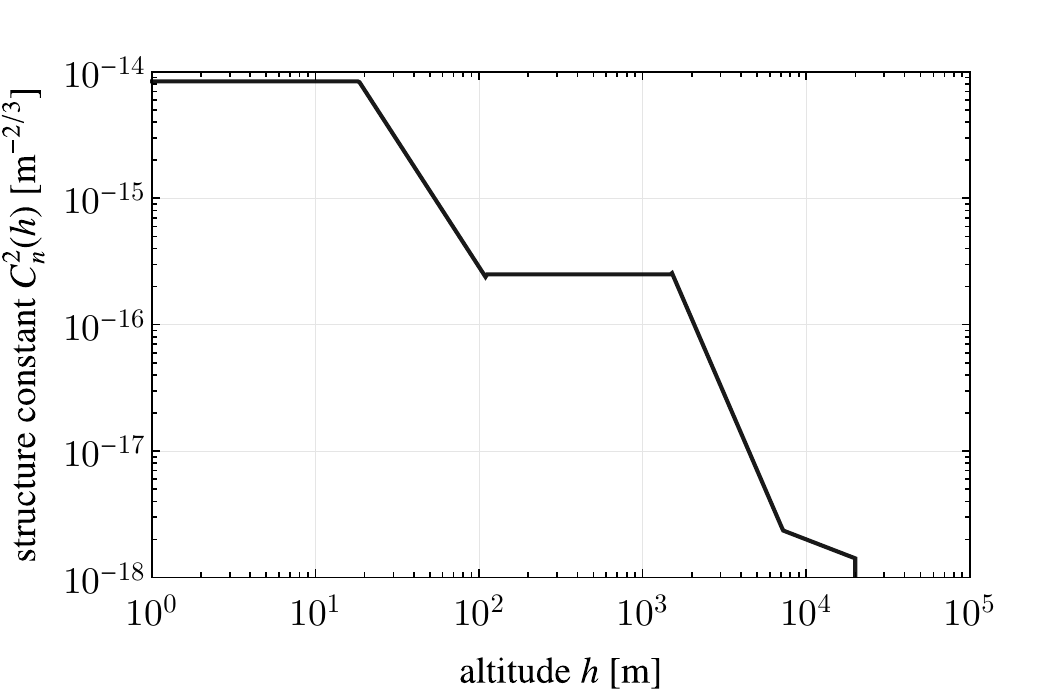}
\caption{Structure constant $C_n^2$ for atmospheric refraction from eq.~\ref{eq:Dn} as a function of altitude $h$ in the SLC night model.}\label{fig:SLC}
\end{figure}

We derive the mean square doubly-differential refraction phase from atmospheric aberrations (eq.~\ref{eq:sigma_phi_atm} in the main text). We find:
\begin{align}
&\left( \sigma_\phi^\text{atm} \right)^2\equiv \Big \langle \left[\Delta\widetilde{\phi}^{(1,2)}(\overline{k}, \hat{\vect{\theta}}_a,\hat{\vect{\theta}}_b)\right]^2 \Big \rangle_\mathrm{atm} \nonumber\\
&= \overline{k}^2\iint \dd s \, \dd s' \Big\langle  \Big(n[\vect{x}_{a1}(s)] + n[\vect{x}_{b2}(s)]-n[\vect{x}_{a2}(s)] - n[\vect{x}_{b1}(s)]\Big) \nonumber \\
&\phantom{= k^2\iint \dd s \, \dd s' \Big\langle} \times \Big(n[\vect{x}_{a1}(s')] + n[\vect{x}_{b2}(s')]-n[\vect{x}_{a2}(s')] - n[\vect{x}_{b1}(s')]\Big)\Big \rangle_\mathrm{atm} \nonumber \\
&= \overline{k}^2\iint_0^{L_\mathrm{atm} \sec \gamma} \dd s \, \dd s'\, \frac{-C_n^2(s \cos \gamma)}{2} \Big\lbrace\\
& +\big|\vect{x}_{a1}(s) - \vect{x}_{a1}(s')\big|^{2/3} + \big|\vect{x}_{b1}(s) - \vect{x}_{b1}(s')\big|^{2/3} + \big|\vect{x}_{a2}(s) - \vect{x}_{a2}(s')\big|^{2/3} + \big|\vect{x}_{b2}(s) - \vect{x}_{b2}(s')\big|^{2/3} \nonumber \\
& - \big|\vect{x}_{a1}(s) - \vect{x}_{b1}(s')\big|^{2/3} - \big|\vect{x}_{b1}(s) - \vect{x}_{a1}(s')\big|^{2/3} - \big|\vect{x}_{a2}(s) - \vect{x}_{b2}(s')\big|^{2/3} - \big|\vect{x}_{b2}(s) - \vect{x}_{a2}(s')\big|^{2/3} \nonumber \\
& -\big|\vect{x}_{a1}(s) - \vect{x}_{a2}(s')\big|^{2/3} - \big|\vect{x}_{b1}(s) - \vect{x}_{b2}(s')\big|^{2/3} - \big|\vect{x}_{a2}(s) - \vect{x}_{a1}(s')\big|^{2/3} - \big|\vect{x}_{b2}(s) - \vect{x}_{b1}(s')\big|^{2/3} \nonumber \\
&  + \big|\vect{x}_{a1}(s) - \vect{x}_{b2}(s')\big|^{2/3} + \big|\vect{x}_{b1}(s) - \vect{x}_{a2}(s')\big|^{2/3} + \big|\vect{x}_{a2}(s) - \vect{x}_{b1}(s')\big|^{2/3} + \big|\vect{x}_{b2}(s) - \vect{x}_{a1}(s')\big|^{2/3} \Big\rbrace.\nonumber
\end{align}
To get to the second equality, we plugged in eq.~\ref{eq:phi_atm}. To get to the last equality, we used eqs.~\ref{eq:Dn} and~\ref{eq:Bn}. We integrate to the ``height'' $L_\mathrm{atm}$ of the atmosphere, though technically  $C_n^2(h)$ drops to zero smoothly. In the last four lines, we have expanded the 16 correlation functions into suggestive groups. The last two lines are not correct if $d \gg L_0$, and we will ignore them in what follows; in that regime, the sum of the last two lines (which involve cross-correlations between paths to \emph{different} detectors) is very strongly suppressed. Each of the four terms in the first line are equal to $[(s-s')^2]^{1/3}$, while those in the second line are of the form $[(\hat{\vect{\theta}}_a s - \hat{\vect{\theta}}_b s')^2]^{1/3}$. Putting this all together gives:
\begin{align}
\left( \sigma_\phi^\text{atm} \right)^2 & \simeq 2 \overline{k}^2\iint_0^{L_\mathrm{atm} \sec \gamma} \dd s \, \dd s'\, C_n^2(s \cos \gamma) \Big\lbrace (s^2 - 2 s s' \cos\theta_{ba} + s'^2)^{1/3} - (s-s')^{1/3} \Big\rbrace \nonumber \\
&\simeq 2 \overline{k}^2\int_0^{L_\mathrm{atm} \sec \gamma}\di s
\, C_n^2(s \cos \gamma)  \underbrace{\int_{-s}^{+s} \di x\, \Big \lbrace \big[x^2 + s (s+x)\theta_{ba}^2 \big]^{1/3} - \big[x^2\big]^{1/3} \Big \rbrace}_{\sqrt{\pi}\frac{ \Gamma(-5/6)}{\Gamma(-1/3)} \theta_{ba}^{5/3} - 2 \theta_{ba}^2 + \mathcal{O}(\theta_{ba}^{11/3}) }.
\end{align}
To get to the last line, we used the small-angle approximation, and used the fact that most of the integral has support at $|x| = |s-s'| \lesssim s$ to separate the integrals. This result, combined with the definition of the isoplanatic patch angle in eq.~\ref{eq:isoplanatic}, gives eq.~\ref{eq:sigma_phi_atm}. For reference, we show the dependence of the structure constant $C_n^2(h)$ on altitude $h$ in figure~\ref{fig:SLC}.

\section{Wave Optics}
\label{app:waveoptics}

In this appendix, we carry out an analysis using scalar wave optics for the propagation of starlight within each telescope. Emphasis is given on analytic results to understand the requirements for an EPIC intensity interferometer. We include finite telescope apertures and finite pixel sizes in our analysis, as well as atmospheric and instrumental aberrations. Some simplifications and idealizations are necessarily made, and we leave a detailed numerical study to future work.

The structure of this appendix follows the depiction of our setup in figure~\ref{fig:telescope}: section~\ref{sec:collimation} treats the primary and secondary mirrors, section~\ref{sec:time-delay} the optical elements of the geometric path-delay system, section~\ref{app:aberrations} the effect of aberrations on image formation, and finally section~\ref{app:grating} presents the details of the calculation of section~\ref{sec:grating} concerning the seeing angle suppression for large telescopes and fast photodetectors resulting from the spectroscopic instrument.

\subsection{Primary and secondary mirrors}
\label{sec:collimation}

The focus of this section is the derivation of the spatial profile of the electric field, after collimation from the secondary mirror (as depicted in figure~\ref{fig:telescope_1}), in the presence of mirror distortions. Our main result here is the functional dependence of the electric field profile on the focal ratio of the two mirrors, which magnifies both stellar separations and wavefront aberrations induced by the primary mirror.

Let two point-like stars, $a$ and $b$, emit monochromatic spherical waves of wavenumber $k$. Right before incidence on the parabolic mirror of a telescope, the electric field is given by
\begin{equation}
\begin{split}
    E_k(\bold{x},t)&=A e^{i\omega t} e^{-i k |\bold{r}_a+\bold{x}|+i\phi^\text{em}_a+i\tilde{\phi}^\text{atm}_a(\bold{x},t)}+Be^{i\omega t}e^{-i k |\bold{r}_b+\bold{x}|+i\phi^\text{em}_b+i\tilde{\phi}^\text{atm}_b(\bold{x},t)}\\
    &\simeq e^{i\omega t}\pare{A e^{-i k( r_a+ \bold{x}\cdot \hat\bth_a) +i\phi^\text{em}_a+i\tilde{\phi}^\text{atm}_a(\bold{x},t)}+Be^{-i k (r_b+ \bold{x}\cdot \hat\bth_b)+i\phi^\text{em}_b+i\tilde{\phi}^\text{atm}_b(\bold{x},t)}},
    \end{split}
\end{equation}
where $A$ and $B$ are the amplitudes of the electric field of each star on the telescope, $\bold{r}_i$ is the physical location of star $i$ on the sky, $\bx$ a 2D vector on a plane parallel to the telescope's aperture (denoted $\vect{b}$ in \Sec{sec:theory}), centered at the vertex of the parabolic mirror, and $\hat{\bth}_i\equiv \bold{r}_i/r_i$ the unit vector pointing to the position of star $i$. Because $\bold{x}$ is a 2D vector, we can trade $\hat{\bth}_a$ for its component parallel to $\bx$, which we denote $\bth_a$ without the hat. Clearly the telescope is pointed such that $|\bth_a|\ll 1$. The phase $\phi_i^\text{em}$ is the random phase of the emitter $i$ and $\tilde{\phi}^\text{atm}_i(\bold{x},t)$ is the space- and time-dependent phase the wave from emitter $i$ inherits from passing through turbulent layers of the atmosphere. In the second line we neglected wavefront curvature effects $\sim k |\bx|^2/r\ll 1$. In what follows we will also drop the explicit time-dependence $e^{i\omega t}$.

We take the primary mirror to have diameter $D$ and centered at $\bx=0$, with a surface given by
\begin{equation}
z_\text{primary}(\bold{x})=\frac{|\bold{x}|^2}{4f}+\frac{1}{2}\tilde{z}^\text{inst}(\bold{x}), \quad |\bx|\in[0,D],
\label{eq:surface_primary}
\end{equation}
where the first term corresponds to an ideal spherical paraboloid of focal length $f$ and $\tilde{z}^\text{inst}$ describes its deformations, which induce \textit{instrumental} (as opposed to atmospheric) aberrations on the wavefront. We will assume that their effect is achromatic.

Using the Kirchhoff-Helmholtz diffraction integral~\cite{BornWolf}, the electric field from star $a$ at a distance $z\gg (D/2)^2/(4f)$ from the primary mirror will be given by
\begin{equation}
\begin{split}
E_{a,k} (\bold{x}',z) =& A\frac{-ik}{2\pi z}  \exp\parea{-ikr_a+i\phi^\text{em}_a +ik\pare{z+\frac{x'^2}{2z}}}\\
&\times\int\di^2 x \, \exp\left\{ik\parea{-\frac{x^2}{2}\pare{\frac{1}{f}-\frac{1}{z}}-\bold{x}\cdot\pare{\frac{\bold{x}'}{z}+\bth_a}+\tilde{z}^\text{inst}(\bold{x})}+i\tilde{\phi}_a^\text{atm}(\bold{x},t)\right\},
\end{split}
\end{equation}
where $\bx'$ is the 2D vector on a surface perpendicular to propagation. The spatial profile, as described by the second line, is intuitive: the first term describes smearing due to defocusing, which vanishes at the focal plane $z=f$, while the second term would give the usual diffraction limited electric field profile centered at $-\bth_a$ on the focal plane, in the absence of aberrations.

Analogous considerations apply for the secondary parabolic mirror that collimates the beam. We find at a distance $z$ from the vertex of the secondary mirror
\begin{equation}
\begin{split}
E_{a,k}^{\text{coll}} (\bold{x}',z) &= A\exp\parea{-ikr_a+i\phi^\text{em}_a +ik(z+f-f')}\frac{f}{f'}\frac{kF^{-1}(z)}{\pi}\\
&\int\di^2 x\exp\left\{ik \parea{F^{-1}(z)\pare{\bx-\frac{f}{f'}\bold{x}'}^2+ \bold{x}\cdot\bth_a+\tilde{z}^\text{inst}(\bold{x})}+i\tilde{\phi}_a^\text{atm}(\bx,t)\right\},
\end{split}
\label{eq:aftersecondary}
\end{equation}
where $F^{-1}(z)\equiv \frac{f'}{2f}\frac{1}{z\frac{f}{f'}+f-f'}$ and $\bx'$ is, again, the 2D vector on a surface perpendicular to propagation. In deriving this result we have assumed for simplicity that the secondary mirror is an ideal paraboloid, large enough to collect all the light from the primary and not induce any further diffraction effects. In what follows, we suppress the atmospheric phase for brevity, as it can formally be treated in the same way as instrumental aberrations. 

We wish to gain some analytic understanding of the electric field profile that exits the two-mirror system, in the presence of potentially large (compared to the wavelength) aberrations, i.e.~in the limit $k\tilde{z}^\text{inst}\gg1$. This can be done with the stationary phase approximation.

We define the exponent to be $g(\bx)\equiv \parea{F^{-1}(z)\pare{\bx-\frac{f}{f'}\bold{x}'}^2+ \bold{x}\cdot\bth_a+\tilde{z}^\text{inst}(\bold{x})}$. In the ideal mirror case $\tilde{z}^\text{inst}(\bold{x})=0$ and the equations $\partial_x g=0$ and $\partial_y g=0$ have the simple solutions (critical points) $x=x'f/f'$ and $y=y' f/f'$. Since $|\bx|\in[0,D/2]$, it follows that $|\bx'|\in[0,Df'/(2f)]$, so the collimated beam exists over a smaller spatial extent, given by the geometric optics width of the beam at the plane of the secondary mirror. 

We may then approximate 
\begin{equation}
    \int \di^2x\, \exp[ikg(\bold{x})]\simeq \begin{cases}
    \pi F(z)\exp\pare{ikf\bx'\cdot\boldsymbol{\theta}_a/f'}/k & \text{for~} |\bx|\in\parea{0,\frac{f}{f'}\frac{D}{2}}\\
    0 & \text{elsewhere}
    \end{cases},
\end{equation}
so that the collimated beam has modulus $|E(\bx',z)|=Af/f'$ and its extent is $|\bx'|\in[0,Df'/(2f)]$. The key takeaway is that, on the $\bx'$ plane, the angle $\bth_a$ appears magnified by the magnification $f/f'$, as expected from geometric optics. Due to the prefactor $f/f'$ of eq.~\ref{eq:aftersecondary}, the energy of the beam is conserved.

In the presence of mirror imperfections $\tilde{z}^\text{inst}$, critical points cannot always be found analytically, but we may work in the limit where they induce small corrections to the shape of the collimated beam. As an example, in this limit the critical points in the $x$ direction are
\begin{equation}
\begin{split}
x\simeq \frac{f}{f'}x'-\frac{F(z)}{2}\parea{\theta_{a}^{(x)}+\left.\frac{\partial \tilde{z}^\text{inst}}{\partial x}\right|_{x=\frac{f}{f'}x'}},
\end{split}
\end{equation}
where we took $\partial_x \tilde{z}^\text{inst}$ to be a perturbation on the zeroth order solution $x\simeq x^{(0)}=x'f/f'$ and $\theta_{a}^{(x)}$ is the $x$-coordinate of $\bth_a$. One can work out optical tolerances for the primary mirror due to this approximation, but they turn out to be less tight compared to those of section~\ref{app:aberrations} that concern pixel sizes. Thus, we do not discuss them here. 

Having established the above, we can approximate the collimated beam for both stars as
\begin{equation}
\begin{split}
E_{k,\text{coll}} (\bold{x}',z) &\simeq \frac{f}{f'} \exp\parea{ik(2(f-f')-z)}\exp\left\{ik \parea{\frac{F(d)}{4}\left.\pare{\nabla \tilde{z}^\text{inst}}^2\right|_{\bx=\frac{f}{f'}\bx'}+\tilde{z}^\text{inst}\pare{\frac{f}{f'}\bold{x'}}}\right\}
\\
&\bigg[A\exp\parea{-ikr_a+i\phi^\text{em}_a +i\tilde{\phi}_a^\text{atm}\pare{\frac{f}{f'}\bx',t}}\exp\left\{ik \parea{\frac{f}{f'}\bold{x'}\cdot\bth_a-\frac{F(d)\bth_a^2}{4}}\right\}\\
&+B\exp\parea{-ikr_b+i\phi^\text{em}_b +i\tilde{\phi}_b^\text{atm}\pare{\frac{f}{f'}\bx',t}}\exp\left\{ik \parea{\frac{f}{f'}\bold{x'}\cdot\bth_b-\frac{F(d)\bth_b^2}{4}}\right\}\bigg],
\end{split}
\label{eq:aftersecondary2}
\end{equation}
where $|\bx'|\in[0,Df'/(2f)]$, $z$ is the coordinate distance with the primary mirror's vertex being at $z=0$, and $d=f-f'-z$ is the (positive) distance from the secondary mirror. In deriving this expression we have dropped the dependence of $\tilde{z}^\text{inst}$ on $\bth_a,\,\bth_b$, as it is small. To simplify notation hereafter we will denote the full wavefront aberration imprinted by the primary-secondary mirror system as $\tilde{z}^\text{inst}\pare{f\bx'/f'}$, instead of the full exponent in the first line above, and drop the $\propto\bth^2$ terms.

The most important feature of our result, eq.~\ref{eq:aftersecondary2}, is the appearance of the magnification factor $f/f'$ multiplying the physical positions of the stars and the arguments of the aberration functions, both instrumental and atmospheric. This is not a surprising result: just as relative stellar separations $\bth_a-\bth_b$ are magnified by $f/f'$, so does the seeing angle due to the rescaled argument of $\tilde{\phi}^\text{atm}$, and the mirror-induced wavefront aberrations due to $\tilde{z}^\text{inst}$.

\subsection{The EPIC optical system}
\label{sec:time-delay}

The EPIC optical delay module has the form of a Mach-Zehnder interferometer of unequal path lengths, achieved by positioning the first beamsplitter at an angle $\gamma\neq45^\text{o}$, as shown in figure~\ref{fig:telescope_2} and described in section \ref{sec:theory}. This system's role is twofold: it both induces the geometric path delay that expands the field of view, and corrects for the wavefront tilt between the images of the two stars. The latter suppresses fringe contrast if uncorrected, an effect known since the advent of intensity interferometry~\cite{hbtII} and relevant when each telescope \emph{itself} resolves the angular separation being measured. 

In section~\ref{sec:beamsplitters} we study the induced phases by the multiple reflections within an idealized beamsplitter and technical requirements when operating it with broadband light. In section~\ref{sec:geometricdelay} we generalize eqs.~\ref{eq:ell} and~\ref{eq:ellopt2} for the geometric path delay and  wavefront correction for a general 2D relative stellar separation and derive expressions for the required two-dimensional wavefront correction angle. We conclude by commenting on tolerances relevant to the EPIC module.

\subsubsection*{Fourier decomposition}
The analysis in this section will be carried out with plane waves for clarity. This retains all generality because any wave can be decomposed into plane waves via a Fourier transform, and the effect of all optical elements is linear so that no two wavelengths mix. As such, any electric field $E(\bold{r})$, where $\bold{r}=(x,y,z)$, can be decomposed as
\begin{equation}
\begin{split}
E(\bold{r})\equiv \int\di^3p\, \tilde{E}(\bold{p})e^{-i \bold{p}\cdot \bold{r}},\quad
\tilde{E}(\bold{p})=\int\frac{\di^3 x}{(2\pi)^3}E(\mathbf{r})e^{+i\bold{p}\cdot\bold{r}}.
\end{split}
\end{equation}

Each Fourier component has a fixed direction and so it is easier to calculate how it transforms after propagation through the EPIC optical system of figure~\ref{fig:telescope_2}. Formally, the direction will be modified such that
\begin{equation}
\tilde{E}(\bold{p})e^{-i\bold{p}\cdot\bold{r}}\to \tilde{E}(\bold{p})e^{-i \bold{M}(\bold{r)}\cdot \bold{p} },
\label{eq:resumedE}
\end{equation}
for some function $\bold{M}(\bold{r})\equiv \pare{M_1(x,y,z),M_2(x,y,z),M_3(x,y,z)}$ that depends on the optical instruments the wave encounters.
We can resum these plane waves into $E'$ at some point after beam-recombination at the exit of the EPIC module so that
\begin{equation}
E'(x,y,z)=E(M_1, M_2,M_3),
\end{equation}
which therefore corresponds to translations and distortions of the shape of the incoming collimated beam.

\subsubsection*{Beamsplitters}
\label{sec:beamsplitters}

Here we study an idealized beamsplitter, consisting of a single slab of material with index of refraction $n_2$. Though this study is necessarily incomplete, we wish to highlight some non-trivial requirements for these optical elements when operated with broadband light. Our main result is that, when accounting for multiple reflections, there are irreducible phases that may extinguish the intensity of certain wavelengths. We defer more complete modeling to future work.

Let a ray propagate in a medium of index of refraction $n_1$ (air) and be incident on a semi-transparent plate of index of refraction $n_2$, at an angle $\gamma$. If the angle of refraction is $\theta$, such that $n_1\sin\gamma=n_2\sin\theta$ by Snell's law, the reflection and transmission coefficients at each surface of the plate for the parallel\footnote{We will neglect polarization effects throughout the discussion here and derive expressions only for the parallel component.} component of the electric field are~\cite{BornWolf}
\begin{equation}
       \text{$n_1\to n_2$:}\quad t=\frac{2n_1\cos\gamma}{n_2\cos\gamma+n_1\cos\theta},\quad\quad r=\frac{n_2\cos\gamma-n_1\cos\theta}{n_2\cos\gamma+n_1\cos\theta}
\end{equation}
\begin{equation}
   \text{$n_2\to n_1$:}\quad
        t'=\frac{2n_2\cos\theta}{n_1\cos\theta+n_2\cos\gamma}=\frac{n_2\cos\theta}{n_1\cos\gamma}t,\quad\quad r'=\frac{n_1\cos\theta-n_2\cos\gamma}{n_1\cos\theta+n_2\cos\gamma}=-r,
\end{equation}
where $r^2+tt'=1$, but $r^2+t^2\neq 1$. Naively, the transmitted beam suffers a $tt'$ suppression, compared to the reflected which suffers only an $r$ suppression. Consistency with energy conservation for elements with similar $r$ and $t$ requires taking into account multiple reflections within the beamsplitter~\cite{BornWolf}. If $E^{(i)}$ is the amplitude of the incident field, then the reflected and transmitted amplitudes are, respectively
\begin{equation}
    \begin{split}
        E^{(r)}=\sqrt{\mathbb{R}}\frac{1-e^{i\delta}}{1-\mathbb{R}e^{i\delta}}E^{(i)},\quad \quad
        E^{(t)}=\frac{\mathbb{T}e^{i\Phi}}{1-\mathbb{R}e^{i\delta}}E^{(i)},
    \end{split}
\end{equation}
where $\mathbb{R}\equiv r^2$ and $\mathbb{T}\equiv tt'$, so that $\mathbb{R+T}=1$, and the phases accumulated are $\delta=2k\ell n_2\cos\theta$, where $\lb$ is the width of the beamsplitter, and $\Phi=k\lb(n_2\cos\theta-n_1\cos\gamma)$. 

The reflected and transmitted intensities are
\begin{equation}
    \begin{split}
        \left|\frac{I^{(r)}}{I^{(i)}}\right|=\frac{F\sin^2\frac{\delta}{2}}{1+F\sin^2\frac{\delta}{2}}\equiv |\tilde{\mathbb{R}}|,\quad\quad
        \left|\frac{I^{(t)}}{I^{(i)}}\right|=\frac{1}{1+F\sin^2\frac{\delta}{2}}\equiv|\tilde{\mathbb{T}}|,
    \end{split}
    \label{eq:int_beamsplitter}
\end{equation}
where $F\equiv\frac{4\mathbb{R}}{(1-\mathbb{R})^2}$, and $|\tilde{\mathbb{R}}|+|\tilde{\mathbb{T}}|=1$ recovers energy conservation. In the main text we denoted $|\mathbb{\tilde{T}}|\equiv\epsilon^2$ for notational simplicity. The intensity of the recombined beam will scale as $|\tilde{\mathbb{R}}|^2|\tilde{\mathbb{T}}|^2=\parea{|\tilde{\mathbb{R}}|(1-|\tilde{\mathbb{R}}|)}^2$, which is maximized for $|\tilde{\mathbb{R}}|=1/2$ as shown in the main text.

\begin{figure}[h]
\begin{subfigure}[b]{0.5\textwidth}
    \centering
    \includegraphics[width=1\textwidth]{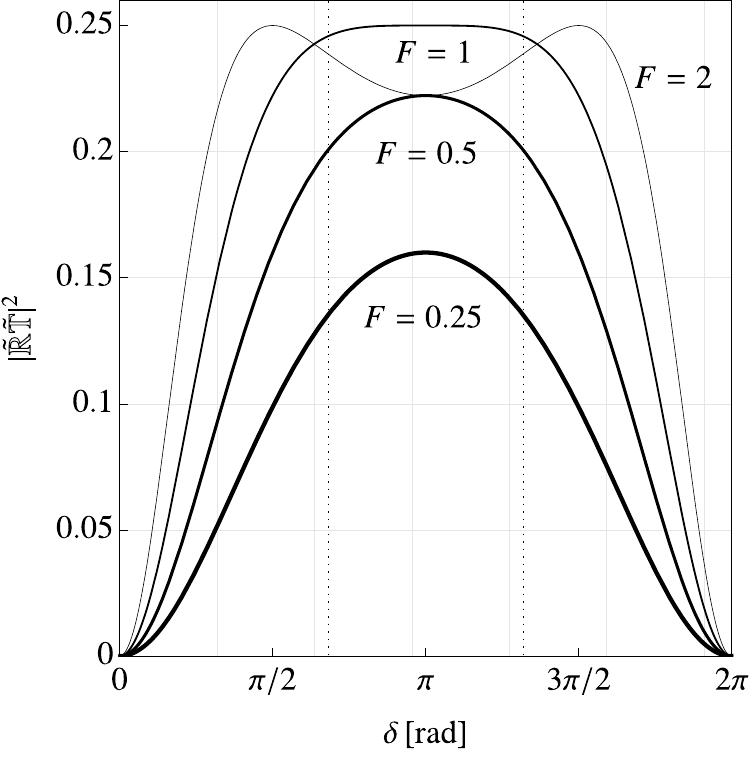}
\end{subfigure}
\begin{subfigure}[b]{0.5\textwidth}
    \centering
    \includegraphics[width=1\textwidth]{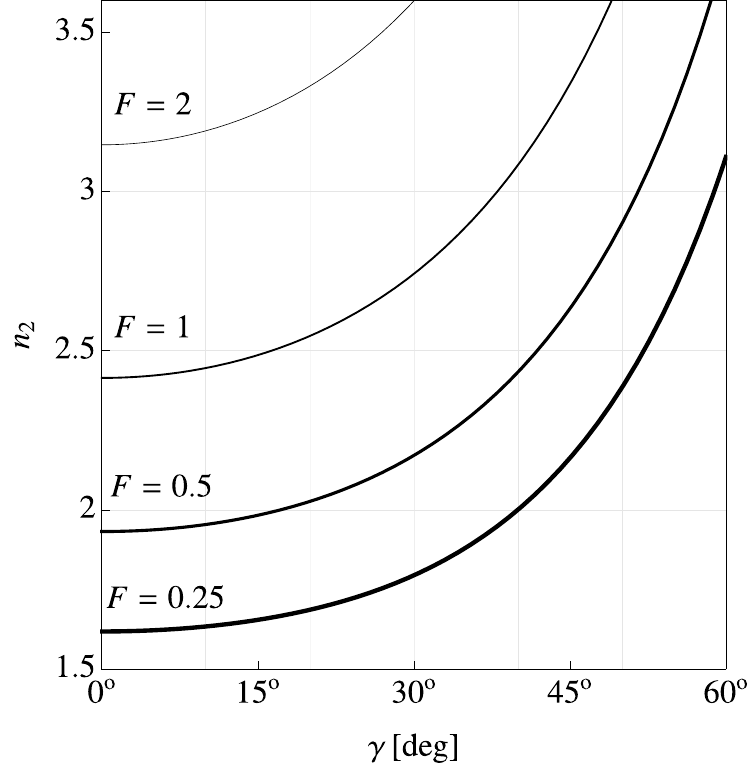}
\end{subfigure}
    \caption{\textit{Left:} Intensity of the main EPIC fringe normalized to total incident intensity, when accounting for multiple reflections within each beamsplitter, as a function of the phase $\delta=2k\ell_\text{b}n_2\cos\theta$ accumulated after one crossing of the beamsplitter. 
The product of the reflection and transmission coefficients plotted on the vertical axis is another definition of the quantity $\epsilon^2(1-\epsilon^2)$ used in e.g. eq.~\ref{eq:I2}. The lines, in order of diminishing thickness, correspond to increasing values of the ratio $F\equiv 4\mathbb{R}/(1-\mathbb{R})^2$, as defined below eq.~\ref{eq:int_beamsplitter}. The vertical dotted lines correspond to the angles $\pi\pm1$, within which there is a coverage of a factor of 2 in wavelength. An $F=1$ beamsplitter has optimal response to broadband light, as proven in the text. \textit{Right:} Required index of refraction $n_2$ as a function of the incidence angle $\gamma$, for different values of $F$. The index of refraction of air is $n_1=1$ and of the beamsplitter $n_2$. The thickness grading is the same as in the left figure.}
\label{fig:beamsplitter}
\end{figure}

However, the intensity of the reflected and transmitted beams depends on the wavelength through the phase $\delta$, so that $|\tilde{\mathbb{R}}|=1/2$ can strictly be achieved for only one wavelength. In order to maximize the intensity in as broad a range of wavelengths as possible, we require both $|\tilde{\mathbb{R}}|^2|\tilde{\mathbb{T}}|^2=1/4$ at a particular wavelength, and that the second derivative $\frac{\di^2}{\di\delta^2}|\tilde{\mathbb{R}}|^2|\tilde{\mathbb{T}}|^2$ vanishes, so that the change around the preferred $\delta$ is as ``slow'' as possible.  This can be achieved formally for $F=1$, since the second derivative scales as $\propto (1-F)$. As seen in the left panel of figure~\ref{fig:beamsplitter}, for $F=1$ the final intensity is equal to $1/4$ for a factor of 2 in wavelength, but decreases significantly outside this range. If we choose $\delta=\pi$ to correspond to the central wavelength we are recording, e.g. $\delta(450\text{nm})=\pi$, then we can cover the region $300-600$nm with intensity loss $\lesssim 1.7\%$.

Three comments are in order in light of the the above results. First, and most importantly, the beamsplitter will be large compared to the wavelength, so that $\ell_\text{b}\gg \lambda$. Figure~\ref{fig:beamsplitter} (left) is periodic in $\delta$ with period $2\pi$, so we can choose in general $\delta(450\text{nm})=(2m+1)\pi$, $m\in\mathbb{N}$, and then the width of the plate needs to be 
\begin{equation}
    \lb=\frac{2m+1}{n_2\sqrt{1-\frac{n_1^2}{n_2^2}\sin^2\gamma}}\frac{\lambda}{4},\quad m\in\mathbb{N},
\end{equation}
where we have neglected the wavelength dependence of the index of refraction. If one chooses $m>1$ for $\lambda$, the above considerations no longer apply as wavelengths $\lambda'$ which are $\mathcal{O}(1)$ different from $\lambda$ (as needed for a broadband setup) will acquire a $\mathcal{O}(1)$ different phase $\delta(\lambda')$, which is not within $\pi$ of $\delta(\lambda)$: for instance $\lambda/2$ acquires a phase $m\pi/2$ which can be very much different from $m\pi-\pi/2$ for large $m$. The broadband response of wide beamsplitters will have to be determined experimentally and might entail some $\mathcal{O}(1)$ loss of light. In the limit $\ell_\text{b}\gg \lambda$, approximately $83\%$ of the spectrum will exit the beamsplitter with more than half its incident intensity and the total intensity lost will be at most $35\%$. Alternatively, by energy conservation, we can retain the full intensity if we made use of both readouts of the second beamsplitter.  

Second, the choice of $F$ imposes a specific combination of the incidence angle and the index of refraction, since $F$ is a proxy for the reflection coefficient. In particular, $ r^2=1-\frac{2}{F}\pare{\sqrt{1+F}-1},$ and so the required incidence angle as a function of the indices of refraction and $F$ is given by
\begin{equation}
    \cos\gamma = \parea{\frac{1-\frac{n_1^2}{n_2^2}}{\frac{n_2^2}{n_1^2}\pare{\frac{1-r}{1+r}}^2-\frac{n_1^2}{n_2^2}}}^{1/2}.
\end{equation}

On the right panel of figure~\ref{fig:beamsplitter} we are plotting the required index of refraction of the plate $n_2$ as a function of the incidence angle $\gamma$, for different values of $F$ and assuming $n_1=1$ for air. From this figure it is clear that, at least for this toy model, there might be a need for materials with large indices of refraction.   

Third, the above discussion concerned intensities, but it is also necessary to know the $\lambda$-dependent phase shift of the electric fields due to each beamsplitter, as it may induce smearing of the intensity fringe we are observing. We find
\begin{equation}
    \begin{split}
    E^{(r)}&=\pare{\frac{F\sin^2\frac{\delta}{2}}{1+F\sin^2\frac{\delta}{2}}}^{1/2}e^{i\Phi_{r}}E^{(i)},\quad \tan\Phi_r\equiv -\frac{(1-\mathbb{R})\sin\delta}{(1+\mathbb{R})(1-\cos\delta)}\\
     E^{(t)}&=\frac{e^{i(\Phi_{t,1}+\Phi_{t,2})}}{\pare{1+F\sin^2\frac{\delta}{2}}^{1/2}}E^{(i)},\quad \tan\Phi_{t,1}\equiv \frac{\mathbb{R}\sin\delta}{1-\mathbb{R}\cos\delta}\quad\text{and}\quad \Phi_{t,2}=\frac{\delta}{2}-k\ell_\text{b}n_1\cos\gamma.
    \end{split}
\end{equation}
Assuming that the beamsplitters are identical, the electric field that exits the EPIC module will have accumulated a phase $\Phi_\text{total}=\Phi_r+\Phi_{t,1}+\Phi_{t,2}$. For $m>1$ this phase oscillates rapidly for a finite range of $\lambda$ and may reduce fringe contrast. The strictest tolerance concerns the size of the beamsplitter, which needs to be such that $\sigma_k\ell_\text{b}\lesssim1$, so that fringe contrast is not lost within each pixel, and  must be known with sub-wavelength precision, so that the relative phase delay \textit{between} pixels can be accounted for. The former sets $\ell_\text{b}\sim$mm and the latter can be determined experimentally.

To summarize, in this section we have studied the broadband response of a beamsplitter made entirely of a single index of refraction. We have identified an irreducible phase contribution that may reduce the intensity of the beam exiting the EPIC module by a third at most, but can be recovered if both readouts of the EPIC module are used. Given our simplistic modeling, we anticipate that experimental work will be needed to optimize these optical elements.

\subsubsection*{Geometric delay and wavefront correction}
\label{sec:geometricdelay}

In this section, we derive the general expression for the geometric path delay in the case of a relative 2D wavefront angle. We derive the expressions for the required angles of the EPIC system's mirrors, and discuss their tolerances.

In order to achieve the wavefront correction described in section~\ref{sec:theory}, we can tune the upper mirror of the path-extension system at an angle $\gamma'\neq\gamma$, as shown in figure~\ref{fig:telescope_2}. Because the upper mirror might need to correct for a 2D angle, we must include a small angle $\delta'$ with the axis that in figure~\ref{fig:telescope_2} is perpendicular to the page. We define as $\ell_\perp$ the distance between the first beamsplitter and the mirror of the upper part of the EPIC delay system when $\gamma'=\gamma$ and $\delta'=0$. This is an unambiguous length convenient for this calculation. It can be shown that $w_p=\ell_\perp\sin2\gamma/\cos\gamma$, where $w_p$ is the length shown on figure~\ref{fig:telescope_2}. 
 
In the limit $|\gamma-\gamma'|\equiv|\varepsilon|\ll1$\footnote{Not to be confused with the transmission coefficient $\epsilon^2$ we defined in section~\ref{sec:theory}.} and $\delta'\ll 1$ we find for the $\bold{M}(\bold{r})$ function of eq.~\ref{eq:resumedE}
\begin{equation}
    \begin{split}
    M_1 &\simeq x+y\delta'\sin2\gamma+2z(\gamma-\gamma')-2\ell_\perp\parea{(\gamma-\gamma')\cos\gamma+\sin\gamma}\\
    M_2 &\simeq -x\delta'\sin2\gamma+ y -z\delta'(1+\cos2\gamma)+2\ell_\perp\delta'\cos\gamma\\
    M_3 &\simeq -2(\gamma-\gamma')x+y\delta'(1+\cos2\gamma)+z-2\ell_\perp\parea{\cos\gamma-(\gamma-\gamma')\sin\gamma}.
    \end{split}
\end{equation}
These expressions have a simple geometric meaning: for a plane wave only in the $z$-direction and for $\gamma=\gamma'$ and $\delta'=0$, the center of the beam is shifted by $2\ell_\perp\sin\gamma$ in the $x$-direction, and gains a phase because of the geometric path between points with the same $z$-coordinate, $2\ell_\perp\cos\gamma$.

Neglecting instrumental and atmospheric aberrations, the recombined collimated beam exiting the second beamsplitter is then
\begin{equation}
\begin{split}
E_{k}^{\text{recomb}}(\bx,z)\simeq&\tilde{\mathbb{R}}\tilde{\mathbb{T}}\frac{f}{f'}\exp\parea{ik(2(f-f')-z)}\\
&\bigg[A e^{-ikr_a+i\phi^\text{em}_a}e^{ikx(2\varepsilon+\frac{f}{f'}\theta_{a,x})}e^{iky(-2\delta'\cos^2\gamma+\frac{f}{f'}\theta_{a,y})}e^{2ik\ell_\perp(\cos\gamma-\varepsilon\sin\gamma-\frac{f}{f'}\theta_{a,x}\sin\gamma)}\\
&+B e^{-ikr_b+i\phi^\text{em}_b}e^{ikx(2\varepsilon+\frac{f}{f'}\theta_{b,x})}e^{iky(-2\delta'\cos^2\gamma+\frac{f}{f'}\theta_{b,y})}e^{2ik\ell_\perp(\cos\gamma-\varepsilon\sin\gamma-\frac{f}{f'}\theta_{b,x}\sin\gamma)}\bigg]\\
&+e^{2ik\ell_\perp(\sin\gamma+\varepsilon\cos\gamma)}\bigg[A e^{-ikr_a+i\phi^\text{em}_a} e^{ik\frac{f}{f'}\bx\cdot\bth_a}e^{-2ik\ell_\perp\frac{f}{f'}\theta_{a,x}\sin\gamma}\\
&\hspace{3.5cm}+B e^{-ikr_b+i\phi^\text{em}_b} e^{ik\frac{f}{f'}\bx\cdot\bth_b}e^{-2ik\ell_\perp\frac{f}{f'}\theta_{b,x}\sin\gamma}\bigg],
\end{split}
\label{eq:delayed}
\end{equation}
where we have expanded the exponents to first order in $\varepsilon$ and $\delta'$. The overall coefficient $\tilde{\mathbb{R}}\tilde{\mathbb{T}}$ has to do with the beamsplitters and expressions for a simple model can be found in section~\ref{sec:beamsplitters}. We assume that these are identical for both beamsplitters.

The principal EPIC fringe arises from the terms in the third and fourth line of eq.~\ref{eq:delayed}. The relative phase between these two terms is
\begin{equation}
\Delta\phi = ik\parea{x\pare{2\epsilon+\frac{f}{f'}\theta_{ba,x}}+y\pare{-2\delta\cos^2\gamma+\frac{f}{f'}\theta_{ba,y}}}+ik\ell_p
\end{equation}
where we defined $\bth_{ba}=(\theta_{ba,x},\theta_{ba,y})\equiv \bth_b-\bth_a$ and $\ell_p$ is the geometric delay, the tuning of which allows for the expansion of the field of view (see section~\ref{sec:epic}). It is given by
\begin{equation}
    \ell_p=\frac{\ell_\perp\cos(\gamma-\gamma')}{\cos\gamma'}\parea{\cos2\gamma+\cos2(\gamma-\gamma')-\sin2\gamma-\sin2(\gamma-\gamma')}-2\ell_\perp\frac{f}{f'}\theta_{{ba,x}}\sin\gamma,
    \label{eq:ellp_general}
\end{equation}
where the last term $\propto\theta_{ba,x}$ is in fact $\propto\varepsilon$, so this expression reduces to eq.~\ref{eq:ell} for $\gamma=\gamma'$. The wavefront correction for large apertures and large stellar separations amounts to canceling the $x$- and $y$- dependence of $\Delta\phi$ by choosing
\begin{equation}
\varepsilon\equiv \gamma-\gamma'=-\frac{f}{2f'}\theta_{ba,x}\quad \text{and}\quad \delta' = \frac{f}{2f'\cos^2\gamma}\theta_{{ba,y}}.
\end{equation}

\subsubsection*{Wavefront correction and telescope pointing tolerances}

The residual angle $\varphi$ from the aforementioned tuning has to be within the diffraction limit of the telescope, so that $\varphi\lesssim (kD/2)^{-1}\simeq  8\,$mas at $\lambda=500\,$nm and for a $4\,$m diameter primary. The angle has to be both known and tuned to within this precision. The former is satisfied by the first step of the source localization protocol outlined in~\ref{sec:design}, while the latter sets an instrumental tolerance which should be achievable.

Furthermore, we have found a term proportional to (a component of) the magnified relative angle of the two stars in eq.~\ref{eq:ellp_general}. We remind the reader that $\ell_p$ must be tuned close to $\bold{d}\cdot\bth$, where $\bold{d}$ is the baseline vector and $\bth$ the relative stellar separation, to within a wavelength, as discussed in sections~\ref{sec:epic} and in particular~\ref{sec:design} and~\ref{sec:observation}. Even though the magnification factor $f/f'$ can be large, $\ell_\perp$ will only be on the order of tens of cm. As a result, this factor does not induce any stricter tolerances than what we have discussed so far.

Finally, we factored out an overall $\bx$ dependent factor $\exp\pare{ik f\bx\cdot\bth_a/f'}$. This factor includes possible telescope pointing errors and, although similar is form to the relative wavefront tilt, it does \emph{not} lead to fringe suppression. Pointing of the telescope affects the light paths of both stars in the same way, so, in fact, it is largely irrelevant: all it does is shift the overall position of the image on the photodetector array, without any suppression of correlation. The only tolerance on pointing is that any error remains within the angle subtended by each pixel. Seeing fluctuations demand that this angle be at least $1$ arcsec, as we will see in section~\ref{app:grating}, and stable pointing with such a precision is achievable even with commercial star-guiding systems for amateur astronomy.

\subsection{Optical aberrations}
\label{app:aberrations}

In this section, we derive tolerance conditions for our optical systems. We first sketch a proof of the well-known fact that a periodic grating produces an identical image of the source at every wavelength, so it is sufficient to neglect the dispersion of the grating and, rather, study the effects of aberrations on a single image of a point-like source. We derive tolerances on primary aberrations analytically and show that the requirements for intensity interferometry in general, and for EPIC in particular, can be orders of magnitude looser than those of traditional imagers and amplitude interferometers. Towards the end, we revisit atmospheric aberrations, which we studied in sections~\ref{sec:atm} and~\ref{app:atm}, this time focusing on the size of the finite seeing disk on the pixel array. We defer a complete numerical study to future work.

\subsubsection*{Multiple image formation by a diffraction grating}

The presence of the diffraction grating complicates the analytic understanding of the 2D image formation at the focal plane in the presence of optical aberrations. Fortunately, we can leverage the well-known fact of spectroscopy that a grating produces a copy of the image at every wavelength and the angular separation of these images on the focal plane is the same as that of the aberration-free images of point-like sources. 

We may consider, for simplicity, a 1D grating of periodicity $d_\text{g}$ and linear size along the ``spectroscopic direction'' $w=Nd_\text{g}$, where $N$ is the number of illuminated slits
of width $s<d$, and a wave $E_k\propto \exp\lbrace ik[z+\tilde{z}(x)]\rbrace$ at normal incidence with wavefront aberrations $\tilde{z}$. Here, $z$ is the direction of propagation and $x$ is the transverse coordinate, namely the one on the plane of the grating. The outgoing light is focused on the photodetector array using a camera of focal length $\fc$. The amplitude of the wave $E_k$ at the position $\theta_\text{F}=x_\text{F}/\fc$ on the focal plane is
\begin{equation}
\begin{split}
E_k(\theta_\text{F})\propto&\sum_{n=-N/2+1}^{N/2}\int_{(2n-1)\frac{d_\text{g}}{2}-\frac{s}{2}}^{(2n-1)\frac{d_\text{g}}{2}+\frac{s}{2}}\di \xi\, e^{-ik\xi\theta_\text{F}} e^{ik\tilde{z}(\xi)}\\
 &=\sum_n e^{ik\theta_\text{F}(2n-1)d_\text{g}/2}\int_{-\frac{s}{2}}^{\frac{s}{2}} \di \xi'\,e^{-ik\parea{\xi'\theta_\text{F}-\tilde{z}\pare{\xi'+(2n-1)d_\text{g}/2}}}.
 \end{split}
\end{equation}
Because $d>s$ we can ignore the $\xi'$ dependence in the argument of $\tilde{z}$ to get
\begin{equation}
\begin{split}
E_k(\theta_\text{F})\propto&s\frac{\sin\pare{k\theta_\text{F}\frac{s}{2}}}{k\theta_\text{F}\frac{s}{2}}\sum_n e^{ik\theta_\text{F}(2n-1)d_\text{g}/2}e^{ik\tilde{z}\pare{(2n-1)d_\text{g}/2}}.
 \end{split}
\end{equation}
Neglecting the single-slit prefactor, which is effectively constant for $\theta_\text{F}\ll 1/(ks)$, the remainder is periodic in $\theta_\text{F}$ with period $T=4\pi/(kd_\text{g})$. Its modulus has half that period, i.e $|E(\theta_\text{F}+mT/2)|=|E(\theta_\text{F})|$. Thus the profile centered around any diffraction order $m\in\mathbb{Z}$ is the same. 

Within a period, we can take the limit $d_\text{g}\to 0$. 
Then we can approximate $(2n-1)d_\text{g}/2\simeq x$, such that $x\in[-Nd_\text{g}/2,Nd_\text{g}/2]=[-w/2,w/2]$, for $N\gg 1$. We specialize to the first diffraction peak $m=1$, to find
\begin{equation}
E_k(\theta_\text{F})\propto\int_{-w}^{w} \di x\,e^{-ikx\pare{\theta_\text{F}-\frac{2\pi}{kd_\text{g}}}}e^{ik\tilde{z}(x)}.
\label{eq:image_grating}
\end{equation}

This is the image of the source in the focal plane as seen through an aperture of width $w$ which imprints aberrations $\tilde{z}$. The salient feature of this expression is that the angle where each wavelength is focused at, $2\pi/(kd_\text{g})$, depends on the wavelength as one expects from a dispersion grating, but the functional form of the wave profile is the same around that point. Prefactors can be determined from energy conservation.

The argument presented here disentangles the effects of aberrations from the dispersion of the grating and allows us to consider only the aberrated intensity distribution of a monochromatic wave in what follows.

\subsubsection*{Tolerances on primary aberrations}

\begin{figure}[h]
\centering
        \includegraphics[width=0.7\linewidth]{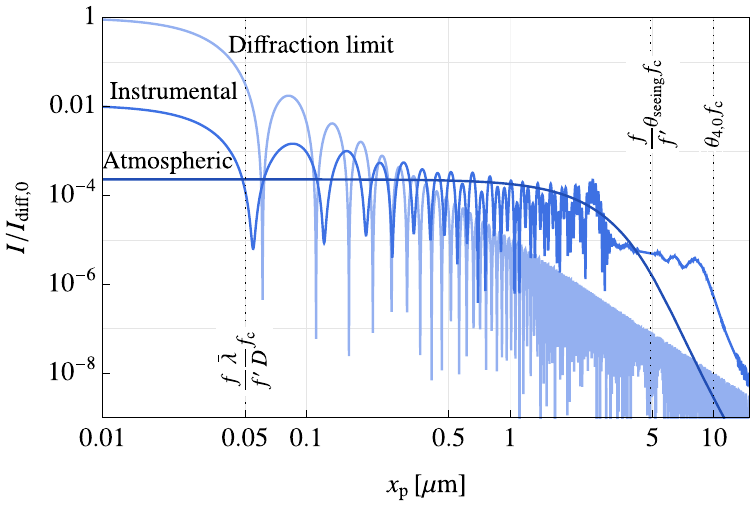}
          \caption{Intensity plotted as a function of the physical pixel coordinate, where have normalized the intensities to the peak of the diffraction limited image. All curves correspond to the same total intensity. We have chosen a $D=10\,\mathrm{m}$ primary, $f/f'=10$, $\fc=100\,$mm at $\lambda=500\,$nm. The lightest blue line corresponds to the diffraction limited image of a point source (Airy pattern), whose characteristic (magnified) diffraction limit $(f/f')(\bar{\lambda/D})\fc$ we have indicated with the leftmost vertical dotted line. The line labelled ``Instrumental'' corresponds to the produced image in a system with primary spherical aberration $\alpha_{4}'=50$. We have indicated the boundary of the region within which most of the intensity resides as $\theta_{4,0}\fc$, which agrees with the analytic estimate described in the main text, see e.g.~\ref{eq:primary_ab}. The line titled ``Atmospheric'' is the time-averaged image of a point source after accounting for atmospheric fluctuations, as described by eq.~\ref{eq:seeing_intensity}. We have chosen a Fried parameter $r_0=10$cm, which corresponds to a $1$as seeing disk at $500$nm, which we have indicated with the middlemost dotted line. The Mar\'{e}chal criterion for a diffraction limited system imposes that the maximum intensity of the distorted image (normalized to the diffraction limited peak) be $\geq0.8$, which shows how much more distortion EPIC can tolerate.}
    \label{fig:sphere_aber}
\end{figure}

Intensity interferometry is not sensitive to the precise functional form of the intensity distribution of an image. Instead, the only requirement is that all the light is collected by the photodetector. In the simplest setup, we allocate one pixel per $\sigma_k$ wavenumbers, so, according to the proof of the previous subsection, we require the spread of a single image to be within this one pixel. We provide a general analytic framework for deriving the spread of the distorted image and calculate tolerances on Seidel aberrations. Our discussion follows ref.~\cite{BornWolf} and is technical. We refer the reader interested in how (primary) tolerances for EPIC compare to those of regular diffraction-limited telescopes to figure~\ref{fig:sphere_aber} and especially table~\ref{tab:seidel}.

Let $\bx=(x,y)$ be the 2D cartesian vector on the plane of the photodetector, which lies at $z=z_\text{p}$ from a lens of width $w=Nd_\text{g}$ and focal length $\fc$ centered at $(\bx,z)=(\bold{0},0)$.\footnote{The lens here is a proxy for the grating. In the case of a grating there is the additional distance from the grating to the camera mirror/lens.} The width of the lens is related to the aperture of the primary by the magnification factor $w\equiv D f'/f$, as we have proven in section~\ref{sec:collimation}, e.g. see eq.~\ref{eq:aftersecondary2}. We define the corresponding polar coordinates $(r,\phi)$ for the focal plane and $(\rho,\theta)$ for the lens. The wave at the plane of the photodetector will be, according to eq.~\ref{eq:aftersecondary2}:
\begin{equation}
E_k(r,\phi)\propto\int_0^\frac{w}{2}\di \rho\int_0^{2\pi}\di\theta \,\rho\exp\parea{ik\pare{\frac{\rho^2}{2}\pare{\frac{1}{z_\text{p}}-\frac{1}{\fc}}-\frac{r}{z_\text{p}}\rho\cos(\theta-\phi)+\tilde{z}\pare{\frac{f}{f'}\rho,\theta}}}.
\label{eq:image_ab}
\end{equation}
The terms proportional to $\rho$ and $\rho^2$ in the exponent are known as \textit{distortion} and \textit{curvature of field}, respectively~\cite{BornWolf}. Both shift the 3D position of the Gaussian focus\footnote{We remind the reader that this corresponds to the position of maximum intensity for a point-like source, i.e.~the center of the Airy pattern. Distortion displaces this point horizontally, within the focal plane where the Airy pattern is, and curvature of field vertically compared to said plane.} and the new position of maximum intensity is referred to as the \textit{diffraction focus}. Crucially, if only such terms exist, they do not change the intensity profile \textit{around} the diffraction focus compared to the profile around the Gaussian focus for an aberration-free system. Therefore, we do not need to set tolerances on these.

It is well known that for a polynomial aberration of the form $\tilde{z}\propto \rho^n$, one can add aberrations of a lower order to maximize the intensity at the Gaussian focus~\cite{BornWolf}. Because different aberrations shift the diffraction focus by different amounts, it is convenient to expand the aberration function in polynomials that have maximum intensity at the same point and, in particular, at the Gaussian focus. We can directly compare the spread of the image with the aberration-free Airy pattern. This expansion is standard and is done in terms of Zernike polynomials, so that 
\begin{equation}
\phiab\pare{\frac{f}{f'}\rho,\theta}=\sum_{l,n,m} \epsilon_{nm}\alpha_{lnm}\bar{\lambda} R^m_n\pare{\frac{\rho}{w/2}}\cos m\theta,
\end{equation}
where $n\geq m$, $n-m$ is even, $\epsilon_{nm}=1/\sqrt{2}$ if $m=0$, $n\neq0$ and $1$ otherwise, and $\alpha_{nlm}$ are dimensionless constants. Because EPIC uses a broadband setup, it is useful to extract a reference length scale $\bar{\lambda}$, which we choose to be $500$nm. The polynomials $R_n^m$ can be found in standard references \cite{BornWolf}. Terms with $2l+m+n=4$ correspond to Seidel or primary aberrations.

If the aberration is small and the induced phase $k\tilde{z}\lesssim1$, tolerances are derived in standard references~\cite{BornWolf}. If this phase is large, which can be the case for intensity interferometry, we can use the stationary phase approximation to evaluate the integrals of eq.~\ref{eq:image_ab}, by identifying the critical points of the exponent. In fact, the precise functional form is not needed, but rather only the area in which most of the intensity lies in. Critical points will exist only for a finite range of $\rho$ and one can prove analytically that the wave profile decays much faster outside this range. Our criterion is that \textit{the pixel must be at least as large as the area in which critical points exist}. We have verified this numerically, as can be seen in figure~\ref{fig:sphere_aber}. The intensity lost using this criterion is $\mathcal{O}(1\%)$, but looser requirements, such as the collection of half the intensity, will give parametrically the same tolerances.

As an illustrative example, we will now calculate the tolerance on the primary (Seidel) spherical aberration for the primary mirror.\footnote{Technically, Seidel aberrations correspond to simple monomials, in contrast to the Zernike polynomials. Because the Mar\'{e}chal criterion for a well-corrected system refers to the diffraction focus (that the rms wavefront error should be less than $\lambda/14$), it is simpler to first derive tolerances of the Zernike polynomial of the same order $n$ as the corresponding Seidel aberration, and then match the coefficients~\cite{BornWolf}.} Essentially identical calculations can be carried out for primary coma and astigmatism. We collect our results in table~\ref{tab:seidel}. 

For cylindrically symmetric Zernike polynomials $m=0$, the integral is independent of $\phi$ and depends only on the ratio $|\bth_\text{F}|\equiv r/\fc$, so that
\begin{equation}
E_k(|\bth_\text{F}|,\phi)=-\frac{ike^{ik\fc}}{\fc}\int_0^{w/2}\di \rho \,\rho J_0\pare{k|\bth_\text{F}|\rho}\exp\parea{ik\phiab\pare{\frac{f}{f'}\rho}}.
\end{equation}
Since the spreading due to aberrations is going to be larger than the diffraction limited spread $|\bth_\text{F}|\sim \lambda/w$,\footnote{The physical angle of a diffraction limited image is $\lambda/w$, where $w=f'D/f$. This is not in conflict with the fact that the diffraction limit of the telescope is $\lambda/D$ since any \textit{relative} angle on the sky will be magnified by $f/f'$.} we can use the asymptotic expansion of the Bessel function and then approximate the integral using the stationary phase approximation. 

The $(n,m)=(4,0)$ Zernike polynomial $R_4^0(x)=6x^4-6x^2+1$ corresponds to the primary spherical aberration $\tilde{z}(\rho)=\alpha_4'\bar{\lambda}(\rho/(w/2))^4$. Critical points exist only for $|\bth_F|\leq2(6/\sqrt{2})\alpha_{4,0}\bar{\lambda}/(w/2)\equiv\theta_{4,0}$. This we must match with the angle subtended by a pixel of diameter $\Delta x_\text{p}$ at Gaussian focus a distance $\fc$ from the camera lens. Rescaling $\alpha_4$ to correspond to the Seidel coefficient $\alpha'_4$, we find
\begin{equation}
\alpha'_4\lesssim 50 \estim{\Delta x_\text{p}}{20\mu\text{m}}\estim{100\text{mm}}{\fc}\estim{D}{10\text{m}}\estim{500\text{nm}}{\bar{\lambda}}\estim{10}{f/f'}.
\label{eq:primary_ab}
\end{equation}

This should be contrasted to the Mar\'{e}chal criterion (the rms wavefront error of a sphere centered at the diffraction focus should be smaller than $\lambda/14$ for a diffraction limited system) which for a primary spherical aberration yields $\alpha_4'\bar{\lambda}\lesssim 0.94\lambda$. For this choice of parameters, the rms wavefront error is $3.75\bar{\lambda}$, which corresponds to twice the tolerable surface rms error for the primary mirror\footnote{
The rms wavefront error $\Delta\tilde{z}$ of the aberration function $\tilde{z}$ is defined as 
    \begin{equation} 
    (\Delta\tilde{z})^2\equiv \frac{1}{\pi}\int_0^1\int_0^{2\pi}\pare{\tilde{z}-\avg{\tilde{z}}}^2\rho\,\di\rho\di\theta, \quad \text{where}\quad\avg{\tilde{z}}\equiv\frac{1}{\pi}\int_0^1\int_0^{2\pi}\tilde{z}\rho\,\di\rho\di\theta.
    \end{equation} 
    }. 
As such, even in the presence of small pixels, intensity interferometry can tolerate aberrations an order of magnitude larger than diffraction limited designs. Because the surface rms error is about a micron, rather than significantly sub-wavelength, many of the expensive and time-consuming mirror polishing processes could be skipped~\cite{bely2003design}.  Larger pixels, allocation of multiple pixels to a single aberrated image and smaller magnifications can all further loosen this tolerance.

\begin{table}[tp]
    \centering
    \renewcommand{\arraystretch}{1.5} 
    \begin{tabular}{l | c | c | c || c}
   \hline \hline

	Type of aberration & Form of $\tilde{z}(\rho,\theta)$ & Tolerance & \makecell{Wavefront \\ error (rms)} & \makecell{Mar\'{e}chal \\ criterion} \\
	\hline
	Spherical aberration  & $\alpha_4' \bar{\lambda} \pare{\frac{\rho}{D/2}}^4$ & $\alpha_4'\lesssim 50$ & $\sim 3.8\bar{\lambda}$ & $\alpha_4'\bar{\lambda}\lesssim 0.94\lambda$ \\
	Coma  & $\alpha_{3,1}' \bar{\lambda}\pare{\frac{\rho}{D/2}}^3\cos\theta$ & $\alpha_{3,1}'\lesssim 43$& $\sim5\bar{\lambda}$ &  $\alpha_{3,1}'\bar{\lambda}\lesssim 0.60\lambda$  \\
	Astigmatism  & $\alpha_{2,2}' \bar{\lambda}\pare{\frac{\rho}{D/2}}^2\cos^2\theta$ & $\alpha_{2,2}'\lesssim 100$ & $\sim20\bar{\lambda}$&  $\alpha_{2,2}'\bar{\lambda}\lesssim 0.35\lambda$    \\
	Curvature of field  & $\alpha_{2}\bar{\lambda}\pare{\frac{\rho}{D/2}}^2$ & - & - & - \\
	Distortion  & $\alpha_{1,1}\bar{\lambda}\pare{\frac{\rho}{D/2}}\cos\theta$ & - & - & -\\
    \hline\hline
    \end{tabular}
    \caption{Tolerances on Seidel aberrations assuming that all the intensity of a single spectral channel is recorded by a \textit{single} pixel. In the third column the tolerance is set by imposing that all the intensity (see text for a precise definition) falls within a pixel of linear size of $20\mu\,$m, for a telescope with $f/f'=10$, a $D=10\,$m primary and a $\fc=100\,$mm camera focal length, at the reference wavelength of $\bar{\lambda}=500\,$nm. The fourth column relates the tolerance to the rms wavefront error or, equivalently, to twice the rms surface error of the primary mirror. In the last column we give the corresponding Mar\'{e}chal criterion for a diffraction limited system from~\cite{BornWolf} for comparison. Larger pixels, allocation of more pixels per spectral channel and smaller magnifications can further alleviate these requirements.}
	\label{tab:seidel}
\end{table}

\subsubsection*{Atmospheric aberrations}
\label{app:atm_seeing}

The atmosphere is another source of image aberrations that broadens the area into which the light is focused on the photodetector. We have established in sections~\ref{sec:atm} and~\ref{app:atm} that sources within the isoplanatic angle suffer the same stochastic image variations from the atmosphere, so that the contrast of the main EPIC fringe is not diminished. Here we are concerned with the size of the distorted image over long timescales, which is known as the \textit{seeing} disk. As with the instrumental aberrations discussed above, the seeing disk should fit within each pixel.

Following the modeling of atmospheric fluctuations based on Kolmogorov turbulence, the intensity on the focal plane over long timescales is given by \cite{ThorneBlandford2017}
\begin{equation}
I(\bth_\text{F})\propto \int\di^2x\, e^{-ik\bth_\text{F}\cdot\bx} e^{-3.44\pare{\frac{f}{f'}\frac{|\bx|}{r_0}}^{5/3}},
\label{eq:seeing_intensity}
\end{equation}
where $r_0$ is the Fried parameter. The FWHM of the seeing disk is found to be 
$\theta_\text{seeing}\sim 0.98\lambda/r_0$, which is given by eq.~\ref{eq:thetaseeing}. Here the seeing angle is magnified at the plane of the photodetector by $f/f'$. The Fried parameter $r_0\sim 10-20$cm, which corresponds to a (unmagnified) seeing disk of $\sim 0.5-1$ arcsec for blue light. We plot the radial profile in figure~\ref{fig:sphere_aber}, to compare it with the spread due to instrumental aberrations. In particular, the size of the magnified seeing disk is the same as that of an image with primary spherical aberration for $\alpha_4'\sim25$.

\subsection{Dispersive element}
\label{app:grating}

In this section we fill in the details of the calculation presented in section ~\ref{sec:grating}, concerning the reduction of fringe contrast for large telescopes, large pixels and fast photodetectors. We stress that the presence of aberrations, instrumental or atmospheric, do not modify this effect. We thus study a diffraction limited system here for clarity.

The analysis here is one-dimensional and the optical configuration consists of a grating of size $Nd_\text{g}$, a camera lens of focal length $\fc$ at a distance $z_\text{l}$ from the grating, and a photodetector array at the focal plane of the lens. A plane monochromatic wave incident on the grating has a profile at the focal plane given by
\begin{equation}
E_k(\theta_\text{F},t)\propto A_ke^{ik(\fc+z_\text{l}-ct)}\exp\parea{ik\frac{x_\text{F}^2}{\fc+z_\text{l}}}\frac{\sin\pare{k\theta_\text{F}\frac{Nd}{2}}}{\sin\pare{k\theta_\text{F}\frac{d}{2}}}\frac{\sin\pare{k\frac{s}{2}\theta_\text{F}}}{k\frac{s}{2}\theta_\text{F}}e^{-ikct},
\label{eq:Ephotodetector_phase}
\end{equation}
where $\theta_\text{F}\equiv x_\text{F}/\fc$ is the ratio of the physical distance on the pixel plane over the camera focal length, $d_\text{g}$ is the periodicity of the grating and $s$ is the width of each slit. The factor $\fc+z_\text{l}$ is common for every wavelength and can be absorbed into a shift of the clock, so will be neglected below. Similarly, the quadratic in $x_\text{F}$ term is a path difference for rays with different wavelengths due to being focused at different points. We will neglect this for now but will comment on it towards the end. These phase factors were neglected in section~\ref{sec:grating} for brevity.

We are concerned with the spatial distribution of different wavelengths within a pixel, which comes about from the term of the form $\sin(Ny)/\sin(y)$. We specialize to the pixel in whose center the focused wavenumber is $k=\bar{k}$. All other wavelengths within the pixel will have wavenumbers $k\equiv\bar{k}+\sigma\in[\bar{k}-\frac{\sigma_k}{2},\bar{k}+\frac{\sigma_k}{2}]$, where $\sigma_k\sim\bar{k}/\mathcal{R}\ll \bar{k}$, $\mathcal{R}$ being the required resolution. The first diffraction peak of the wavelength $\lambda=2\pi/k$ occurs at $\theta_{k}=\frac{\lambda}{d_\text{g}}$. We can shift $\theta_\text{F}$ such that $\theta_\text{F}=\theta_\text{F}'-\frac{\bar{\lambda}}{d}$, bringing the pixel's center at $\theta_\text{F}'=0$, where $\bar{\lambda}$ is focused at. Then $k\theta_\text{F}\frac{Nd_\text{g}}{2}= k\theta_\text{F}'\frac{Nd_\text{g}}{2} + \pi N\frac{k}{\bar{k}}$ and similarly for the denominator. Now within the pixel of interest $\theta_\text{F}'\in\pare{-\frac{\bar{\lambda}}{2d_\text{g}\mathcal{R}},\frac{\bar{\lambda}}{2d_\text{g}\mathcal{R}}}$, so that $k\theta_\text{F}'d_\text{g}\ll 1$ and we can approximate the denominator as $\sin\pare{k\theta_\text{F}\frac{d_\text{g}}{2}}=-\sin(k\theta_\text{F}\frac{d_\text{g}}{2}-\pi)=-\sin\parea{k\theta_\text{F}'\frac{d_\text{g}}{2} + \pi\pare{ \frac{k}{\bar{k}}-1}}\simeq-\parea{k\theta_\text{F}'\frac{d_\text{g}}{2}+ \pi\pare{ \frac{k}{\bar{k}}-1}} $. Similarly the numerator is $-\sin\parea{k\theta_\text{F}'\frac{Nd_\text{g}}{2}+ \pi N\pare{ \frac{k}{\bar{k}}-1}}$. As proven in the beginning of section~\ref{app:aberrations}, the resulting profile is the image of the point source at the wavelength $k$, as if we had an aperture of size $Nd_\text{g}$ without the grating. Dropping the prime on $\theta_\text{F}'$ from here on, we recover eq.~\ref{eq:Ephotodetector2}:
\begin{equation}
E_k(\theta_\text{F},t)\propto A_k\frac{\sin\parea{k\theta_\text{F}\frac{Nd_\text{g}}{2}+ \pi N\pare{ \frac{k}{\bar{k}}-1}}}{k\theta_\text{F}\frac{Nd_\text{g}}{2}+ \pi N\pare{ \frac{k}{\bar{k}}-1}}e^{-i
k ct}.
\tag{\ref{eq:Ephotodetector2}}
\end{equation}
A straightforward calculation gives the numerator and denominator of the excess fractional intensity correlation
\begin{equation}
\begin{split}
\mathcal{N}\propto &\left<\int\di k\int\di k' |A_k|^2|A_{k'}|^2 e^{-i(k-k')ct}\right.\\
&\times\int\di\theta_\text{F}^{(1)}\,N\frac{\sin\parea{k\theta_\text{F}^{(1)}\frac{Nd_\text{g}}{2}+ \pi N\pare{ \frac{k}{\bar{k}}-1}}}{k\theta_\text{F}^{(1)}\frac{Nd_\text{g}}{2}+ \pi N\pare{ \frac{k}{\bar{k}}-1}}N\frac{\sin\parea{k'\theta_\text{F}^{(1)}\frac{Nd_\text{g}}{2}+ \pi N\pare{ \frac{k'}{\bar{k}}-1}}}{k'\theta_\text{F}^{(1)}\frac{Nd_\text{g}}{2}+ \pi N\pare{ \frac{k'}{\bar{k}}-1}}\\
&\left.\times\int\di\theta_\text{F}^{(2)}\,N\frac{\sin\parea{k\theta_\text{F}^{(2)}\frac{Nd_\text{g}}{2}+ \pi N\pare{ \frac{k}{\bar{k}}-1}}}{k\theta_\text{F}^{(2)}\frac{Nd_\text{g}}{2}+ \pi N\pare{ \frac{k}{\bar{k}}-1}}N\frac{\sin\parea{k'\theta_\text{F}^{(2)}\frac{Nd_\text{g}}{2}+ \pi N\pare{ \frac{k'}{\bar{k}}-1}}}{k'\theta_\text{F}^{(2)}\frac{Nd_\text{g}}{2}+ \pi N\pare{ \frac{k'}{\bar{k}}-1}}\right>_{\sigma_t}\\
\mathcal{D}\propto &\left|\int\di\theta_\text{F}\int\di k \,\left|A_k\frac{\sin\parea{k\theta_\text{F}\frac{Nd_\text{g}}{2}+ \pi N\pare{ \frac{k}{\bar{k}}-1}}}{k\theta_\text{F}\frac{Nd_\text{g}}{2}+ \pi N\pare{ \frac{k}{\bar{k}}-1}}\right|^2\right|^2,
\end{split}
\label{eq:numeratorapp}
\end{equation}
where $\mathcal{N}\equiv\avg{I_1 I_2}-\avg{I_1}\avg{I_2}$ and $\mathcal{D}\equiv\avg{I_1}\avg{I_2}$, as in eqs.~\ref{eq:gratingN} and~\ref{eq:gratingD}, and $\avg{...}_{\sigma_t}$ denotes averaging over time with a normalized gaussian kernel of standard deviation $\sigma_t$. We emphasize that the terms involving the focal plane of each telescope, $\theta_\text{F}^{(i)}$, $i=1,2$ contain \emph{different} wavenumber $k$ and $k'$ in the numerator. Consequently, if these are focused too far apart spatially, then this term averages out and fringe contrast can be lost.

There is no obstacle in evaluating these formulae numerically, but we can understand the source of the suppression better in the limit where the pixel is much larger than the width of these sinc functions (Airy patterns). Then we can take the integrations over $\theta_\text{F}^{(i)}$ to infinity in these expressions, which is a good approximation if there are several distinct ``Airy patterns'' within each pixel. We have also verified this fact numerically.

One can show that
\begin{equation}
\begin{split}
\int_{-\infty}^{+\infty}\di\theta_\text{F}^{(i)}\,&N\frac{\sin\parea{k\theta_\text{F}^{(i)}\frac{Nd_\text{g}}{2}+ \pi N\pare{ \frac{k}{\bar{k}}-1}}}{k\theta_\text{F}^{(i)}\frac{Nd_\text{g}}{2}+ \pi N\pare{ \frac{k}{\bar{k}}-1}}N\frac{\sin\parea{k'\theta_\text{F}^{(i)}\frac{Nd_\text{g}}{2}+ \pi N\pare{ \frac{k'}{\bar{k}}-1}}}{k'\theta_\text{F}^{(i)}\frac{Nd_\text{g}}{2}+ \pi N\pare{ \frac{k'}{\bar{k}}-1}}\\
&\simeq\frac{2\pi N}{\bar{k}d_\text{g}}\frac{\sin\pare{N\pi\frac{k-k'}{\bar{k}}}}{N\pi\frac{k-k'}{\bar{k}}}.
\end{split}
\end{equation}
This should be contrasted to the term $\avg{I_1}\avg{I_2}$, which involves integrals of the same wavelengths at each telescope, in which case $\avg{I_1}=\avg{I_2}\propto 2\pi N/{\bar{k}d_\text{g}}$. Neglecting this prefactor, we have
\begin{equation}
\begin{split}
\mathcal{N}\propto\avg{\int\di k\int\di k' |A_k|^2|A_{k'}|^2 e^{-i(k-k')ct}\parea{\frac{\sin\pare{N\pi\frac{k-k'}{\bar{k}}}}{N\pi\frac{k-k'}{\bar{k}}}}^2}_{\sigma_t}.
\end{split}
\end{equation}
This shows where the ``sinc-squared filter'' we encountered in eq.~\ref{eq:Grating_correlation} stems from. It restricts the integral only to those wavenumbers that are $\sim (N\pi)^{-1}$ close. Because $N\gg \mathcal{R}$ in order to accommodate optical aberrations but, most importantly, seeing fluctuations, this term, if uncorrected, will reduce the fringe contrast.

The starlight spectrum has a natural width $\sigma_n$ and a peak at $k_0$. By writing $|A_k|^2=\frac{1}{\sqrt{2\pi}\sigma_n} \exp\parea{-\frac{(k-k_0)^2}{2\sigma_n^2}}$, where $\sigma_n\sim k_0\gg\sigma_k$, we can approximate within the integral $|A_k|^2\simeq|A_{\bar{k}}|^2$, which is a constant and can be dropped in ratios. We find:
\begin{equation}
\begin{split}
\frac{\mathcal{N}}{\mathcal{D}}\simeq\avg{\int_{\bar{k}-\frac{\sigma_k}{2}}^{\bar{k}+\frac{\sigma_k}{2}}\di k\int_{\bar{k}-\frac{\sigma_k}{2}}^{\bar{k}+\frac{\sigma_k}{2}}\di k'\, e^{-i(k-k')ct}\parea{\frac{\sin\pare{N\pi\frac{k-k'}{\bar{k}}}}{N\pi\frac{k-k'}{\bar{k}}}}^2}_{\sigma_t}.
\end{split}
\end{equation}
In the limit of no dispersion (such as in the case where we used a filter of bandwidth $\sim\bar{k}/\mathcal{R}$ instead), this becomes
\begin{equation}
\avg{\int_{\bar{k}-\frac{\sigma_k}{2}}^{\bar{k}+\frac{\sigma_k}{2}}\di k\int_{\bar{k}-\frac{\sigma_k}{2}}^{\bar{k}+\frac{\sigma_k}{2}}\di k'\, e^{-i(k-k')ct}}_{\sigma_t}=\frac{1}{c^2\sigma_t^2}\parea{-2+2e^{-c^2\sigma_k^2\sigma_t^2/2}+\sqrt{2\pi}c\sigma_k\sigma_t\,\text{erf}\pare{\frac{c\sigma_k\sigma_t}{\sqrt{2}}}},
\end{equation}
which, in the limit $c\sigma_t\ll\sigma_k^{-1}$, along with the prefactor $\propto\sigma_n^{-2}$ of the $|A_k|^2$ terms, gives $\propto\pare{\frac{\sigma_k}{\sigma_n}}^2\frac{1}{c\sigma_k\sigma_t}$. The first quadratic term is a suppression coming from taking only part of the total spectrum within the pixel, and the second term is the usual suppression due to a finite time-resolution. This is the standard result used throughout section~\ref{sec:theory}.

The integral in the presence of dispersion cannot be calculated analytically, but we can bring it in a form that is more transparent. We define the function $h(\sigma)=e^{-i\sigma ct}\parea{\frac{\sin\pare{\sigma/\sigma_d}}{\sigma/\sigma_d}}^2$, where we defined $\sigma_d\equiv \bar{k}/{\pi N}$ as the ``diffraction limited'' width. By changing variables to $\sigma\equiv k+k'$ and $\sigma'\equiv k-k'$, the double integral can be made one-dimensional
\begin{equation}
\begin{split}
\avg{\int_{\bar{k}-\frac{\sigma_k}{2}}^{\bar{k}+\frac{\sigma_k}{2}}\di k\int_{\bar{k}-\frac{\sigma_k}{2}}^{\bar{k}+\frac{\sigma_k}{2}}\di k'\, e^{-i(k-k')ct}\parea{\frac{\sin\pare{N\pi\frac{k-k'}{\bar{k}}}}{N\pi\frac{k-k'}{\bar{k}}}}^2}_{\sigma_t}&=\avg{\int_0^{\sigma_k}\di\sigma\,(\sigma_k-\sigma)\parea{h(\sigma)+h(-\sigma)}}_{\sigma_t}\\
=2&\int_0^{\sigma_k}\di\sigma\,(\sigma_k-\sigma)e^{-\frac{c^2\sigma^2\sigma_t^2}{2}}\parea{\frac{\sin\pare{\sigma/\sigma_d}}{\sigma/\sigma_d}}^2,
\end{split}
\end{equation}
which concludes the proof of eq.~\ref{eq:Grating_correlation}. This function is plotted in figure~\ref{fig:grating_suppression} for relevant parameters of the different proposed phases of EPIC interferometers. We refer the reader to section~\ref{sec:grating} for the implications of these results and for potential ways to alleviate the resulting suppression.

We have carried out the same calculation in the presence of a prism instead of a grating. The result remains the same as that of \Eq{eq:numeratorapp}, with the only difference that the quantity $N$ in the $\theta_\text{F}$-independent term (the resolving power if the system was diffraction limited) is replaced by $N_\text{prism}\equiv b\, \di n/\di\lambda$, where $b$ is the length of the base of the prism and $\di n/\di \lambda$ the derivative of the index of refraction with respect to the wavelength. The requirement of including the seeing angle within each pixel is the same as in eq.~\ref{eq:grating_oversize}, but with $N\to N_\text{prism}$.

\bibliographystyle{unsrt}
\bibliography{biblio}

\end{document}